\documentclass[aps,prd,reprint,superscriptaddress,twocolumn,showpacs,showkeys,floatfix]{revtex4-2}
\usepackage{graphicx}
\usepackage{amsmath}
\usepackage{bm}
\usepackage{dcolumn}
\usepackage{multirow}
\usepackage{color}
\usepackage{mathtools}
\newcommand{\pder}[2]{\ensuremath{\frac{\partial{#1}}{\partial{#2}}}}
\newcommand{\pdder}[2]{\ensuremath{\frac{\partial^2{#1}}{\partial{#2}^2}}}
\newcommand{\Dg}{\ensuremath{\Delta g}}
\newcommand{\Dgphi}{\ensuremath{\Delta \gamma_{\phi}}}
\newcommand{\DDr}{\ensuremath{\Delta \left( \Delta_{14} \right)}}
\newcommand{\DDl}{\ensuremath{\Delta \left( \Delta_{23} \right)}}

\begin{document}

\title{NanoNewton electrostatic force actuators for femtoNewton-sensitive measurements: system performance test in the 
LISA Pathfinder mission}
% Authorlist PSD glitch 2022

\author{M~Armano}\affiliation{\addressa}
\author{H~Audley}\affiliation{\addressb}
\author{J~Baird}\affiliation{\addressca}
\author{M~Bassan}\affiliation{\addressgg}
\author{P~Binetruy}\thanks{Deceased.}\affiliation{\addressc} %30 March 2017
\author{M~Born}\affiliation{\addressb}
\author{D~Bortoluzzi}\affiliation{\addressf}
\author{E~Castelli}\affiliation{\addressu}
\affiliation{\addressi}
\author{A~Cavalleri}\affiliation{\addressk}
\author{A~Cesarini}\affiliation{\addresso}
\author{V\,Chiavegato}\thanks{Corresponding authors:}\affiliation{\addressi}
\author{A\,M~Cruise}\affiliation{\addressj}
\author{D\,Dal Bosco}\affiliation{\addressi}
\author{K~Danzmann}\affiliation{\addressb}
\author{M~De Deus Silva}\affiliation{\addressa}
\author{R~De Rosa}\affiliation{\addresshh}
\author{L~Di Fiore}\affiliation{\addressii}
\author{I~Diepholz}\affiliation{\addressb}
\author{G~Dixon}\affiliation{\addressj}
\author{R~Dolesi}\affiliation{\addressi}
\author{L~Ferraioli}\thanks{Corresponding authors:}\affiliation{\addressl}
\author{V~Ferroni}\affiliation{\addressi}
\author{E\,D~Fitzsimons}\affiliation{\addressm}
\author{M~Freschi}\affiliation{\addressa}
\author{L~Gesa}\thanks{Deceased.}\affiliation{\addressn} % 29 May 2020
\author{D~Giardini}\affiliation{\addressl}
\author{F~Gibert}\affiliation{\addressee}\affiliation{\addressi}
\author{R~Giusteri}\affiliation{\addressb}
\author{A~Grado}\affiliation{\addresskk}
\author{C~Grimani}\affiliation{\addresso}
\author{J~Grzymisch}\affiliation{\addressh}
\author{I~Harrison}\affiliation{\addressp}
\author{M\,S~Hartig}\affiliation{\addressb}
\author{G~Heinzel}\affiliation{\addressb}
\author{M~Hewitson}\affiliation{\addressb}
\author{D~Hollington}\affiliation{\addressd}
\author{D~Hoyland}\affiliation{\addressj}
\author{M~Hueller}\affiliation{\addressi}
\author{H~Inchausp\'e}\affiliation{\addresscb}\affiliation{\addressca}
\author{O~Jennrich}\affiliation{\addressh}
\author{P~Jetzer}\affiliation{\addressq}
\author{B~Johlander}\affiliation{\addressa}
\author{N~Karnesis}\affiliation{\addressbb}
\author{B~Kaune}\affiliation{\addressb}
\author{N~Korsakova}\affiliation{\addressca}
\author{C\,J~Killow}\affiliation{\addressr}
\author{L~Liu}\affiliation{\addressi}
\author{J\,A~Lobo}\thanks{Deceased.}\affiliation{\addressn} % 30 September 2012
\author{J\,P~L\'opez-Zaragoza}\affiliation{\addressn}
\author{R~Maarschalkerweerd}\affiliation{\addressp}
\author{D~Mance}\affiliation{\addressl}
\author{V~Mart\'{i}n}\affiliation{\addressn}
\author{L~Martin-Polo}\affiliation{\addressa}
\author{F~Martin-Porqueras}\affiliation{\addressa}
\author{J~Martino}\affiliation{\addressca}
\author{P\,W~McNamara}\affiliation{\addressh}
\author{J~Mendes}\affiliation{\addressp}
\author{L~Mendes}\affiliation{\addressa}
\author{N~Meshksar}\affiliation{\addressl}
\author{J~Moerschell}\affiliation{\addressff}
\author{M~Nofrarias}\affiliation{\addressn}
\author{S~Paczkowski}\affiliation{\addressb}
\author{M~Perreur-Lloyd}\affiliation{\addressr}
\author{A~Petiteau}\affiliation{\addressc}\affiliation{\addressca}
\author{E~Plagnol}\affiliation{\addressca}
\author{C~Praplan}\affiliation{\addressff}
\author{J~Ramos-Castro}\affiliation{\addresss}
\author{J~Reiche}\affiliation{\addressb}
\author{F~Rivas}\affiliation{\addresscc}\affiliation{\addressi}
\author{D\,I~Robertson}\affiliation{\addressr}
\author{G~Russano}\affiliation{\addressx}\affiliation{\addressi}
\author{L~Sala}\affiliation{\addressi}
\author{P~Sarra}\affiliation{\addressaa}
\author{S\,L~Schule-Walewski}\affiliation{\addressll}
\author{J~Slutsky}\affiliation{\addressu}
\author{C\,F~Sopuerta}\affiliation{\addressn}
\author{R~Stanga}\affiliation{\addressjj}
\author{T~Sumner}\affiliation{\addressd}
\author{J~ten Pierick}\affiliation{\addressl}
\author{D~Texier}\affiliation{\addressa}
\author{J\,I~Thorpe}\affiliation{\addressu}
\author{D~Vetrugno}\affiliation{\addressi}
\author{S~Vitale}\affiliation{\addressi}
\author{G~Wanner}\affiliation{\addressb}
\author{H~Ward}\affiliation{\addressr}
\author{P~Wass}\affiliation{\addressd}\affiliation{\addressdd}
\author{W\,J~Weber}\thanks{Corresponding authors:}\affiliation{\addressi}
\author{L~Wissel}\affiliation{\addressb}
\author{A~Wittchen}\affiliation{\addressb}
\author{C~Zanoni}\affiliation{\addressi}
\author{P~Zweifel}\affiliation{\addressl}

%\collaboration{LISA Pathfinder Collaboration}\email[Corresponding authors:\\  ]
%{vittorio.chiavegato@unitn.it\\luigi.ferraioli@erdw.ethz.ch\\williamjoseph.weber@unitn.it}
\collaboration{LISA Pathfinder Collaboration}\email[]
{vittorio.chiavegato@unitn.it\\luigi.ferraioli@erdw.ethz.ch\\williamjoseph.weber@unitn.it}

\def\addressa{European Space Astronomy Centre, European Space Agency, Villanueva de la
Ca\~{n}ada, 28692 Madrid, Spain}
\def\addressb{Albert-Einstein-Institut, Max-Planck-Institut f\"ur Gravitationsphysik und Leibniz Universit\"at Hannover,
Callinstra{\ss}e 38, 30167 Hannover, Germany}
\def\addressc{IRFU, CEA, Universit\'e Paris-Saclay, F-91191 Gif-sur-Yvette, France}
\def\addressca{Universit\'e Paris Cit\'e, CNRS, Astroparticule et Cosmologie, F-75013 Paris, France}
\def\addresscb{Institut f\"ur Theoretische Physik, Universit\"at Heidelberg, Philosophenweg 16, 69120 Heidelberg, Germany}
\def\addressd{Physics Department, Blackett Laboratory, High Energy Physics Group, Imperial College London, Prince Consort Road, London SW7 2BW, United Kingdom}
\def\addresse{Dipartimento di Fisica, Universit\`a di Roma ``Tor Vergata'',  and INFN, sezione Roma Tor Vergata, I-00133 Roma, Italy}
\def\addressf{Department of Industrial Engineering, Universit\`a di Trento and Trento Institute for 
Fundamental Physics and Application / INFN, I-38123 Povo, Trento, Italy}
\def\addressg{Airbus Defence and Space, Claude-Dornier-Strasse, 88090 Immenstaad, Germany}
\def\addressh{European Space Technology Centre, European Space Agency, 
Keplerlaan 1, 2200 AG Noordwijk, The Netherlands}
\def\addressi{Dipartimento di Fisica, Universit\`a di Trento and Trento Institute for 
Fundamental Physics and Application / INFN, I-38123 Povo, Trento, Italy}
\def\addressj{The School of Physics and Astronomy, University of
Birmingham, B15 2TT Birmingham, United Kingdom}
\def\addressk{Istituto di Fotonica e Nanotecnologie, CNR-Fondazione Bruno Kessler, 
    I-38123 Povo, Trento, Italy}
\def\addressl{Institut f\"ur Geophysik, ETH Z\"urich, Sonneggstrasse 5, CH-8092 Z\"urich, Switzerland}
\def\addressm{The UK Astronomy Technology Centre, Royal Observatory, Edinburgh, Blackford Hill, Edinburgh EH9 3HJ, United Kingdom}
\def\addressn{Institut de Ci\`encies de l'Espai (CSIC-IEEC), Campus UAB, Carrer de Can Magrans s/n, 08193 Cerdanyola del Vall\`es, Spain}
\def\addresso{DISPEA, Universit\`a di Urbino Carlo Bo, Via S. Chiara, 27 61029 Urbino/INFN, Italy}
\def\addressp{European Space Operations Centre, European Space Agency, 64293 Darmstadt, Germany }
\def\addressq{Physik Institut, Universit\"at Z\"urich, Winterthurerstrasse 190, CH-8057 Z\"urich, Switzerland}
\def\addressr{SUPA, Institute for Gravitational Research, School of Physics and Astronomy, University of Glasgow, Glasgow G12 8QQ, United Kingdom}
\def\addresss{Department d'Enginyeria Electr\`onica, Universitat Polit\`ecnica de Catalunya,  08034 Barcelona, Spain}
\def\addressu{Gravitational Astrophysics Lab, NASA Goddard Space Flight Center, 8800 Greenbelt Road, Greenbelt, MD 20771 USA}
\def\addressx{INAF Osservatorio Astronomico di Capodimonte, I-80131 Napoli, Italy}
\def\addressy{INFN-Sezione di Napoli, I-80126, Napoli, Italy}
\def\addressz{Dipartimento di Fisica ed Astronomia, Universit\`a degli Studi di Firenze and INFN-Sezione di Firenze, I-50019 Firenze, Italy}
\def\addressaa{OHB Italia S.p.A, Via Gallarate, 150 - 20151 Milano, Italy}
\def\addressbb{Department of Physics, Aristotle University of Thessaloniki, Thessaloniki 54124, Greece}
\def\addresscc{Universidad Loyola Andaluc\'ia, Department of Quantitative Methods, Avenida de las Universidades s/n, 41704, Dos Hermanas, Sevilla, Spain}
\def\addressdd{Department of Mechanical and Aerospace Engineering, MAE-A, P.O. Box 116250, University of Florida, Gainesville, Florida 32611, USA}
\def\addressee{isardSAT SL, Marie Curie 8-14, 08042 Barcelona, Catalonia, Spain}
\def\addressff{Haute Ecole Vallaisanne, CH-1950 Sion, Switzerland}
\def\addressgg{Dipartimento di Fisica, Università di Roma ``Tor Vergata'' / INFN-Sezione Roma Tor Vergata, I-00133 Roma, Italy}
\def\addresshh{Dipartimento di Fisica, Università di Napoli ``Federico II'' and INFN-Sezione di Napoli, I-80126, Napoli, Italyy}
\def\addressii{INFN-Sezione di Napoli, I-80126 Napoli, Italy}
\def\addressjj{Dipartimento di Fisica ed Astronomia, Universit\`{a} degli Studi di Firenze and INFN-Sezione di Firenze, I-50019 Firenze, Italy}
\def\addresskk{INAF Osservatorio Astronomico di Capodimonte, I-80131 Napoli, Italy and INFN-Sezione di Napoli, I-80126 Napoli, Italy}
\def\addressll{Thales Alenia Space in Switzerland, CH-8052 Z\"urich, Switzerland}

%\author{LPF science collaboration, plus selected hardware providers\\
%(Joseph and possibly others at HEV, others at ETHZ, some from RUAG, ???)}
\date{\today}

\begin{abstract}

Electrostatic force actuation is a key component of the system of geodesic reference test masses (TM) for the LISA orbiting gravitational wave observatory and in particular for performance at low frequencies, below 1~mHz, where the observatory sensitivity is limited by stray force noise.  The system needs to apply forces of order 10$^{-9}$~N while limiting fluctuations in the measurement band to levels approaching 10$^{-15}$~N/Hz$^{1/2}$.  We present here the LISA actuation system design, based on audio-frequency voltage carrier signals, and results of its in-flight performance test with the LISA Pathfinder  test mission.  In LISA, TM force actuation is used to align the otherwise free-falling TM to the spacecraft-mounted optical metrology system, without any forcing along the critical gravitational wave-sensitive interferometry axes.  In LISA Pathfinder, on the other hand, the actuation was used also to stabilize the TM along the critical $x$  axis joining the two TM, with the commanded actuation force entering directly into the mission's main differential acceleration science observable.  The mission allowed demonstration of the full compatibility of the electrostatic actuation system with the LISA observatory requirements, including dedicated measurement campaigns to amplify, isolate, and quantify the two main force noise contributions from the actuation system, from actuator gain noise and from low frequency ``in band'' voltage fluctuations.   These campaigns have shown actuation force noise to be a relevant, but not dominant, noise source in LISA Pathfinder and have allowed performance projections for the conditions expected in the LISA mission.  
  
\end{abstract}

%abstract: 
%key subsystem for the LISA GW detector 
%in LPF contains low frequency science signal along x
%in LISA x forces not needed but phi torques needed to align TM to IFO
%LPF diff accel noise measurement, including dedicated actuation noise experiments, have measured 
%key components of the actuation acceleration noise contribution, for gain fluctuations (aud freq carriers)
%and in band additive voltage noise.  Gain noise is up to 50 ppm/sqrt(Hz) at 0.1 mHz (7 ppm/Hz1/2 at 1~mHz).  
%In band noise 50 $\mu$V/Hz$^{1/2}$ at 0.1~mHz (15 at 1 mHz).  projections to LISA.

\maketitle

\section{Introduction}

The ESA mission LISA Pathfinder \cite{LPF_prl_2016} (LPF), which launched on December~3,~2015 and completed science operations in July~2017, measured the differential acceleration between two free-falling test masses.  The experiment was sensitive to stray forces acting on a TM, which introduce noise into their otherwise geodesic orbits and ultimately limit the sensitivity of a future space observatory for gravitational waves in the 20~$\mathrm{\mu}$Hz-1~Hz band, such as the proposed LISA mission 
\cite{LISA_proposal_2017}. 

\begin{figure} [t]
\includegraphics{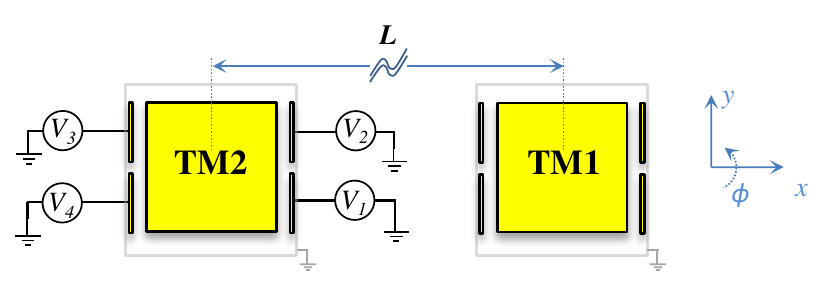}
\caption{\label{fig_act_scheme} Cartoon of LPF, with two TM 
along the sensitive $x$ axis, inside their respective GRS electrode housings.  The numbering
scheme for the sensor $X$-face electrodes, used for actuation and sensing of the  $x$ and $\phi$ (rotation around $z$) degrees of freedom, is shown in the actuation generators for TM2.}
\end{figure}

The main LPF experimental observable, $\Delta g \equiv \frac{f_2}{m_2} - \frac{f_1}{m_1}$, 
is a gravity gradiometer signal, 
the differential force per unit mass on two TM separated by $L = 37.6$~cm.  
The spacecraft (SC) was ``drag-free'' controlled, with precision cold-gas thrusters, to follow one 
TM (TM1) along the sensitive $x$ measurement axis (see Fig.~\ref{fig_act_scheme}). 
The second TM (TM2) was 
forced to follow TM1: any non-zero $\Delta g$ had to be compensated with applied
forces to avoid accelerating TM2, over time, into the surrounding 
SC apparatus.  The required actuation force (per unit mass) 
on TM2, $g_{\mathrm{2c}}$, had to be accurately calibrated 
as part of the signal used to construct $\Delta g$, and any additional 
force noise introduced by the actuator voltages contributed noise in  
$\Delta g$.  Critical sources of low frequency force noise \cite{LPF_noise_budget_cqg_2011} include 
 additive voltage noise mixing with TM charge and stray electrostatic  
fields, and multiplicative ``gain noise'' in the actuator voltage amplitudes,  
scaling with the applied force levels.

LISA Pathfinder was designed for a
differential acceleration ``dynamic range'' of roughly 1~nm/s$^2$, 
while aiming to resolve fluctuations
at the 30~fm/s$^2$/Hz$^{1/2}$ level at~mHz frequencies.  At 
the L1 Lagrange point, far from the $\mu$m/s$^2$ differential gravity 
for a similarly sized gradiometer in low earth orbit, mechanical 
tolerances in the mass balancing of the local spacecraft 
``self-gravity'' \cite{grav_valerio_cqg_2016} 
set this nm/s$^2$ design range.   
Translated into forces on a 2~kg LPF TM, the actuators must give 
force authority of order several nN while allowing resolution of femtoNewton (fN) 
level force variations on time scales of 1000~s.  

Actuation forces were required to balance the differential acceleration between the two TM for the three translational axes plus three angular accelerations for each TM.  The LPF actuation system also had to coexist with a TM position readout at the 2~nm/Hz$^{1/2}$ level\cite{prd_2017_cap_sens}.  Sensing and actuation capabilities were needed on all 6 degrees of freedom (DOF), though a higher precision interferometer \cite{prl_ifo,prd_ifo} substituted capacitive sensing on the critical science measurement axis.  For LPF and LISA, as in geodesy missions like GOCE \cite{goce_rummel} and tests of gravity like GPB \cite{GPB_cqg_2015} and MICROSCOPE \cite{microscope_2017},  sensing and force actuation are combined into a single electrostatic sensor with the conducting TM surrounded by a conducting housing with an array of electrodes.  The LPF ``gravitational reference system'' or GRS \cite{grs_SPIE_2002,grs_cqg_2003} -- the TM, electrode housing and associated sensing / actuation electronics --  differs from the electrostatic accelerometers used in the missions cited above because of its larger mass, larger TM-electrode separations -- 4~mm on the most sensitive axis -- and, except for GPB, absence of any discharge wire. These design choices were all imposed by the extremely low force noise requirements for LISA and LPF.    

The required 2~nm/Hz$^{1/2}$ capacitive sensing has been demonstrated with the LPF TM held in place before release into free-fall \cite{prd_2017_cap_sens}, while the role of the GRS as a sensor and actuator has been validated in the overall LPF ``drag-free and attitude control system'' (DFACS) \cite{lpf_dfacs_ads, lpf_dfacs_platform, lpf_cold_gas,cal_paper}. This article addresses the 
details of the electrostatic force actuation system for LISA, as designed and tested in the LISA 
Pathfinder mission.  

The article first presents, in Sec.~\ref{sec_system_overview}, the actuation system design, 
from its specific features to limit electrostatic force noise and its conceptual design to the high-level circuit implementation.  This section includes an overview of the in-flight use conditions encountered 
in LISA Pathfinder and some key elements of its calibration and verification as part of the LPF 
differential acceleration measurement chain.    

The following two sections form the experimental core of this paper and present the models and measurement
campaigns used to quantify two critical force noise sources arising in the electrostatic actuation system: actuation gain fluctuations (Sec.~\ref{sec_act_gain_noise}) and low-frequency additive voltage noise (Sec.~\ref{sec_inband_noise}).  
Tests were performed in the true in-orbit LPF conditions, which included a background $\Delta g$ that was always within 25~pm/s$^2$ of zero, more than an order of magnitude below the nm/s$^2$ level discussed above, due to successful spacecraft gravitational balancing.  This allowed lowering the actuation force ``authority'', which, as will be discussed in this paper, was of fundamental importance in pushing the ultimate differential acceleration noise down to the 2~fm/s$^2$/Hz$^{1/2}$ level at ~mHz frequencies \cite{LPF_prl_2018}.  Quantifying the actuator noise required application of larger, balancing forces to increase the effect of  actuation voltage fluctuations.  Likewise, ``in-band'' additive voltage noise was quantified by measuring acceleration noise with an intentionally charged TM. 
 
We note that the conversion of actuation voltage fluctuations into force noise, described in Secs.~\ref{sec_act_gain_noise} and \ref{sec_inband_noise}, is relatively straightforward.  Additionally, the actuation electronics was subject to pre-flight electronic noise tests on ground.  However, a full test measuring force and torque noise from the actuation system, in the complete flight conditions with two TM in multi-axis free-fall, allows a direct validation including correlations and other possible effects escaping the model.  Additionally, as for other precision experiments in space, a large time -- roughly 6 years for the LPF electronics -- and a launch separates the ground tests from flight performance, making \textit{in situ} measurements, such as those presented here, a key part of a reliable experimental noise model.     
 
In Sec.~\ref{sec_projections} we use the results of these measurement campaigns to make a projection of the contribution of actuation noise to the LPF differential acceleration measurement $\Delta g$ and to the acceleration noise of a single TM in experimental scenario of the LISA mission.  
In contrast with LPF, LISA does not require force actuation along the interferometer $x$ ``science axis'' used for measuring the gravitational wave-induced tidal acceleration.  It does however need angular torque control applied with electrostatic fields along the critical $x$ axis, and these are a potentially important source of TM acceleration noise that can impact the mission science return.  

The article ends with some final comments on the role of actuator noise in the overall acceleration
noise performance for LPF and LISA, as well as a consideration on the uniqueness of LISA Pathfinder as a
testbench for the measurement of small forces and torques.

\section{GRS electrostatic actuation system design and role in LPF $\Delta g$ measurement}
\label{sec_system_overview}
The LPF observable $\Delta g$ is constructed from Newtonian dynamics and telemetry for the commanded forces and interferometric readouts for the relative TM displacement, $\Delta x \equiv x_2 - x_1$, and the relative TM1-SC displacement, $x_1$ \cite{LPF_prl_2016}:
\begin{equation}
\Delta g = \Delta \ddot{x} - \lambda g_{\mathrm{2c}} 
+ \Delta \omega^2 x_1 + \omega_2^2 \, \Delta x 
\label{eqn_Dg}
\end{equation}
The leading terms are the measured acceleration $\Delta \ddot{x}$ and commanded force $g_{\mathrm{2c}}$, scaled by calibration factor $\lambda$, which  dominates at frequencies below the roughly 1~mHz unity-gain controller bandwidth. Smaller corrections due to coupling of the two TM to the SC motion are approximated as elastic with effective spring constants $m_1 \omega_1^2$ and $m_2 \omega_2^2$ -- with differential ``stiffness'' $\Delta \omega^2 \equiv \omega_2^2 - \omega_1^2$. Inertial force terms included in the standard LPF analysis \cite{LPF_prl_2016, LPF_prl_2018} are omitted here for simplicity. 

The accuracy, linearity, and stability of the actuator gain -- factor $\lambda$ in the subtraction 
of the applied force -- set the accuracy with which $\Delta g$ is measured, most 
significantly at low frequencies below 1~mHz.  
The applied electrostatic forces also depend on the TM position, with the resulting force gradients contributing
to the stiffness $\omega_1^2$ and $\omega_2^2$; these must be reliably known and stable
for accurate calculation of $\Delta g$. Finally, and most importantly,
unmodelled force noise associated with the actuation fields -- due to actuation gain fluctuations 
or to additive circuit voltage noise that mixes with stray electrostatic 
fields -- contributes directly to the noise 
in $\Delta g$ along with any other external force noise on the TM.  

\subsection{Actuation conceptual design}
The actuation design employs audio frequency voltages, in the 60-270~Hz band, to create the needed DC and slowly varying electrostatic forces needed for dynamical control of the TM - spacecraft system. This exploits the quadratic force-voltage dependence, $F \propto \Delta V\:^2$, to give a DC force that depends only on the carrier amplitude plus a force at twice the carrier frequency, decades above the mHz~LPF measurement band (the electrostatic force model and actuation algorithm are presented in Appendix A).  This is chosen to limit the force errors and low frequency force noise from stray fields from TM charge and surface ``patch'' potentials 
\cite{imperial_charge_2005,asr_dc_bias,charge_DC_bias_prl_2012,tim_charge_lisa}:  any steady or slowly varying stray potential difference  mixes with the applied audio carrier to produce a force, and force noise, safely outside the LISA band around the carrier frequency.  Considering stray potentials of order 100~mV, and their noise, use of AC carriers is a necessary design innovation for LPF.  

\begin{figure} [bt]
\includegraphics[width=3.3in]{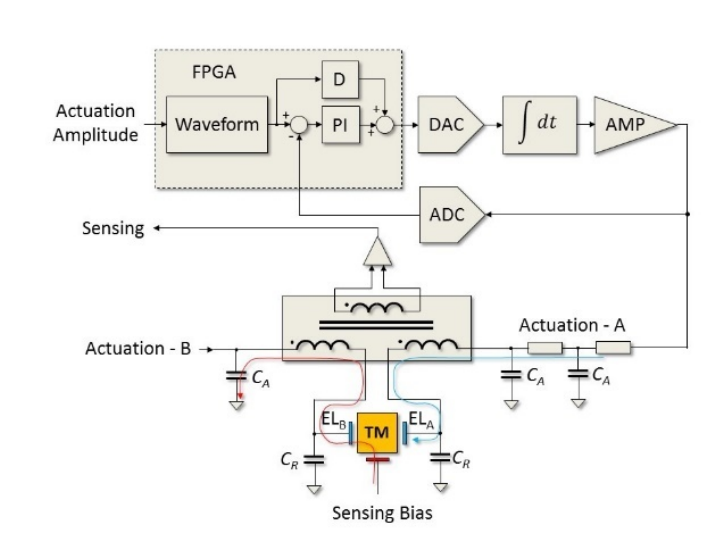}
\caption{\label{fig_fee_act} Circuit block diagram featuring sensing and actuation elements for two opposing electrodes (a single sensing channel).  Red and blue curves indicate current paths for, respectively 100~kHz sensing and audio / DC actuation.}
\end{figure}

Referring to the $X$ electrode configuration in Fig.~\ref{fig_act_scheme}, the LPF actuation scheme follows
\begin{eqnarray}
V_{1c} \left( t \right) & = & 
V_{1x} \sin \omega_x t + V_{1\phi} \sin \omega_{\phi} t
\nonumber \\
V_{2c} \left( t \right) & = &  
-V_{1x} \sin \omega_x t + V_{2\phi} \cos \omega_{\phi} t
\nonumber \\
V_{3c} \left( t \right) & = & 
V_{2x} \cos \omega_x t - V_{1\phi} \sin \omega_{\phi} t
\nonumber \\
V_{4c} \left( t \right) & = & 
-V_{2x} \cos \omega_x t - V_{2\phi} \cos \omega_{\phi} t
\label{eqn_applied_V}
\end{eqnarray}
with similar expressions for the $Y$ electrodes (degrees of freedom $y$ and $\theta$) and 
$Z$ electrodes (DOF $z$ and $\eta$).  
The subscript ``c'' is employed here in $V_{j\mathrm{c}} \left( t \right)$ to indicate
the \textit{commanded} voltage on electrode $j$.

%The measurements in the 
%following sections quantify the forces induced when this \textit{ideal}
%waveform is replaced with the true imperfect waveform, including gain noise ($\alpha$)
%and random additive noise $v$, 
%\begin{equation}
%V_{j} = V_{jc} \left[ 1 + \alpha_j \left( t \right) \right] + v_j \left( t \right) 
%\label{eqn_V_real}
%\end{equation} 

The applied $x$ actuation voltages $V_{1x}$ and $V_{2x}$ yield a total time-average force proportional to $\left( V_{1x}^2 - V_{2x}^2 \right)$ and a force gradient proportional to $\left( V_{1x}^2 + V_{2x}^2 \right)$.  The LPF ``constant stiffness'' actuation algorithm fixes $V_{1x}^2 + V_{2x}^2 \equiv V_{\mathrm{MAX}_x}^2$, allowing application of force per unit mass $g_{\mathrm{c}}$ in the range $\pm g_0 = \pm \frac{1}{2 m} \pder{C_X^{\star}}{x} V_{\mathrm{MAX}_x}^2$, with 
\begin{equation}
V_{1x/2x} 
= 
\left[ \frac{M \left(g_0 \pm g_{\mathrm{c}} \right)}{\left| \pder{C_X{^\star}}{x} \right|} \right]^{1/2}
\label{eqn_F_to_V}
\end{equation}
where $\pder{C_X{^\star}}{x}$ is the relevant capacitance derivative for an $X$ electrode (see App.~A) and $M$ the mass of the nominally identical test masses. The desired force is thus produced by unbalancing the electrostatic forces pulling on opposing sides of the TM. The resulting $x$ axis stiffness contribution is independent of the applied force $g_{\mathrm{c}}$,
\begin{equation}
\omega_{xx}^2 = - g_0 \frac{ \left| \pdder{C^{\star}}{x} \right|}{ \left| \pder{C^{\star}}{x} \right| } 
\approx -\frac{2 g_0}{d_x}
\label{eqn_stiffness}
\end{equation}
where $d_x$ is the gap between the $X$ electrodes and the TM, assumed equal on opposing sides.   

Expressions analogous to Eqns.~\ref{eqn_F_to_V} and \ref{eqn_stiffness} are obtained for the voltages $V_{1\phi/2\phi}$ 
and stiffness associated with electrostatic actuation for the $\phi$ rotational degree of freedom, described by the 
commanded torque (per unit moment of inertia) $\gamma_{\phi_c}$ within the torque authority range of 
$\pm \gamma_{\phi_0}$, both with units of angular acceleration, $\mathrm{/s^2}$.  Both $x$ and $\phi$ actuation 
contribute individually to the stiffness along both the $x$ and $\phi$ degrees of freedom (DOF). The constant stiffness 
algorithm allows a fixed and calculable (negative) elastic coupling in the control dynamics and eliminates a first-order cross-
coupling between $\phi$ torques and $x$ acceleration for an off-center TM.  

Voltage waveforms are applied with opposite phase ($\pm V_{1x}$ on electrodes 1 and 2, for instance, see Eqn.~\ref{eqn_applied_V}), to eliminate any induced TM potential, at least for a centered TM. Orthogonal waveforms (cos / sin) and different frequencies ($\frac{\omega_x}{2 \pi} = 60$~Hz and $\frac{\omega_{\phi}}{2 \pi} =270$~Hz) avoid cross-talk between the different actuation degrees of freedom.

% tab_act_configs

\subsection{Actuation circuit implementation}

The actuation circuit implementation is sketched as part of the sensing / actuation ``front-end electronics'' in Fig.~\ref{fig_fee_act}. Sensing and actuation operate simultaneously with the same electrodes, with currents sharing the primary windings of the sensing differential transformer \cite{grs_SPIE_2002}.  Capacitive sensing uses a ``contact free'' injection of a 100~kHz bias on the TM, with the difference of current flowing through opposing pairs of electrodes measured by a resonant differential transformer bridge followed by transimpedance amplifier and homodyne detection scheme to give six gap-sensing displacement readouts.

Audio frequency actuation voltages are applied to the electrodes through the primary windings, with two passive $RC$ stages (with $C_A$) used to limit interference with the 100~kHz position readout. Digital ``target'' actuation voltages are generated in a Field Programmable Gate Array (FPGA), where audio waveforms are synthesized from commanded peak amplitudes and then summed with the DC voltages, with DC and audio amplitudes updated at the 10~Hz experimental sampling rate, sufficient for the force controller loops.  The actuation outputs are stabilized by a $\Sigma-\Delta$ loop using a DAC, integrator, 96~kHz ADC, and digital PID controller, effectively tying the actuation waveform generator stability to that of the ADC voltage reference, which is the same for groups of 4 actuation channel circuits located on a single board, such as for the 4 electrodes used for sensing and actuating $x$ and $\phi$ electrodes, as illustrated in Fig.~\ref{fig_act_scheme}.  

Based on the DC force balancing requirements \cite{grav_valerio_cqg_2016}, the LPF actuation system was designed for an $x$ actuation authority of $g_0 = 1.15$~nm/s$^2$ (2.2~nN) for TM2 -- with no TM1 $x$~actuation -- and $\phi$ authority $\gamma_{\phi_0} \approx 15$~nrad/s$^2$ (10.4~pNm) for both TM. These are given also in  Tab.~\ref{tab_act_configs} as the ``nominal'' configuration, used at the start of the mission and in the second actuation noise trial described in Sec.~\ref{sec_act_gain_noise}.  Given the LPF GRS $X$ electrodes, with surface area 530~mm$^2$ and gap $d_x = 4$~mm, this corresponds (Eqns.~\ref{eqn_F_to_V} and \ref{eqn_stiffness}) to $V_{\mathrm{MAX}_x} \approx 4.5$~V and $V_{\mathrm{MAX}_{\phi}} \approx 2.9$~V (peak amplitudes), with associated $x$ axis electrostatic stiffness contributions of approximately $-660 \times 10^{-9}$/s$^2$ and $-270 \times 10^{-9}$/s$^2$.  The maximum possible total instantaneous voltage, 7.4~V, fits comfortably into a roughly 10~V envelope allowed by the actuation electronic science mode AC voltage range.  The actuation circuitry also allowed for several volt DC voltages applied to any sensing / actuation electrode, for compensating stray DC voltages, measuring TM charge, and biasing the UV discharge 
\cite{LPF_prl_charge, prd_2018_uv_discharge}.  
  
The actuation nominal bit resolution was 153~$\mu$V, yielding an effective force quantization of order 100~fm/s$^2$ with this nominal force authority (1.15~nm/s$^2$).  To smooth the resulting ``force bit'' steps in the commanded force, a software $\Sigma-\Delta$ loop in the TM2 dynamic control loop was implemented in the on-board computer, which dithers the commanded voltage between adjacent voltage levels, reducing the effective bit size by a factor 30 on 100~s time scales.  

% nb calculate with 4.5 V on each electrode
% 4*VMAX/sqrt(2)/2*DV*dCx_dx_ACT / M

\subsection{Actuation use, functionality, and performance in LISA Pathfinder}

Due to the superior gravitational balancing actually achieved in the as-flown LPF, the actual needed force and torque actuation levels were considerably below the ``nominal'' levels, most notably between -25~pm/s$^2$ to +12~pm/s$^2$ in the TM2 $x$ force (except during the actuation noise campaign described in the next section).  This allowed lowering the force authority in $x$ to $g_0 \approx 26$~pm/s$^2$ ($V_{\mathrm{MAX}_x} \approx 0.7$~V).  The $\phi$ controller authorities were also lowered as allowed by the smaller levels of torque needed, with the typical TM1 and TM2 authorities of roughly 2.2 and 1.5~nrad/s$^2$ ($V_{\mathrm{MAX}_{\phi}} \approx 0.9$~and~$0.7$~V). The resulting total electrostatic $x$~axis stiffness from actuation was roughly $-40 \times 10^{-9}$/s$^2$ for both TM.  This configuration, known as UURLA and described on line 1 of Tab.~\ref{tab_act_configs}, was used for the main noise measurements in LPF \cite{LPF_prl_2016, LPF_prl_2018}.  

Periodically, and upon change in the actuation authorities, the $x_2$ actuation 
calibration factor ($\lambda$) and the stiffness coefficients ($\omega_2^2$ and $\Delta \omega^2$)
were calibrated with a ``system identification'' experiment in which the control setpoints for the 
$x$ positions of the two TM were commanded to oscillate sinusoidally over a range 
of frequencies from 1 to 50~mHz and with typical amplitudes up to 10~nm \cite{cal_paper}.  
Parameters $\lambda$, $\omega_2^2$, and $\Delta \omega^2$
are then extracted by fitting the time series of $g_{\mathrm{2c}}$, $x_1$, and $\Delta x$ to the model 
in Eqn.~\ref{eqn_Dg}.  The gain factor $\lambda$ has been found to be stable over the year of data analyzed, at a level approaching 0.01\% and with a mean value within 1\% of that calculated from the voltage source design and the electrostatic 
force model.  The measured dependence of the stiffness on force and torque authorities $g_{0}$ and $\gamma_{\phi 0}$, independent of the applied forces and torques,  verifies the ``constant stiffness'' 
algorithm and corresponds with the electrostatic model to within roughly 5\% \cite{cal_paper, 
nico_lisa_symp}.  The LPF in-flight dynamical calibration of $\Delta g$, 
including the actuator calibration and stiffness, 
has been addressed in detail in a dedicated paper \cite{cal_paper}.

% no distinction of x actuation, should be explained ... 
% cross talk, 

\begin{figure} [t]
\includegraphics[width=3.3in]{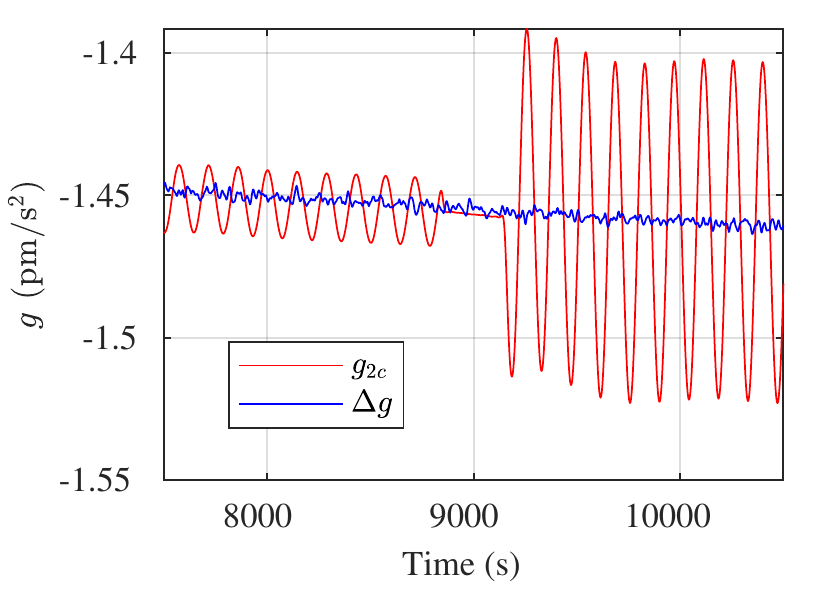}
\caption{\label{fig_caltone} Time series of applied force $g_{\mathrm{2c}}$ and resulting 
differential acceleration $\Delta g$ during the ``calibration tone'' experiment 
described in the text.}
\end{figure}

An additional ``calibration tone'' experiment has been 
performed (Fig.~\ref{fig_caltone}) in which a sinusoidally oscillating force at 7~mHz was superimposed on the controller force command in $g_{\mathrm{2c}}$, with amplitude of 20~fN and then 100~fN. The presence of the ``out-of-loop'' force should not be visible in an accurately calculated time series of $\Delta g$, including accurate and stable calibration factor $\lambda$ (see Eqn.~\ref{eqn_Dg}):
the commanded oscillation in the force $g_{\mathrm{2c}}$ produces a corresponding oscillating differential acceleration $\Delta \ddot{x}$
that cancels in $\Delta g$, which thus includes only the external, ``out of loop'' residual forces on the TM. This is in fact observed, with the resulting 
trace of $\Delta g$ found to be compatible with the background statistical noise and residuals at the modulation frequency below the fm/s$^2$ level -- and less than 1\% -- in every cycle, demonstrating the fm/s$^2$ accuracy of the differential accelerometer with actuation. We note that obtaining this accurate calibration and linearity was not automatic and required a correction to a subtle but deterministic roundoff error in the actuation DAC circuitry.  This correction is not addressed here but was critical in reaching the best performances obtained in LPF \cite{LPF_prl_2018} and is addressed in detail in a dedicated paper \cite{luigination_2020}.

A final aspect of the actuation performance is force crosstalk between the various electrostatically-actuated degrees of freedom.  The constant stiffness and symmetric waveform design presented here aims to minimize ``leakage'' of applied forces into other degrees of freedom, but residual actuation crosstalk \cite{PETER_prl,PETER_crosstalk} remains, due to geometric imperfections and, most importantly for the critical $\phi$ torque to $x$ force, differences in the voltage gains between the 4 $X$ electrodes.  A measured residual coupling of spacecraft rotational acceleration -- estimated by the applied electrostatic torques, $\gamma_{\phi_{\mathrm{SC}}} \approx \gamma_{\phi_{\mathrm{c}}}$  -- with a cross-coupling coefficient of typically $\beta_{\phi x}\equiv \frac{\Delta g}{\gamma_{\phi_\mathrm{c}}} \approx 150$~$\mu$m \cite{cal_paper,LPF_prl_2018}.  This could be attributed to a gain imbalance, of order 0.5\%, between the different electrode actuation circuits, though such an effect would be essentially indistinguishable from other Euler-force effects in the LPF data.  This gives a rough level for actuation crosstalk effects, possibly relevant also in LISA.      

\section{Actuation gain fluctuations: noise model, experiment design, and result}
\label{sec_act_gain_noise}
% develop the additive noise term, 20200529
% subsections for models (gain, additive) and experiments
% insert model for additive noise
% comment in conclusion about possible role of additive noise 

\subsection{Actuation gain noise model}
\label{subsec_act_gain_model}
In the LPF noise model, the most critical actuation force noise arises in in-band 
gain or multiplicative amplitude fluctuations \cite{LPF_noise_budget_cqg_2011}; even in the event of constant commanded force -- and thus commanded voltage amplitudes $V_{1x}$ and $V_{2x}$, Eqn.~\ref{eqn_F_to_V} -- fluctuations in the true carrier amplitude at the electrode result in a fluctuating force.  With $F \propto V^2$, the force fluctuations caused by an actuator electrode can be described in terms of the relative voltage fluctuation, or gain fluctuation, $\alpha \equiv \frac{\delta V}{V}$ ,
\begin{equation}
\delta F = 2 F \frac{\delta V}{V} = 2 F \alpha  
\label{eqn_dF}
\end{equation} 
If $\alpha \left( t \right)$ 
is the same for all 4 circuits responsible for the TM2 $x$ force, as expected for 
fluctuations in their common DC voltage reference, the resulting noise in the $\Delta g$ measurement 
depends only on the actuator stability and on the commanded \textit{net} force, 
\begin{equation}
S_{\Delta g(ACT)}^{1/2} \approx 2 \left| g_{\mathrm{c}} \right| S_{\alpha}^{1/2}     
\label{eqn_SDg_simp}
\end{equation} 
This was used to set requirements on the LPF self-gravity differential force,
$\Delta g_{DC} \leq 1.3$~nm/s$^2$, and the actuation
gain noise, $S_{\alpha}^{1/2} \leq 2 \times 10^{-6}$/Hz$^{1/2}$ at 
1~mHz \cite{grs_SPIE_2002,cqg_anza_2005}.  

Uncorrelated 
amplitude fluctuations between the different electrodes complicate this picture, as understood later 
in LPF development \cite{LPF_noise_budget_cqg_2011}; Eqn.~\ref{eqn_dF} is 
valid on an \textit{electrode-by-electrode} basis such that the force 
noise depends not only on the net applied force on TM2 $g_{\mathrm{2c}}$ but also on the individual voltage
levels, and thus also on the authority $g_0$ and the applied torque and torque authorities, 
$\gamma_{\phi c}$ and $\gamma_{\phi 0}$ for both TM. For instance, uncorrelated fluctuations in the 
amplitudes of the carrier voltages $\pm V_{1\phi}$ applied to electrodes 1 and 3 to create a positve 
$\phi$ torque, will give rise to asymmetric $x$ force fluctuations that do not cancel and which thus contribute to the noise in $\Delta g$.  

% 8 actuators, with gain fluctuations that in general could have any level of correlation (36 parameters)
% in our analysis use a minimal set that describes our data with gain fluctuation mechanisms that physically 
% are known to exist 
% not unique but effective   
%\color{red}
Considering the eight relevant actuator gains and any possible correlations between them, in general 
there would be 36 relevant cross-spectrum terms at each frequency.  Considering an experimental campaign
with limited duration and number of experimental configurations, we propose here a minimal model, including only 
terms with known physical origin to describe, and then fit, the acceleration noise spectrum.  
This possible, but not unique, parametrization of the multiplicative gain fluctuations for electrode $j$ of TM $i$ with 
commanded voltage $V_{ijc}$ is 
\begin{equation}
V_{ij} \left( t \right) = V_{ijc} \left( t \right) 
\left[ 1 + \alpha \left( t  \right) + \alpha_i \left( t \right) + \alpha_{ij} \left( t \right)  \right]
% add term
+ v_{ij} \left( t \right)
\label{eqn_dV}
\end{equation}
where we include: 
\begin{itemize}
\item $\alpha$, a gain fluctuation common to all 8 $X$ electrodes for the 
two TM, such as a systematic dependence on the GRS FEE box temperature (our experiments however 
will not be sensitive to this term)
\item $\alpha_i$, to become $\alpha_1$ and $\alpha_2$, which is a ``TM correlated'' gain fluctuation, corresponding to fluctuations in the 
single voltage reference voltage common to the 4 circuits used for $x$/$\phi$ actuation 
on a single TM 
\item $\alpha_{ij}$, independent gain fluctuations for electrode $j$ of TM $i$, uncorrelated between the 8 
electrodes  
\end{itemize}
%\color{black}
The role of additive noise $v_{ij}$ mixing with the carrier voltages is discussed separately, in the next subsection.  

In this model the resulting noise in $\Delta g$ is a sum of contributions from 11 independent noise generators , 
\begin{equation}
\Delta g^{ACT} \left( t \right)
=  a \, \alpha \left( t \right)
+ \displaystyle \sum_{i=1,2} a_i \, \alpha_i \left( t \right)
 + \displaystyle \sum_{ij} a_{ij} \, \alpha_{ij} \left( t \right)
\label{eqn_Dg_full}
\end{equation}
with coefficients 
\begin{eqnarray}
a & = &   2 \left( g_{\mathrm{2c}} - g_{\mathrm{1c}} \right) = 2 \left( \Delta g \right)  \nonumber \\
a_1 & = &  -2 g_{\mathrm{1c}} \nonumber \\
a_2 & = &  2 g_{\mathrm{2c}} \nonumber \\
a_{21} & = & \frac{1}{2}
\left( \bar{g}_{2c} + g_{20} + R^{\star}_{\phi} \bar{\gamma}_{\phi_{\mathrm{2c}}} 
+ R^{\star}_{\phi} \gamma_{\phi_20}  \right) 
\label{eqn_act_accel_coeff}
\end{eqnarray}
where we take as example of the independent electrode gain noise terms $a_{21}$ for 
electrode 1 of TM2, which is used to apply positive $x$ forces and $\phi$ torques.   
Here $R^{\star}_{\phi} \equiv \frac{I}{m} \frac{\pder{C^{\star}}{x}}{ \pder{C^{\star}}{\phi} } \approx 32$~mm is an
effective armlength converting angular into linear acceleration ($I$ is the TM moment of inertia). See Appendix B for a more detailed discussion of this model.

If we consider these 11 noise generators mutually uncorrelated, the resulting PSD is
\begin{equation}
S_{\Delta g}^{ACT} 
%  =  4 \left( \Delta g \right) ^2 S_{\alpha} + 
%4 g_{c1}^2 S_{\alpha_1} + 4 g_{c2}^2 S_{\alpha_2} 
%+ \displaystyle\sum_{ij} A_{ij} S_{\alpha_{ij}}
 =  A S_{\alpha} + \displaystyle \sum_{i=1,2} A_i S_{\alpha_i} + \displaystyle\sum_{ij} A_{ij} S_{\alpha_{ij}}
\label{eqn_SDg_full}
\end{equation}
with coefficients
\begin{eqnarray}
A & = & a^2  =   4 \left( g_{\mathrm{2c}} - g_{\mathrm{1c}} \right)^2 = 4 \left( \Delta g \right)^2 = 4 \left( \Delta g^{\mathrm{DC}} \right)^2 
\nonumber \\
A_1 & = & a_1^2 = 4 g_{\mathrm{1c}}^2 \: \: \: \: \left[ = 0 \right] \nonumber \\
A_2 & = & a_2^2 = 4 g_{\mathrm{2c}}^2 \: \: \: \: \left[ = 4 \left( \Delta g^{\mathrm{DC}} \right)^2 \right] \nonumber \\
A_{21} & = & a_{21}^2 = \frac{1}{4}
\left( \bar{g}_{\mathrm{2c}} + g_{20} + R^{\star}_{\phi} \bar{\gamma}_{\phi_{\mathrm{2c}}} 
+ R^{\star}_{\phi} \gamma_{\phi_{20}}  \right)^2 
\label{eqn_act_accel_coeff}
\end{eqnarray}
where $A_{21}$ is given as an example of the eight relevant uncorrelated gain noise $A_{ij}$ coefficients.
Here the numbers offset in block parentheses at right for the ``board'' terms $A_1$ and $A_2$ refer to the typical operating conditions of LPF, where the differential DC acceleration $\Delta g^{DC}$ is balanced by forcing only TM2, with $g_{\mathrm{1c}}= 0$ and $g_{\mathrm{2c}} \approx - \Delta g^{DC}$ \footnote[2]{This ``typical'' condition is not required, as both TM can have a ``common mode'' applied force, as in the fourth and final actuation gain noise test (see Tab.~\ref{tab_act_configs})}.  

Board correlated gain fluctuations, for instance $S_{\alpha_2}$ for TM2 arising from a fluctuating 
reference voltage, indeed couple to the net applied forces to give force noise,
as suggested by Eqn.~\ref{eqn_SDg_simp}. Additionally, however,    
uncorrelated gain fluctuations for individual electrodes, $S_{\alpha_{ij}}$, introduce noise in $\Delta g$ related to commanded force and force authority (ie $\left\{ g_{\mathrm{2c}}, g_{20} \right\}$) as well as commanded $\phi$ torques and torque authorities ($\left\{ \gamma_{\phi_{\mathrm{1c}}} , \gamma_{\phi_{10}}, \gamma_{\phi_{\mathrm{2c}}},\gamma_{\phi_{20}} \right\}$).  It is thus important to limit also the residual DC torques, gravitational or otherwise, and to reduce the force and torque authorities to the minimum levels that still allow compensation of the DC forces with sufficient margin for the system dynamics.

We stress that, while physically motivated, this parametrization in terms of 11 independent, uncorrelated ``noise generators'' is not the only possible model; it is a useful construct for quantifying actuation force noise in LPF and LISA, insofar that it is compatible with the data, which will be discussed shortly.  Any model with actuator gain noise will however have force noise with a PSD increasing quadratically with the forces and torques applied by the single electrodes (Eqn.~\ref{eqn_dF}). 
 
\subsection{Model for mixing of additive actuation noise with actuation carriers}
Additive voltage noise near the actuation frequencies mixes with the carrier voltages to ``down-convert'' into low frequency force noise.  The relevant ``cross-terms'' in the squared actuation voltage have the form, for electrode 1 of a TM as an example, 
\begin{equation}
\delta g \left( t \right) = \frac{1}{M} \left| \pder{C_X{^\star}}{x}  \right|
v_{1} \left( t \right) \left[ V_{1x} \sin \omega_x t + V_{1\phi} \sin \omega_{\phi} t \right] 
\label{eqn_white_time}
\end{equation}
Summing over the 4 $X$ electrodes for that TM and considering the conversion of commanded force/torque into actuation voltage amplitude, the downconverted in-band TM acceleration noise will be
\begin{equation}
S_g = \frac{2 \left| \pder{C_X{^\star}}{x} \right| }{M} 
\left[ 
S_{v_n} \left( \omega_x \right) g_0  
+ 
S_{v_n} \left( \omega_{\phi} \right) R^{\star}_{\phi} \gamma_{\phi_0}  
\right]
\label{eqn_white_noise}
\end{equation}
In contrast with the mHz gain noise, 
the broadband noise is rather easy to model with circuit design and component data.  
In the experiment analysis and noise projections that follow, we calculate and insert 
this acceleration noise contribution, assuming $S_{v_n}^{1/2} \left( \omega_x \right) \approx   
S_{v_n}^{1/2} \left( \omega_{\phi} \right) \approx 2 \: \mathrm{\mu V/Hz^{1/2}}$, based on 
ground measurements.  

The additive audio frequency noise thus gives a white noise, at least in the relevant mHz band, 
with noise power proportional to the applied force and torque authorities $g_0$ and $\gamma_{\phi_0}$, 
while the gain noise terms scale quadraticaly with the forces, as $g_0^2$ and $\gamma^2_{\phi_0}$. 
Gain noise dominates over this additive voltage noise for $\Delta g$ at mHz and sub-mHz frequencies in typical LPF science (UURLA) conditions, even more so 
in the actuation noise test, presented next, with larger force levels.  

\begin{table*}[tb]
\centering
\begin{tabular}{|c|c|c|cc|cc|cc|cc|}
\hline
Expt. & Name & time & $g_{1\mathrm{c}}$ & $g_{10}$ & $\gamma_{\phi_{1\mathrm{c}}}$ & $\gamma_{\phi_{10}}$ 
& $g_{2\mathrm{c}}$ & $g_{20}$ & $\gamma_{\phi_{2\mathrm{c}}}$ & $\gamma_{\phi_{20}}$ 
\\[1ex]
& & hours & \multicolumn{2}{c|}{pm/s$^2$} & \multicolumn{2}{c|}{prad/s$^2$} & \multicolumn{2}{c|}{pm/s$^2$}
& \multicolumn{2}{c|}{prad/s$^2$} \\ 
\hline
1 & UURLA  & 61 & 0 & 0 & -980 & 2170 & -3 & 26 & 130 & 1450   \\[1ex] 
2 & nominal  & 46 & 0 & 0 & -960 & 15040 & -3 & 1140 & 150 & 15040    \\[1ex] 
3 &  big  & 46 & 0 & 2590 & -860 & 15040 & -3 & 2590 & 90 & 15040    \\[1ex] 
4 & big off  & 46 & 2060 & 2590 & -820 & 15040 & 2060 & 2590 & 140 & 15040    \\[1ex] 
%1 & URLA  & 67 & 0 & 0 & -980 & 2170 & -3 & 26 & 134 & 1447   \\[1ex] 
%2 & nom  & 44 & 0 & 0 & -964 & 15042 & -3 & 1144 & 152 & 15042    \\[1ex] 
%3 &  big  & 44 & 0 & 2593 & -860 & 15042 & -3 & 2593 & 91 & 15042    \\[1ex] 
%4 & big off  & 44 & 2064 & 2587 & -818 & 15042 & 2056 & 2587 & 144 & 15042    \\[1ex] 
\hline
\end{tabular}
\caption{Summary of actuation gain noise experiments for four different 
configurations, including measurement duration and average commanded force 
and force authorities -- for instance $g_{1\mathrm{c}}$ and $g_{10}$ -- and analogously for 
torque (eg $\gamma_{\phi_{1\mathrm{c}}}$ and $\gamma_{\phi_{10}}$)\cite{roberta_phd}.  }
\label{tab_act_configs}
\end{table*}

\begin{figure*}[ht]
\includegraphics[width=6.6in]{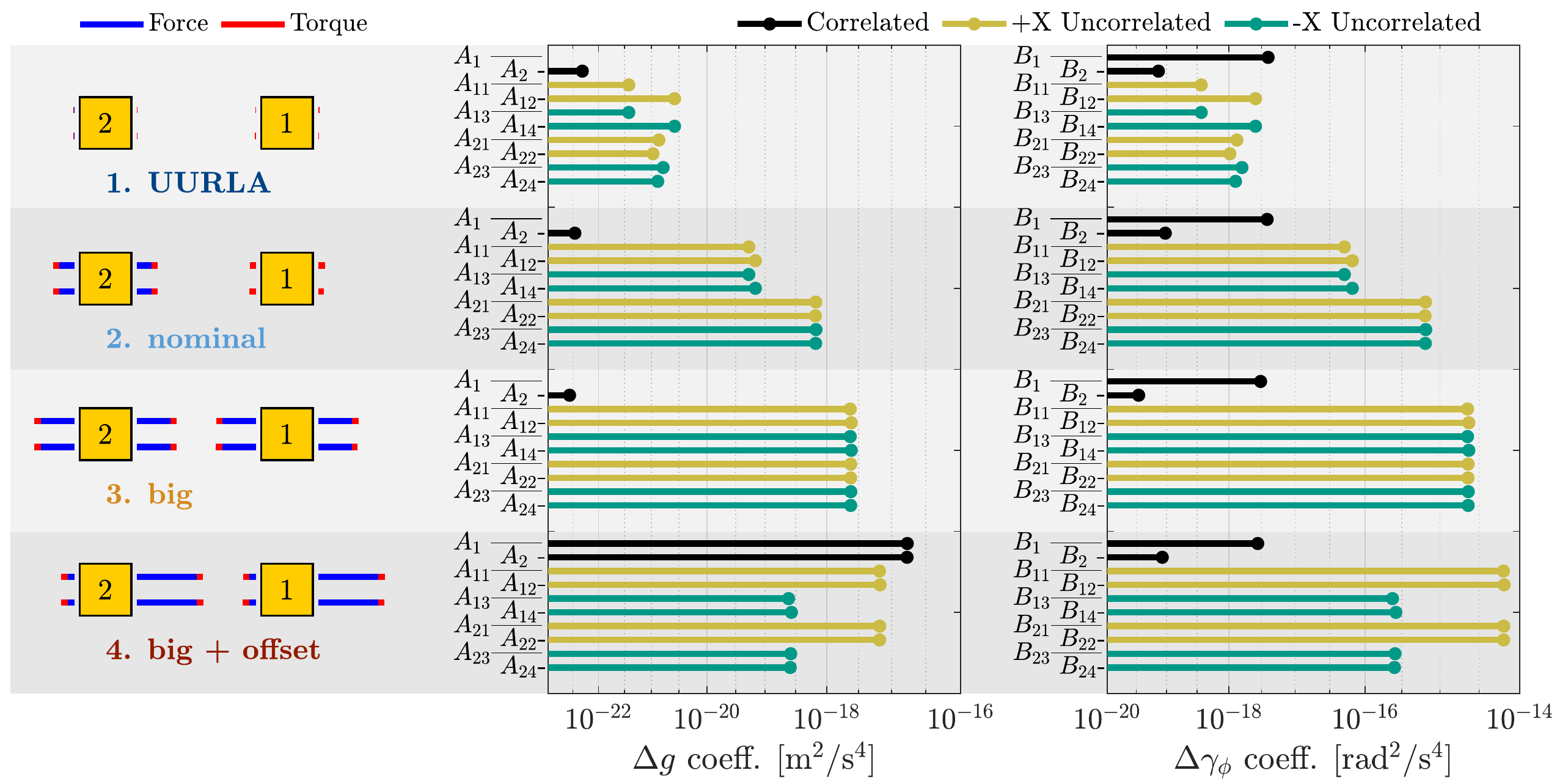}
\caption{\label{fig_illust_AB} 
At left, illustration of the applied force vectors for every electrode in the four tested actuation configurations, including force and torque contributions in, respectively, blue and red. At right are plotted the force and torque sensitivity coefficients for correlated ($A_i$ and $B_i$, black) and uncorrelated actuation gain noise ($A_{ij}$ and $B_{ij}$), with $+X$ (electrodes 1 and 2 in Fig.~\ref{fig_act_scheme}) and $-X$ (electrodes 3 and 4) actuators shown in dark yellow and in green. 
The coefficients are calculated using the averaged commanded forces and torques and force/torque authorities, as described in
the text.  }
\end{figure*}

\subsection{Actuation noise measurement campaign}
Quantifying the actuation gain fluctuations, at least for the $x$/$\phi$ actuators that can give $x$ force noise, is important for the LPF $\Delta g$ noise budget and for a parametric projection to LISA.    We did this in LPF by observing the increase in the differential acceleration noise in a series of tests with increasing forces and torques.  The changes in the applied forces and torques are constrained by the need to maintain the same quasi-static torques, to keep each TM aligned to the spacecraft, and the same \textit{differential} applied forces, to hold the TM separation fixed.  

\begin{figure*}[t]
\includegraphics[width=6.6in]{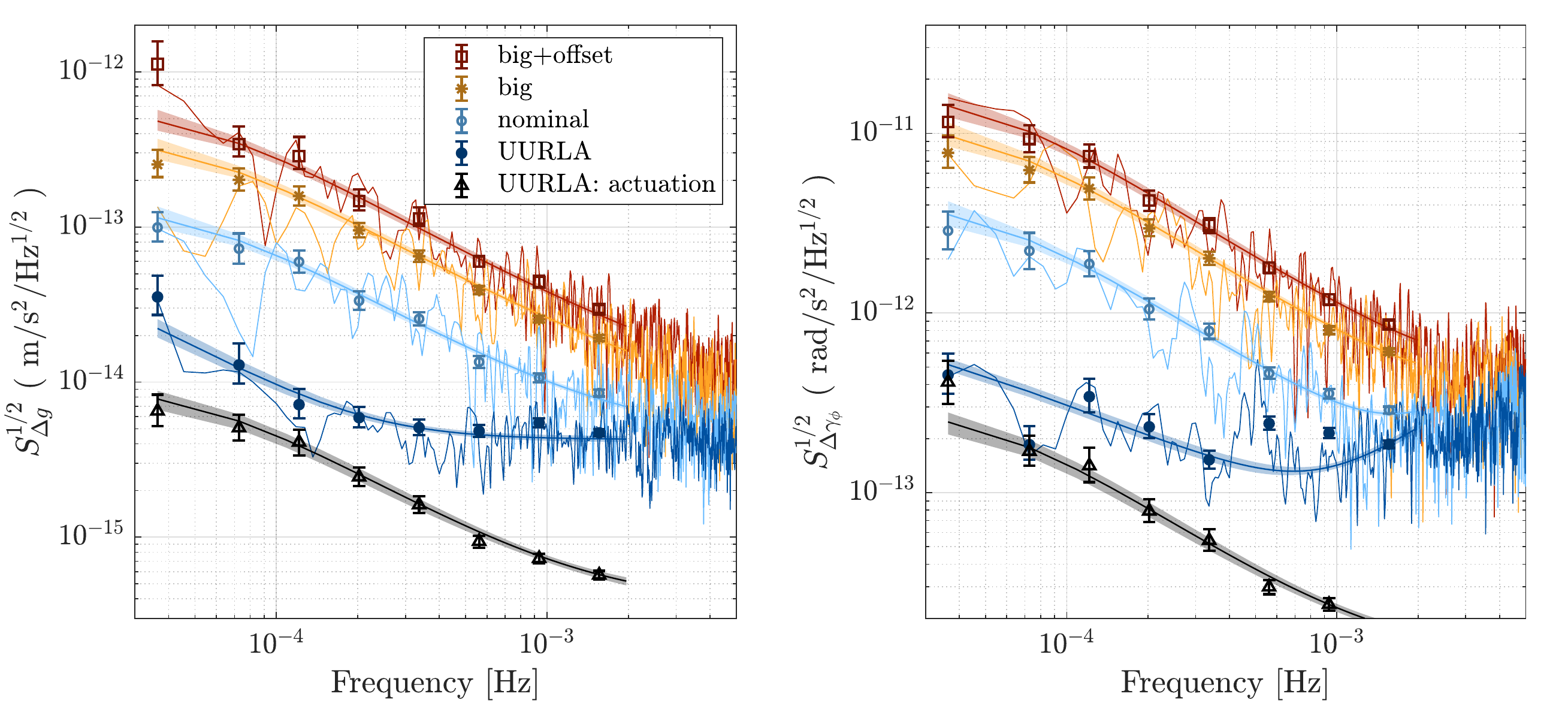}
\caption{\label{fig_s_alpha_exper} Experimental data for noise PSD for $\Delta g$ and 
$\Delta \gamma_{\phi}$ in the 4 different actuation configurations measured, including 
a fit to the actuation noise model (described in the text and in Appendix B), 
performed both for a ``smooth'' model of the noise frequency dependence (colored bands) 
and on a frequency by frequency analysis (individual points with error bars). Both plots contain (dark gray) 
the model prediction for the actuation gain noise contribution in the UURLA low force authority configuration 
used in the published benchmark plots for LPF acceleration noise \cite{LPF_prl_2016, LPF_prl_2018}.}
\end{figure*}

In addition to the differential $x$ acceleration $\Delta g$, our analysis of the tests considers also the measured differential angular acceleration, $\Delta \gamma_{\phi}$, which is sensitive to the same 8 $X$ electrode actuators and thus to the same gain fluctuations $\alpha$, $\alpha_i$, and $\alpha_{ij}$.  This is defined  
\begin{equation}
\Delta \gamma_{\phi} = \ddot{\phi_2} - \ddot{\phi_1} - \gamma_{\phi_{2c}} + \gamma_{\phi_{1c}} 
+ \omega^2_{\phi_2} \phi_2 - \omega^2_{\phi_1} \phi_1
\end{equation}
The noise in $\Delta \gamma_{\phi}$ from actuation gain fluctuations is modelled analogously to that
in $\Delta g$ (Eqns.~\ref{eqn_SDg_full} and \ref{eqn_act_accel_coeff}), 
\begin{equation}
\Delta \gamma_{\phi}^{ACT} \left( t \right)
=  b \, \alpha \left( t \right)
+ \displaystyle \sum_{i=1,2} b_i \, \alpha_i \left( t \right)
 + \displaystyle \sum_{ij} b_{ij} \, \alpha_{ij} \left( t \right)
\label{eqn_act_ang_accel_coeff}
\end{equation}
%\begin{equation}
%S_{\Delta \gamma_{\phi}}^{ACT} 
%%  =  4 \left( \Delta \gamma_{\phi} \right) ^2 S_{\alpha} + 
%%4 g_{c1}^2 S_{\alpha_1} + 4 g_{c2}^2 S_{\alpha_2} 
%%+ \displaystyle\sum_{ij} A_{ij} S_{\alpha_{ij}}
% =  B S_{\alpha} + \displaystyle \sum_{i=1,2} B_i S_{\alpha_i} + \displaystyle\sum_{ij} B_{ij} S_{\alpha_{ij}}
%\label{eqn_SDg_phi_full}
%\end{equation}
with coefficients $b$, $b_i$ and $b_{ij}$ coupling gain noise into angular acceleration.  For instance $b = 2 \: \Delta \gamma_{\phi}^{DC} $ and $b_1 = 2 \: \gamma_{\phi_1c} $, while for the independent fluctuations in the gain of, for instance, electrode 1 of TM2, we find  $b_{21} = R^{\star}_{\phi} a_{21}$ (see Appendix B for a complete description of the actuation noise model and analysis techniques).  

This allows analysis of the differential acceleration noise $S_{\Delta \gamma_{\phi}}$ -- with coefficients $B$, $B_i$, and $B_{ij}$ in direct analogy with the translational differential acceleration noise coefficients  $A$ of Eqn.~\ref{eqn_act_accel_coeff}.  Additionally, it allows an analysis of the cross-spectrum $S_{\Delta g, \Delta \gamma_{\phi}}$ representing the correlation between the fluctuations of differential translational and rotational accelerations.  For instance a gain fluctuation $\alpha_{i1}$ or $\alpha_{i4}$, on an electrode 1 or 4 in Fig.~\ref{fig_act_scheme} -- the ``bottom'' electrodes in this view -- will produce correlated fluctuations in $\Delta g$ and $\Delta \gamma_{\phi}$ with the same sign -- thus
$S_{\Delta g, \Delta \gamma_{\phi}} > 0$ -- while a gain fluctuation on ``top'' electrodes 2 or 3 will produce anti-correlated fluctuations and thus a negative cross spectrum. Including $S_{\Delta g, \Delta \gamma_{\phi}}$ in our analysis thus further helps distinguish between the electrodes creating force noise.      

%For instance $B = 4\left( \Delta \gamma_{\phi}^{DC} \right)^2$ and $B_1 = 4\left( \gamma_{\phi_1c} \right)^2$ 
%(see Appendix B).  

Noise measurements were performed in four different actuation configurations 
over 10 days during May 2016. The net forces/torques and 
authorities are shown in Table~\ref{tab_act_configs}, and the corresponding 
electrode-by-electrode force vectors are shown in Fig.~\ref{fig_illust_AB} along with the 
$A$ and $B$ sensitivity coefficients.  The first experiment (``UURLA'') 
employs the typical LPF science configuration used in the published differential
acceleration noise data \cite{LPF_prl_2016, LPF_prl_2016}, with minimum authorities.  
Configurations 2 and 3 increase the force and torque authorities ($g_0$ and $\gamma_0$), 
first to the 
``nominal'' level (Exp.~2), and then (Exp.~3) a further increase in the force authority to 
2.6~nm/s$^2$ on each TM (``big'' configuration).  
These increased authority experiments essentially leave the net applied forces and 
torques ($g_{ic}$ and $\gamma_{\phi_ic}$) unchanged but increase the forces on the single 
electrodes, increasing sensitivity to 
uncorrelated gain fluctuations (coefficients $A_{ij}$ and $B_{ij}$).
Experiment~4 adds an applied ``out-of-loop'' offset
force ($\approx 2.1$~nm/s$^2$) to TM1 to produce, in closed loop, a common mode force on both TM
(and thus also a net spacecraft acceleration).   This final 
experiment increases sensitivity both to the $+X$ actuators and   
to the board-correlated fluctuations for each TM ($A_{i}$); the degeneracy between these two effects 
in $\Delta g$ is broken by observing the differential angular acceleration noise, 
as the applied net DC torques  -- and thus the sensitivity to board correlated fluctuations through $B_i$ -- are unchanged.  With more experimental time available, further measurements could have employed large negative forces or other configurations to help isolate individual noise contributors. 

The dataset 
can be thought of as a two channel ($x$ and $\phi$) acceleration noise 
test of three ``enhanced-actuation'' configurations producing noise above a background level
measured in the first (UURLA) configuration. The experiments performed are sensitive to all 
of the 11~actuation gain noise generators in the proposed model; some combinations of 
these noise PSD are clearly resolved, while others are found to be compatible with zero to 
within upper limits that place significant experimental bounds on the circuitry gain noise.

The measured acceleration noise levels in $\Delta g$ and $\Delta \gamma_{\phi}$ are shown in Fig.~\ref{fig_s_alpha_exper}.  The solid curves are standard Welch periodograms, with 50\% overlapping 110000~s Blackman-Harris window, while the discrete data points, with error bars, are calculated with a variable window length adapted to the frequency, as in Ref.~\cite{LPF_prl_2018}. For the three ``enhanced actuation'' experiments, two 110000~s windows are used for the minimum frequency point at 36.4~$\mu$Hz, nine~33000~s windows at 121.2~$\mu$Hz, and  76~ 4300~s windows for the point at 0.93~mHz.

The visible progressive increase in the acceleration noise over the four experiments merits comment before discussing a fit to the actuation noise model: 
\begin{itemize}  
\item The noise increase with force authority ($g_0$, $\gamma_0$), clearly resolved from Expt.~1 UURLA (dark blue) to the nominal (Expt.~2, light blue) and ``big'' (Expt.~3, orange) tests, quantifies the key role of uncorrelated gain fluctuations.  A gain fluctuation correlated across all 4 TM1 or TM2 actuators ($\alpha_1$ or $\alpha_2$) would not increase force or torque noise in these tests where the net applied forces and torques are essentially left unchanged. 
\item The measured acceleration noise in the Expt.~2 ``nominal'' configuration would have set the LPF acceleration noise floor had the gravitational balance not allowed lowering the force authorities.  At 0.1~mHz, reducing the authorities to the UURLA configuration improves the overall acceleration noise floor by roughly a factor 50~in noise power, a substantial decrease allowing a much more stringent experimental anchor to the LISA low frequency mission requirements \cite{LISA_proposal_2017}.  
\item  The modest increase in both $\Delta g$ and $\Delta \gamma_{\phi}$ upon application of a large offset force, from Expt.~3 (orange) to Expt.~4 (red), confirms the domination of uncorrelated gain fluctuations over the board-correlated gain fluctuations.  A large correlated noise contribution ($\alpha_1$ or $\alpha_2$) would have produced a more sizable increase in $S_{\Delta g}$ without any effect on $S_{\Delta \gamma_{\phi}}$.  
\end{itemize}

% fit, priors
% do not resolve all - -degeneracies but also the fact that certain noise generators are small -- but can 
% put relevant upper limits on all
% 
\begin{figure*} [thb]
\includegraphics[width=6.6in]{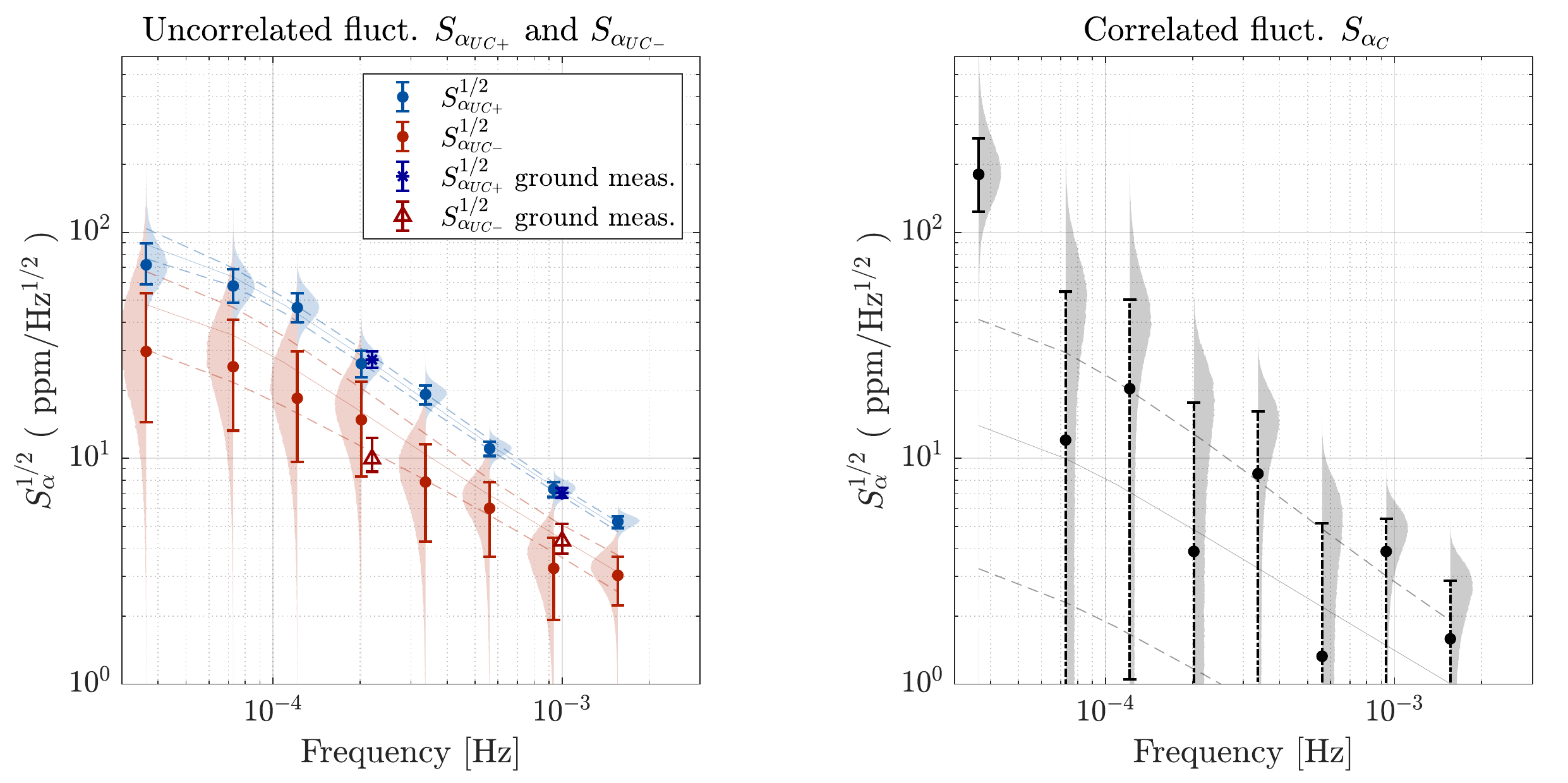}
\caption{\label{fig_S_alpha} Actuation gain noise results, for the uncorrelated single electrode 
noise in the $+X$ and $-X$ actuators at left and, at right, for the (largely unresolved) board correlated actuation gain noise, with posterior distributions for discrete frequencies shown as shaded regions.  The upper limits of all data points represent robust experimental constraints, while dashed error bars represent points for which the lower limit is largely determined by the prior assumptions.  Results for a smooth, analytic frequency-dependent model fit are also shown as solid lines, with dashed lines defining the $\pm \sigma$ confidence intervals. Also shown are the results, at 0.2~mHz and 1~mHz, from a pre-flight ground measurement campaign with the same flight electronics.
  }
\end{figure*}

A fit of the acceleration noise dataset to the actuation noise model described by Eqn.~\ref{eqn_SDg_full} (more generally by 
Eqn.~\ref{eq:act_model} in App.~B) is also shown in Fig.~\ref{fig_s_alpha_exper}, with the fit analysis first performed separately at each frequency (discrete points) and then to a phenomenological analytical frequency dependence model (smooth bands).  The fit, performed using a Bayesian Markov chain Monte Carlo (MCMC) approach, uses all the actuation terms in Eqn.~\ref{eq:act_model}, plus background acceleration noise that is independent of the applied actuation forces and torques.  This background acceleration noise includes also the first actuation term for completely correlated noise $S_{\alpha}$ (Eqn.~\ref{eqn_SDg_full}), as the coupling to acceleration noise --  via $\Delta g^{DC}$ in translation and $\Delta \gamma_{\phi}^{DC}$ for rotational -- is virtually unchanged across the four experiments.  

Not all of the ten actuation noise generators are resolved in this analysis, 
and a fit to a reduced set of parameters would be sufficient to describe measured noise in the four experiments; nonetheless, we include all these terms in the fit in order to predict the actuation noise in other LPF configurations with arbitrary applied force and torque values.Additionally, while we do not resolve the noise in every individual noise generator, we do put relevant experimental upper limits on \textit{all} noise generators; no single noise generator can create more noise than the total noise observed in the experimental data for $\Delta g$ and $\Delta \gamma_{\phi}$, and this constrains the upper limit on the PSD of each single noise generator.   

% chop this:
%, considering the distributions, with proper correlations, for parameters describing the noise in all the different %generators in the model ($S_{\alpha_1}$ and $S_{\alpha_2}$ and eight uncorrelated electrode gain noises  
%$S_{\alpha_{ij}}%$).  

%  $\chi^2$-likelihood characteristic of Welch periodogram PSD
%\color{red} 
The MCMC fitting technique considers a likelihood with the proper statistics for Welch periodogram estimates of 
PSD and cross-PSD (CPSD) for Gaussian noise processes \cite{fit_noise_prd, LorenzPhD} and is described in detail in Appendix B, along with assumed priors on parameters and typical observed posterior distributions.  We note here that the fit is parametrized in terms of averaged levels of board correlated gain noise ($S_{\alpha_C}$) and uncorrelated individual electrode gain noise ($S_{\alpha_{\mathrm{UC}}}$), with secondary parameters ($\mu_l$) describing the division of noise into the specific TM or specific electrodes. We employ an uninformative prior \cite{jeffrey} for the parameters $S_{\alpha_{\mathrm{UC}}}$ and $S_{\alpha_{\mathrm{C}}}$, with a distribution that is uniform in logarithmic space, so as not to constrain the order of magnitude of the gain noises.   In the case of the ``board correlated'' noise $S_{\alpha_{\mathrm{C}}}$, which is poorly resolved in our data, it was necessary to add a lower limit ``cutoff'' to the prior distribution, physically motivated but conservative (see App.~B) to ensure convergence of the Markov chain.

In Fig.~\ref{fig_s_alpha_exper} we also show, as black points, the projection of the actuation only -- without the background acceleration noise -- contributions to acceleration noise in the low authority UURLA configuration.  While in the three ``increased actuation experiments'' the measured acceleration noise is almost entirely due to actuation gain fluctuations, in the UURLA configuration actuation gain noise explains a significant, but not dominant, fraction of the measured noise in $\Delta g$ across the relevant sub-mHz bandwidth.  This important conclusion applies here and to the longer duration ``benchmark'' differential acceleration noise tests that are the published legacy of LISA Pathfinder \cite{LPF_prl_2016,LPF_prl_2018}.  In rotational acceleration, however, the actuation gain contribution would appear to explain all the measured noise in $\Delta \gamma_{\phi}$ at the very lowest frequencies, below 100~$\mu$Hz.

The uncorrelated actuation gain noise, averaged over the 8 individual electrode actuators, is resolved at all frequencies studied.  While we do not resolve every individual electrode gain noise level, the four different experimental configurations and two ``measurement channels''  ($\Delta g$ and $\Delta \gamma_{\phi}$) \textit{do} allow resolution of different combinations of the uncorrelated noise in these actuators.  For instance we resolve
the contributions of the groups of actuators used to apply $+x$ and $-x$ forces,  
\begin{eqnarray}
S_{\alpha_{\mathrm{UC+}}} & \equiv & \frac{1}{4} \left( S_{\alpha_{11}} + S_{\alpha_{12}} 
+ S_{\alpha_{21}} + S_{\alpha_{22}}\right) \nonumber \\
S_{\alpha_{\mathrm{UC-}}} & \equiv & \frac{1}{4} \left( S_{\alpha_{13}} + S_{\alpha_{14}} 
+ S_{\alpha_{23}} + S_{\alpha_{24}}\right) \nonumber \\
\label{eqn_S_UC_ave}
\end{eqnarray}
Distributions for $S_{\alpha_{\mathrm{UC+}}}$ and $S_{\alpha_{\mathrm{UC-}}}$ are evaluated by simple summing of the MCMC chains for the individual noise parameters.  The central (median) and $\pm \sigma$ values for $S_{\alpha_{\mathrm{UC+}}}$ and $S_{\alpha_{\mathrm{UC-}}}$, along with the underlying distributions, are shown in Fig.~\ref{fig_S_alpha}.  In the left panel, the noise in the $+X$ actuators is clearly resolved, with $\pm \sigma$ intervals of roughly $\left[ 40 , 55 \right]$~ppm/Hz$^{1/2}$ at 121~$\mu$Hz and $\left[ 6.7 , 7.9 \right]$~ppm/Hz$^{1/2}$ at 0.94~mHz. The noise in the negative actuators is smaller and more weakly resolved, with $\pm \sigma$ intervals of roughly $\left[ 10, 30 \right]$~ppm/Hz$^{1/2}$ and $\left[ 2, 4.5 \right]$~ppm/Hz$^{1/2}$ at the same two frequencies. The $-X$ actuator noise result is thus weakly detected, at roughly the $2 \sigma$ level in noise power, with the upper limit having more relevance to our conclusions. The experiment is slightly more sensitive to the noise in the $+X$ actuators, due to the large positive forces in the ``big + offset'' test; nonetheless the chance statistical difference between the groups of (nominally identical) $+X$ and $-X$ actuators is significant and resolved across a large frequency range.  

%\color{red} at least considering single points at fixed frequency, only marginally incompatible with zero\color{black}

We note that these results are consistent with estimates from ground measurement campaigns with the same exact actuation circuits performed years before launch.  The results of those tests, which used a lock-in amplifier to measure the differential gain noise between the same electrode on the two TM-- for instance $\left( \alpha_{21} - \alpha_{11} \right)$ -- are added as additional ``ground meas'' points near 0.2~mHz and 1~mHz in Fig.~\ref{fig_S_alpha}.  The results are compatible, with the specific sample of the four $+X$ actuators observed to be systematically noisier than the $-X$ circuits.  

The measurement campaign is also sensitive to correlated gain fluctuations, $S_{\alpha_1}$ and $S_{\alpha_2}$, among the two sets of four actuators used for each TM, in particular in the ``big + offset'' experiment with a large $+x$ force ($\approx 2$~nm/s$^2$) on both TM.  However, the experimental level of correlated gain noise was low enough that the averaged ``board correlated'' gain noise, defined
\begin{equation}
S_{\alpha_{\mathrm{C}}} \equiv \frac{1}{2} \left( S_{\alpha_{1}} + S_{\alpha_{2}} \right) 
\label{eqn_S_C_ave}
\end{equation}
is essentially compatible with zero, for all but the lowest frequency analyzed, with lower limits that are strongly dependent on the prior distribution assumptions.  Such points are indicated with dashed error bars in plot in the right panel in Fig.~\ref{fig_S_alpha}, while the criteria for distinguishing such points is discussed in Appendix~B. The measurements do allow however robust experimental upper limits in the posterior distribution of $S_{\alpha_\mathrm{C}}$.  These upper limits are virtually independent of the prior assumptions and are experimentally constraining, at a level of roughly 50 and 6~ppm/Hz$^{1/2}$ at, respectively, 121~$\mu$Hz and 0.94~mHz. Inclusion of the ``board correlated'' terms ($A_i$, $B_i$ and $C_i$) has little or no impact on the overall fit, and we would recover the same values for the uncorrelated noise to within $1 \sigma$ in a simplified fit without the board-correlated terms.  However, we know the board-correlated noise is present, at least through the voltage reference noise, and thus we keep these terms in our fit in order to place upper limits on this potentially important noise contribution. 

%with lower tails that extend orders of magnitude below the probability peak and are essentially only limited by the \textit{a %priori} distribution assumptions.    

% LT1021 ... 
% 0.013 ppm/sqrt(Hz) at 10 Hz, with 1/f corner at 10 Hz ... 
% thus expect 1.3e-8 * sqrt(10/1e-4) = 1.3e-8 * 300 = 4 ppm/sqrt(Hz)  
% comment on T noise?  
% Davor say: worst case 5.2 ppm/K
% LPF T fluctuations paper, gives 4e-2 K/sqrt(Hz) (worst case) at 0.1 mHz
% this gives 0.2 ppm/sqrt(Hz) 

We note also that our ``non-detection'' of the board-correlated gain noise is consistent with datasheet estimates for the voltage reference used in the actuation circuitry \cite{lt1021}, for which a very rough extrapolation of the $f^{-1}$ noise measured around 10~Hz would give several ppm/Hz$^{1/2}$ in our sub-mHz band.  Additionally, no correlation of noise with platform-level thermometers is observed, and none is expected: considering datasheet values for ``worst case'' voltage reference temperature coefficients (roughly 5~ppm/K) with typical platform-level temperature fluctuations \cite{lpf_T_noise}, including those measured at the FEE,  at 0.1~mHz 
at or below 0.1~K/Hz$^{1/2}$, temperature-driven voltage reference noise would be below the ppm/Hz$^{1/2}$ level (and thus not relevant at our measurement levels).   In any case, our measurements indicate that board-correlated common mode gain noise is a small contributor
to the LPF acceleration noise data, where applied forces were consistently below 20~pm/s$^2$.   
 
\begin{figure*} [thb]
\includegraphics[width=6.6in]{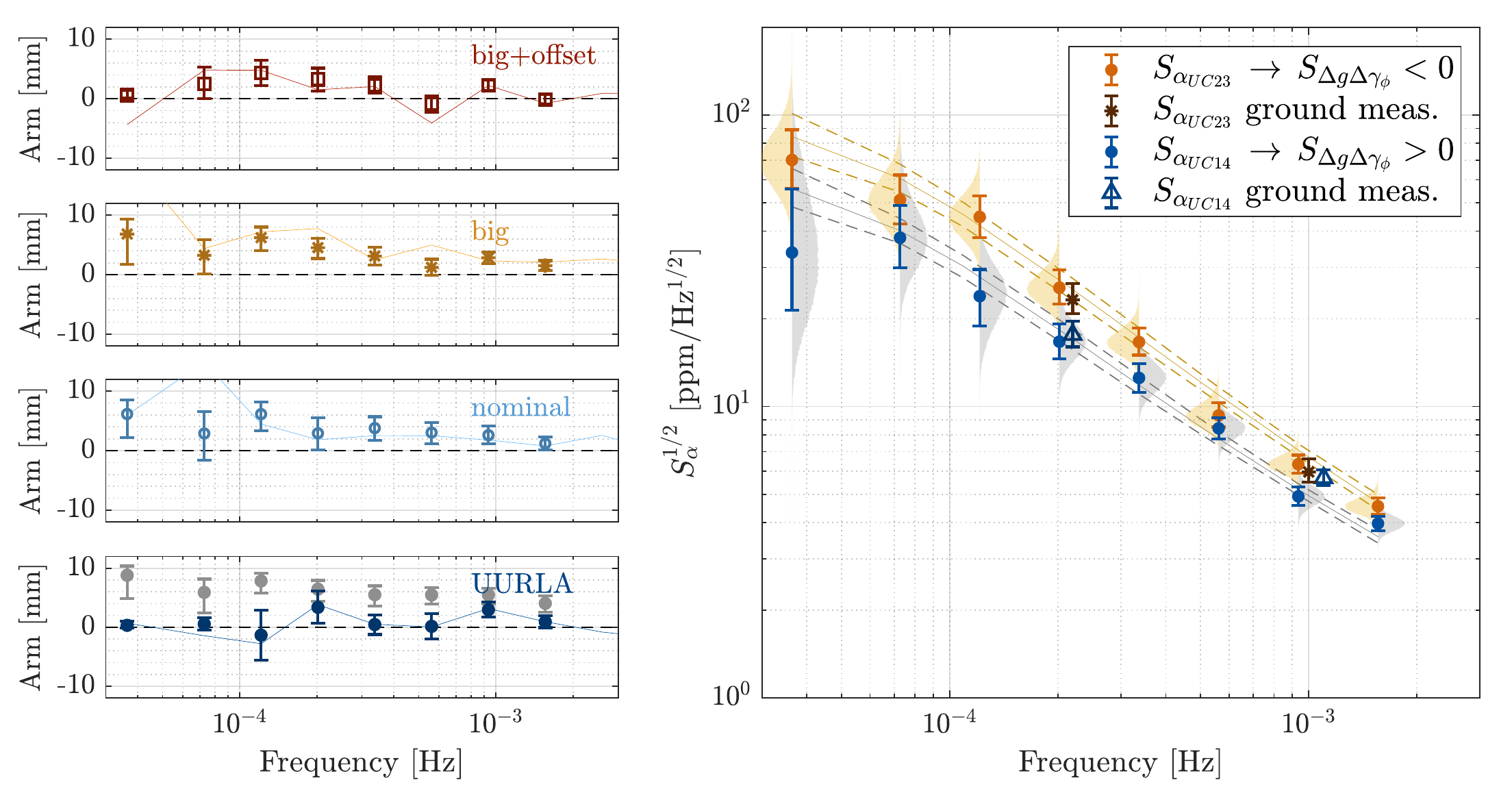}
\caption{\label{fig_cross} At left, solid lines are raw cross spectra between $\Delta g$ and $\Delta \gamma_{\phi}$, expressed as the effective armlength $r_{\phi}$ as defined in the text. Also shown, as discrete points with error bars, are the fit model estimates of the effective arm (for UURLA, where actuation noise is sub-dominant, we also show the armlength extracted for the actuation contribution, in gray). At right are the values of uncorrelated gain noise in the ``top'' electrodes 2/3 and in the ``bottom'' electrodes 1/4, along with estimates from on-ground electronics tests (darker points at 0.2~mHz and 1~mHz).  We note that $S_{\alpha_{\mathrm{UC23}}} > S_{\alpha_{\mathrm{UC14}}}$ is consistent with positive armlength, $r_{\phi}$.}
\end{figure*}

An additional ``raw'' data curve for our actuation gain noise measurement campaign is that of the cross-spectrum between fluctuations in $\Delta g$ and $\Delta \gamma_{\phi}$, which allows us to resolve a different combination of the uncorrelated electrode gain noises.  This is shown for the four experiments at the left in Fig.~\ref{fig_cross}, with solid curves representing the raw periodogram cross-spectral estimates, expressed here in terms of an effective armlength defined
\begin{equation}
r_{\phi} \equiv 
-\frac{I}{M}\frac{S_{\Delta g, \Delta \gamma_{\phi}} }{S_{\Delta g}} \approx -\frac{s^2}{6}\frac{S_{\Delta g, \Delta \gamma_{\phi}} }{S_{\Delta g}}
\label{eqn_arm}
\end{equation}  
where $s = 46$~mm is the TM sidelength.  The sign convention is chosen such that a positive armlength corresponds to a predominance of force noise acting on the $X$ faces of the TM with a force center displaced positively along the $y$ axis.  For instance, if \textit{all} the relevant force noise were coming from actuation gain fluctuations originating in ``top'' electrode 2 or 3 of either TM, the translational and rotational force noise would have perfect negative correlation, with a resulting armlength 
$r_{\phi} \approx \frac{\left| \pder{C_X^{\star}}{x} \right|}{\left| \pder{C_X^{\star}}{\phi} \right| } \approx 11$~mm, half the on-center separation
between adjacent electrodes on the sensor $X$ face.  If instead a ``bottom'' electrode 1 or 4 dominated we would find $r_{\phi} \approx -11$~mm, while for force noise spread equally between different $X$ electrodes $r_{\phi}$ would tend to zero. Including the cross-spectrum data into the global fit helps break degeneracy between gain noise from electrodes 1/4 and from electrodes 2/3.

On the same graphs, at discrete frequencies we show the armlength extracted from the fit parameters.  This is dominated by actuation, except for the low force UURLA test (where we also show both the fit-model actuation-only armlength, in gray).    

We note, in both the raw armlength data and in the fit model prediction, a tendency towards positive armlengths of several mm, across the sub-mHz frequency band.  
%As the applied force vectors display a roughly ``top-bottom'' symmetry -- and thus $A_{i1} \approx A_{i2}$,  $A_{i4} \approx %A_{i3}$, $B_{i1} \approx B_{i2}$, and $B_{i4} \approx B_{i3}$ -- 
With our applied forces, the typically positive values of $r_{\phi}$ indicate that electrodes 2/3 are slightly but consistently noisier than electrodes 1/4. This is reflected in the extracted values for the ``top-bottom'' groupings of the uncorrelated gain noise for the (four) electrodes 1/4 and the group of four electrodes 2/3,  
\begin{eqnarray}
S_{\alpha_{\mathrm{UC14}}} & \equiv & \frac{1}{4} \left( S_{\alpha_{11}} + S_{\alpha_{14}} 
+ S_{\alpha_{21}} + S_{\alpha_{24}}\right) \nonumber \\
S_{\alpha_{\mathrm{UC23}}} & \equiv & \frac{1}{4} \left( S_{\alpha_{12}} + S_{\alpha_{13}} 
+ S_{\alpha_{22}} + S_{\alpha_{23}}\right) \nonumber \\
\label{eqn_S_UC_14_23}
\end{eqnarray}
As with the statistical difference between the $+X$ and $-X$ electrodes, these results are consistent with the ground measurement campaign results measured years before, which also detected higher noise in the 2/3 electrode pairs relative to electrodes 1/4 (see discrete ``ground measurement'' points in Fig.~\ref{fig_cross}). 

%\color{red} note to VC: I removed the comments about electrodes 2/3 also noisier
%as the $2X$ sensing channel, too confusing without further explanation\color{black}
%We note that the electrodes 2/3 compose the $2x$ capacitive position sensing channels, which were also %measured to be slightly noisier then the $1x$ sensing channels, composed of electrode 1/4, for both %LPF~TM \cite{prd_2017_cap_sens}, but this is just an observation and possibly a coincidence, for which %at present we have no explanation.   

Finally, given the observed ``smooth'' frequency dependence of the model fit -- in the total acceleration noise shown in Fig.~\ref{fig_s_alpha_exper} but also that of the fit parameters for the different actuation gain noises, $S_{\alpha_\mathrm{C}}$ and $S_{\alpha_{\mathrm{UC}}}$  and the groups $S_{\alpha_{\mathrm{UC+}}}$/$S_{\alpha_{\mathrm{UC-}}}$ (Fig.~\ref{fig_S_alpha}) and $S_{\alpha_{\mathrm{UC14}}}$/$S_{\alpha_{\mathrm{UC23}}}$ (Fig.~\ref{fig_cross}) -- we also perform a fit with an analytical model of the frequency dependence of the actuation noise PSD, with $f^{-1}$ and $\frac{1}{f^2 + f_{3dB}^2}$ terms.  These are shown as colored bands in Figs.~\ref{fig_s_alpha_exper} and \ref{fig_S_alpha}.  This analytical frequency-dependence does not represent a physically unique model but is chosen empirically based on the results emerging from the analysis performed at discrete frequencies, in order to allow a simple projection to other experiments at arbitrary intermediate frequencies. 

The $f_{3dB}$ term is included to describe the observed flattening of the PSD at low frequencies, 
with an extracted roll-off frequency of $55 \pm 15$~$\mu$Hz.   While this observed noise ``saturation'' is only slightly incompatible with a simple $f^{-2}$, it does indicate that there is no dramatic noise increase just below the 100~$\mu$Hz band.   
\begin{figure*}[t]
\includegraphics[width=6.6in]{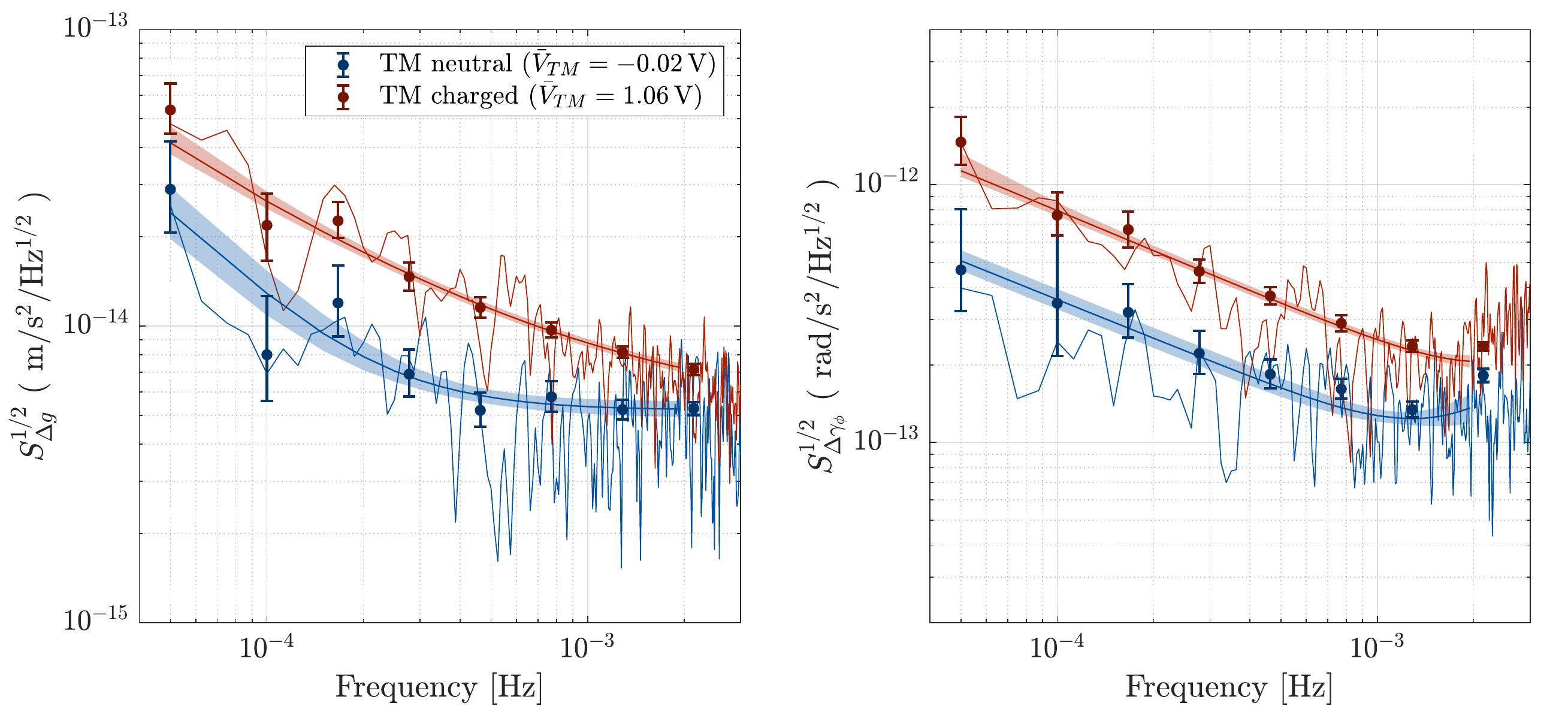}
\caption{\label{fig_SDg_charged} Noise in the differential acceleration (translational $\Delta g$ and rotational
$\Delta \gamma_{\phi}$) measured with the two LPF TM charged to roughly -1.06~V and with the TM nearly neutral.
Results of model fits considering noise in $\Delta \left( \Delta_x \right)$ and $\Delta \left( \Delta_{\phi} \right)$,
with analytic $f^{-1}$ and $f^{-2}$ frequency dependence, are shown as colored bands. 
}
\label{fig_SDg_charge}
\end{figure*}
It is also worth noting that the smooth model represents a fit with many degrees of freedom; with the 110000~s windows used in the fit, we have roughly 600 data points -- 4 experiments with 3 spectrums $S_{\Delta g}$, $S_{\Delta \gamma_{\phi}}$, and $S_{\Delta g, \Delta \gamma_{\phi}}$ and roughly 50 frequency bins -- and many fewer fit parameters, 26 in all.  This offers some chance for a posterior predictive ``goodness of fit'' test \cite{gelman2020bayesian}, performed by using the model noise parameters to predict the distribution of the Welch periodogram spectral estimates in a relatively short -- in this case 2 or 3~110000~s window -- measurements.  For each of the four experiments, assuming stationary Gaussian noise and an accurate model, we would expect to find 683\%  expect to find 68.3\% of the periodograms values in the $\pm \sigma$ interval of our model. For the four experimental runs, we find 59\%, 67\%, 75\% and 70\% of the points falling in the $\pm \sigma$ interval for, respectively, ``UURLA'', ``nominal'', ``big'' and ``big + offset''. While we do not associate the frequency-dependent fit with a rigorous physically motivated model, this goodness of fit indicator, in addition to the observed ``smoothness'' of the extracted noise parameters, shows consistency between the experimental data and the parametric actuation noise model considering Gaussian, stationary noise.

\section{In-band voltage fluctuations: noise model and experimental results with charged TM}
\label{sec_inband_noise}

In-band additive noise in the actuation voltages, described as $v_j \left( t \right)$ for electrode 
$j$ in Eqn.~\ref{eqn_dV},  couple to the mean TM potential, $V_{\mathrm{TM}}$, related mostly to the accumulated 
TM charge -- see line 2 of Eqn.~\ref{eqn_estat_f_act} and Eqn.~\ref{eqn_Delta_x_calc} 
of App.~A -- and to the residual DC biases on the electrodes (line 3 in  
Eqn.~\ref{eqn_estat_f_act}). The TM charge, through the 
mean TM potential $V_{\mathrm{TM}}$, is coupled to the average stray field, described by the 
translational potential difference $\Delta_x$ by (see App.~A or Ref.~\cite{charge_DC_bias_prl_2012}) 
\begin{equation}
F = -V_{\mathrm{TM}} \left| \pder{C_X}{x} \right| \: \Delta_x  
\label{eqn_Delta_x_ACT}
\end{equation}
with a similar expression for the angular acceleration in terms of the ``rotational potential 
imbalance'' $\Delta_{\phi}$ that couples to TM charge.  

These effective differential potentials that couple to TM charge are normalized to a single $X$ electrode capacitance, such that an additive offset in the actuation output voltage $v_j$,  applied homogeneously across each $X$ electrode, contributes in simple fashion to $\Delta_{x}$ and $\Delta_{\phi}$
\begin{eqnarray}
\Delta_x^{ACT} & = &  v_1 + v_2 - v_3 - v_4   \nonumber \\
\Delta_{\phi}^{ACT} & = &  v_1 - v_2 + v_3 - v_4  
\label{eqn_Delta_x_phi_v}
\end{eqnarray} 
for a given TM, with electrode numbering again as in Fig.~\ref{fig_act_scheme}.  While noise in 
$\Delta_x$ and $\Delta_{\phi}$ includes possible contributions from 
intrinsic ``patch field'' potentials on the gold TM and electrode housing surfaces, we 
can interpret the measured noise levels as an upper limit of the contribution from the actuation voltages (and will 
comment on this hypothesis shortly).   

Limiting acceleration noise from this interaction was achieved in LPF by intermittently discharging 
the TM, with UV illumination \cite{prd_2018_uv_discharge}, at intervals of 1-3 weeks.  With 
typical environmental charging 
approximately +25 elementary charges per second, 
the TM potential drifted away from neutral by as much as $V_{\mathrm{TM}} \approx 100$~mV \cite{lpf_charging}, with 
a residual sensitivity to noise in $\Delta_x$ via Eqn.~\ref{eqn_Delta_x_ACT}. 
The DC value of $\Delta_x$ was measured via the change
in force on the TM with a ``step'' change in TM charge\cite{LPF_prl_charge} and then nulled by application 
of DC actuation voltages, thus minimizing the force noise arising from TM charge fluctuations.  
The residual interaction of the noisy actuation voltages $v_j$ 
with the local stray DC biases can not in general be cancelled and remained a potentially 
relevant force noise source. 

\begin{figure*}[t]
\includegraphics[width=6.6in]{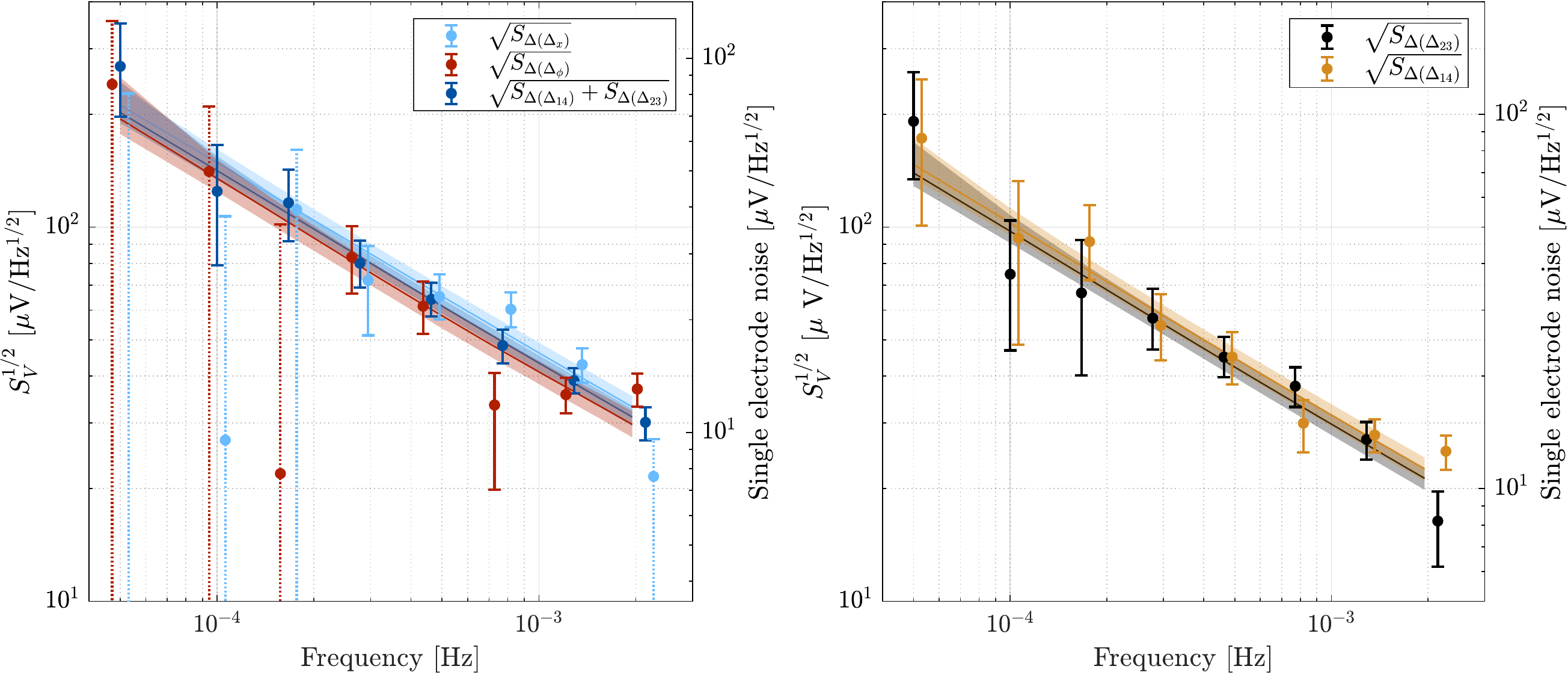}
\caption{\label{fig_Sv}  At left, noise in the actuation voltage combinations, $\Delta \left( \Delta_x \right)$ and 
$\Delta \left( \Delta_{\phi} \right)$, coupling to, respectively, translational and rotational acceleration noise
as estimated from the data in Fig.~\ref{fig_SDg_charge}.  Noise in the combinations $\Delta \left( \Delta_{12} \right)$ and 
$\Delta \left( \Delta_{23} \right)$ are assessed in a joint, correlated analysis of the $\Delta g$ and $\Delta \gamma_{\phi}$,
shown individually on the right and summed on the left.  For both graphs, a righthand axis indicates the effective
single electrode voltage noise inferred from the noise in the various combinations (all of these estimates are mutually 
consistent).}
\end{figure*}

%The possibility in LPF to controlledly change $V_{\mathrm{TM}}$ with the discharge system 
%allows amplifying the TM acceleration noise from the coupling to fluctuations in 
%$\Delta_x$ and $\Delta_{\phi}$ (and thus to the stray actuation 
%voltages $v_i$).

A dedicated experiment with a highly charged TM allowed quantification of the noise in $\Delta_x$ 
and $\Delta_{\phi}$, and thus to the stray ``in-band'' noise in the actuation voltages, $v_j$.   
A preliminary analysis of these tests was presented in Ref.~\cite{LPF_prl_charge}.  
Two new elements merit revisiting this experiment in the analyis presented here: 
\begin{itemize}
\item we calculate and subtract the effect of a deterministic roundoff error in
the average actuation voltage due to inaccuracies in the audio frequency actuation waveforms.  This error, closely tied to the actuation force inaccuracy 
described in Ref.~\cite{luigination_2020}, introduces a noisy voltage offset varying with 
the commanded control forces ($x$) and torques ($\phi$).  We subtract 
the calculated contribution to the TM acceleration to quantify the underlying stochastic voltage noise, 
which is thus
slightly but significantly below that originally estimated in Ref.~\cite{LPF_prl_charge}.      
\item we have added analysis of the measured differential TM angular acceleration, $\Delta \gamma_{\phi}$, 
in addition to $\Delta g$.  This allows 
measurement of a second combination of electrode actuation voltage noise and allows assessment of correlations in the 
voltage noise between different channels.  
\end{itemize}  

% compensating Delta_x

The measurements were performed from 1-4 May in 2016, first with the two TM nearly neutral -- $V_{\mathrm{TM}}$ of -16 and -24~mV
for, respectively, TM1 and 2 -- and then with the TM charged, to roughly -1.066 and -1.058~V.  The analyzed periods for the two experiments
have durations of 39 and 59 hours, respectively.  The measured acceleration noise levels are shown in Fig.~\ref{fig_SDg_charge}.  
A clear increase 
in the acceleration noise is measured in both $\Delta g$ and $\Delta \gamma_{\phi}$, which we can attribute to noise in the 
relevant stray voltage fields $\Delta_x$ and $\Delta_{\phi}$ in the two TM. 

The increase in TM acceleration noise can be translated into an effective voltage noise, considering the differential 
TM accelerations $\Delta g$ and $\Delta \gamma_{\phi}$ following the single TM treatment 
above (Eqn.~\ref{eqn_Delta_x_ACT}): 
\begin{eqnarray}
\delta \Delta g  
& =  &
- \frac{1}{M}  \left| \pder{C_X}{x} \right| \left[  V_{\mathrm{TM}2} \delta \Delta_{x_2} -  V_{\mathrm{TM}1} \delta \Delta_{x_1} \right]
\nonumber \\
& \approx & 
- \frac{1}{M}  \left| \pder{C_X}{x} \right| V_{\mathrm{TM}} \: \: \delta \Delta \left( \Delta_{x} \right) 
\nonumber \\
\delta \Delta \gamma_{\phi}
& \approx & 
- \frac{1}{I}  \left| \pder{C_X}{\phi} \right| V_{\mathrm{TM}} \: \: \delta \Delta \left( \Delta_{\phi} \right) 
\label{eqn_DDx}
\end{eqnarray}
with the approximation $V_{\mathrm{TM}1} \approx V_{\mathrm{TM}2} \equiv V_{\mathrm{TM}}$ valid in these experiments 
with the two TM each charged to roughly 1~V. Here, 
$\Delta g$ is sensitive to the 
difference in the relevant coupling potential 
$\Delta \left( \Delta_x \right) \equiv \Delta_{x_2} - \Delta_{x_1}$ and similar for the rotational acceleration
$\Delta \gamma_{\phi}$ in terms of the differential rotational coupling potential, $\Delta \left( \Delta_{\phi} \right)$.   
This results in acceleration noise given by
\begin{eqnarray}
S_{\Delta g} & =  & S_{\Delta g}^{BGND} + 
\left( \frac{V_{\mathrm{TM}}\pder{C_X}{x}}{M} \right)^2 \: S_{\Delta \left( \Delta_x \right) }   \nonumber \\
S_{\Delta \gamma_{\phi}} & =  & S_{\Delta \gamma_{\phi}}^{BGND} + 
\left( \frac{V_{\mathrm{TM}}\pder{C_X}{\phi}}{I} \right)^2 \: S_{\Delta \left( \Delta_{\phi} \right) }
\label{eqn_S_DDx}  
\end{eqnarray}
We perform independent analyses of $\Delta g$ and $\Delta \gamma_{\phi}$ to obtain $S_{\Delta \left( \Delta_x \right)}$ and 
$S_{\Delta \left( \Delta_{\phi} \right) }$, in both cases using both a single frequency 
analysis and a smooth frequency
dependence --  chosen empirically to 
include $f^{-1}$ and $f^{-2}$ terms for both $S_{\Delta \left( \Delta_{x} \right)}$ and  
 $S_{\Delta \left( \Delta_{\phi} \right) } $, with the observed data consistent with a $f^{-1}$ 
 dependence except perhaps below 100~$\mu$Hz where the $f^{-2}$ contribution could become relevant.  
The results of this analysis are shown in Fig.~\ref{fig_Sv}, with the frequency-smooth fits to the 
acceleration noise model overlayed in Fig.~\ref{fig_SDg_charge}.  We note that the independent 
analyses of the translational and rotational noise makes no assumptions about correlations between
voltage fluctuations on the different electrodes, and we will comment on this shortly.

The smooth fit for $S_{\Delta \left( \Delta_x \right)}$ around 0.1~mHz has a 
$\pm \sigma$ interval of roughly [135, 165]~$\mu$V/Hz$^{1/2}$, compared to [225, 320]~$\mu$V/Hz$^{1/2}$ reported
in Ref.~\cite{LPF_prl_charge}\footnote[1]{The result of [160, 200]~$\mu$V/Hz$^{1/2}$ in Ref.~\cite{LPF_prl_charge} 
was for $S_{\Delta_x}$ for a single TM, assuming that the two TM have uncorrelated, statistically equivalent 
noise in $\Delta_x$. The differential noise in $\Delta_x$ that is truly measured, and compared here, 
is thus simply multiplied by a factor two in power. }. This reduction is related to the deterministic
subtraction of the actuation voltage roundoff error. The results for $S_{\Delta \left( \Delta_{\phi} \right)}$ 
are similar to those for $\Delta_x$, with a $\pm \sigma$ interval [120, 150]~$\mu$V/Hz$^{1/2}$ at 0.1~mHz.  

\begin{figure}[t]
\includegraphics[width=3.3in]{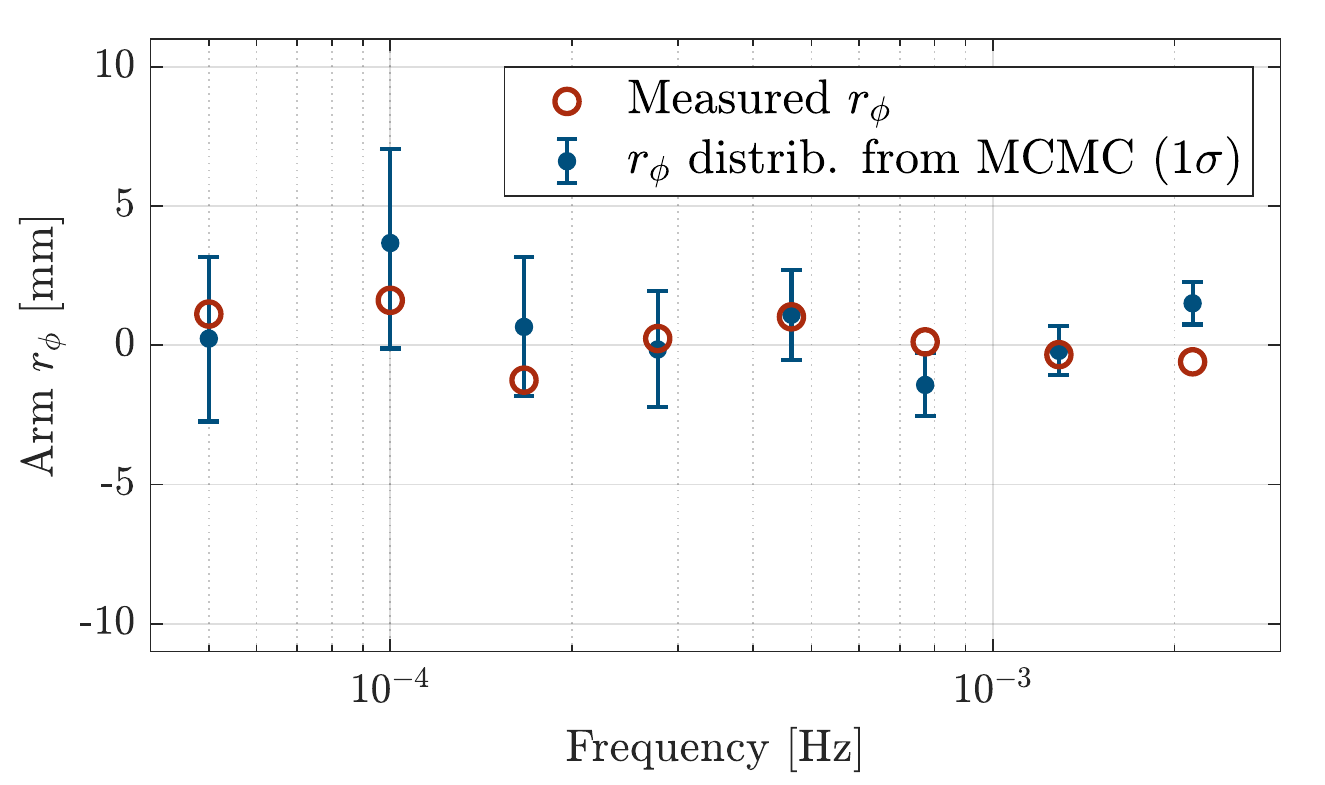}
\caption{\label{fig_cross_acceleration} Effective armlength $r_{\phi}$, based on cross-spectral density 
$S_{\Delta g,\Delta \gamma_{\phi}}$ as defined in Eqn.~\ref{eqn_arm}, 
in the charged TM experiment, raw (red) and reconstructed with $\pm \sigma$ error bars from 
the MCMC fit (blue).  The data are consistent with uncorrelated translational and rotational acceleration noise. }
\label{fig_cross_charge}
\end{figure}

We observe that $S_{\Delta \left( \Delta_{x} \right) } \approx S_{\Delta \left( \Delta_{\phi} \right) }$ 
across the frequency band.  If there were an important correlated fluctuations between actuation voltages 
on different electrodes, these could add or subtract differently in the 
combinations (see Eqn.~\ref{eqn_Delta_x_phi_v}) coupling
into force and torque, resulting in different noise levels in $\Delta_x$ and $\Delta_{\phi}$.  The absence 
of such a noise asymmetry is at least consistent with uncorrelated voltage fluctuations.  

Additionally, $S_{\Delta \left( \Delta_{x} \right) } \approx S_{\Delta \left( \Delta_{\phi} \right) }$ suggests that actuation voltage noise -- rather than fluctuations of ``patch'' potentials on the gold TM and electrode surfaces-- are dominating the interaction with the charged TM.  If patch field effects were dominant, one would expect $S_{\Delta \left( \Delta_{\phi} \right) } > S_{\Delta \left( \Delta_{x} \right) }$, by perhaps a factor 2-3.   This is expected first because the rotational combination $\Delta_{\phi}$ includes significant contributions from the electrode housing $X$ and $Y$ surfaces while $\Delta_x$ mainly involves just the $X$ surfaces. Second, the $Y$ electrode housing surfaces on the $Y$ faces are closer ($d_y=2.9$~mm while $d_x$=4~mm, with the relevant capacitance derivatives scaling as $d^{-2}$).  Thus the rough equivalence $S_{\Delta \left( \Delta_{x} \right) } \approx S_{\Delta \left( \Delta_{\phi} \right) }$is consistent with actuation voltage fluctuations.  

In the hypothesis that coupling to TM charge is dominated by uncorrelated in-band additive noise in the actuation voltages, we can further dissect the contribution of stray actuation 
voltage to the measured differential TM accelerations.  Specifically, 
\begin{eqnarray}
\Delta \left( \Delta_x^{ACT} \right) & \equiv &  \Delta_{x_2}^{ACT} - \Delta_{x_1}^{ACT} 
= \Delta \left( \Delta_{14} \right) + \Delta \left( \Delta_{23} \right)     
\nonumber \\
\Delta \left( \Delta_{\phi}^{ACT} \right) & \equiv &  \Delta_{\phi_2}^{ACT} - \Delta_{\phi_1}^{ACT} 
= \Delta \left( \Delta_{14} \right) - \Delta \left( \Delta_{23} \right)     
\label{eqn_Dg_charged_breakdown}
\end{eqnarray}
where $\Delta \left(  \Delta_{14} \right)$ and $\Delta \left(  \Delta_{23} \right)$ are each 4-electrode
actuation voltage noise differences, 
\begin{eqnarray}
\DDr & \equiv & v_{21} - v_{24} - v_{11} + v_{14}
\nonumber
\\
\DDl & \equiv & v_{22} - v_{23} - v_{12} + v_{13}
\label{eqn_DD14_DD23}
\end{eqnarray}
with $v_{ij}$ for the stray additive actuation voltage on electrode $j$ of the LPF TM $i$.  The sum and 
difference of the voltage combinations \DDr~ and \DDl~thus couple into differential translational
and rotational acceleration $\Delta g$ and $\Delta \gamma_{\phi}$.  
In the hypothesis that the noise in \DDr~ and \DDl~
are uncorrelated, we can see from Eqns.~\ref{eqn_DDx} and \ref{eqn_Dg_charged_breakdown} that the noise in
\Dg~ and \Dgphi~will each be proportional to the sum, $\left( S_{\DDr} + S_{\DDl} \right) $, while the cross-spectrum 
$S_{\Delta g, \Delta \gamma_{\phi}}$ will be proportional to the difference, 
$\left( S_{\DDr} - S_{\DDl} \right) $.  
Fitting to this model, including $\left\{ S_{\Dg} , S_{\Dgphi}, S_{\Dg,\Dgphi} \right\}$
in a single analysis, we obtain 
estimates for $S_{\DDr}$ and $S_{\DDl}$ (at right in Fig.~\ref{fig_Sv}).  
Consistent with a model of uncorrelated \DDr~ and \DDl, their sum is compatible 
with the individual analyses for $S_{\Delta \left( \Delta_x \right)}$ and 
$S_{\Delta \left( \Delta_{\phi} \right)}$ (shown on the left in Fig.~\ref{fig_Sv}).

%\begin{figure}[t]
%\includegraphics[width=3.3in]{../figs/cross_charge_fully_zoom.pdf}
%\includegraphics[width=3.3in]{../figs/single_electrode_noise.pdf}
%\caption{\label{fig_single}An equivalent uncorrelated single electrode voltage noise, obtained from the 
%values estimated for the $x$ axis stray field noise, $S_{\Delta \left( \Delta_x \right)}$ in Fig.~\ref{}.  \color{red} Correct this ...
%do we want to include estimates from $S_{\Delta \left( \Delta_x \right)}$, $S_{\Delta \left( \Delta_{\phi} \right)}$ and 
%$S_{\DDr} + S_{\DDl}$?. Or we %could just include this as a right axis label on the left figure 
%in Fig.~\ref{fig_Sv}.\color{black}
%}
%\label{fig_cross_charge}
%\end{figure}

Additionally, the cross-correlation between acceleration noise 
$\Delta g$ and $\Delta \gamma_{\phi}$ is compatible with zero across the frequency band studied (Fig.~\ref{fig_cross_charge}). 
If the increase in TM acceleration noise were dominated by a single ``bad'' noisy electrode, we would expect
a full force/torque correlation -- with armlength $r_{\phi} \approx \pm 11$~mm, depending on the electrode -- but that is 
not observed here.  Null cross-correlation is consistent with $S_{\DDr} \approx S_{\DDl}$, indicating that 
the summed noise in the two groups of 4 electrodes (Eqn.~\ref{eqn_DD14_DD23}) are statistically equivalent 
at our level of measurement resolution.  

\begin{figure*}[t]
\includegraphics[width=6.2in]{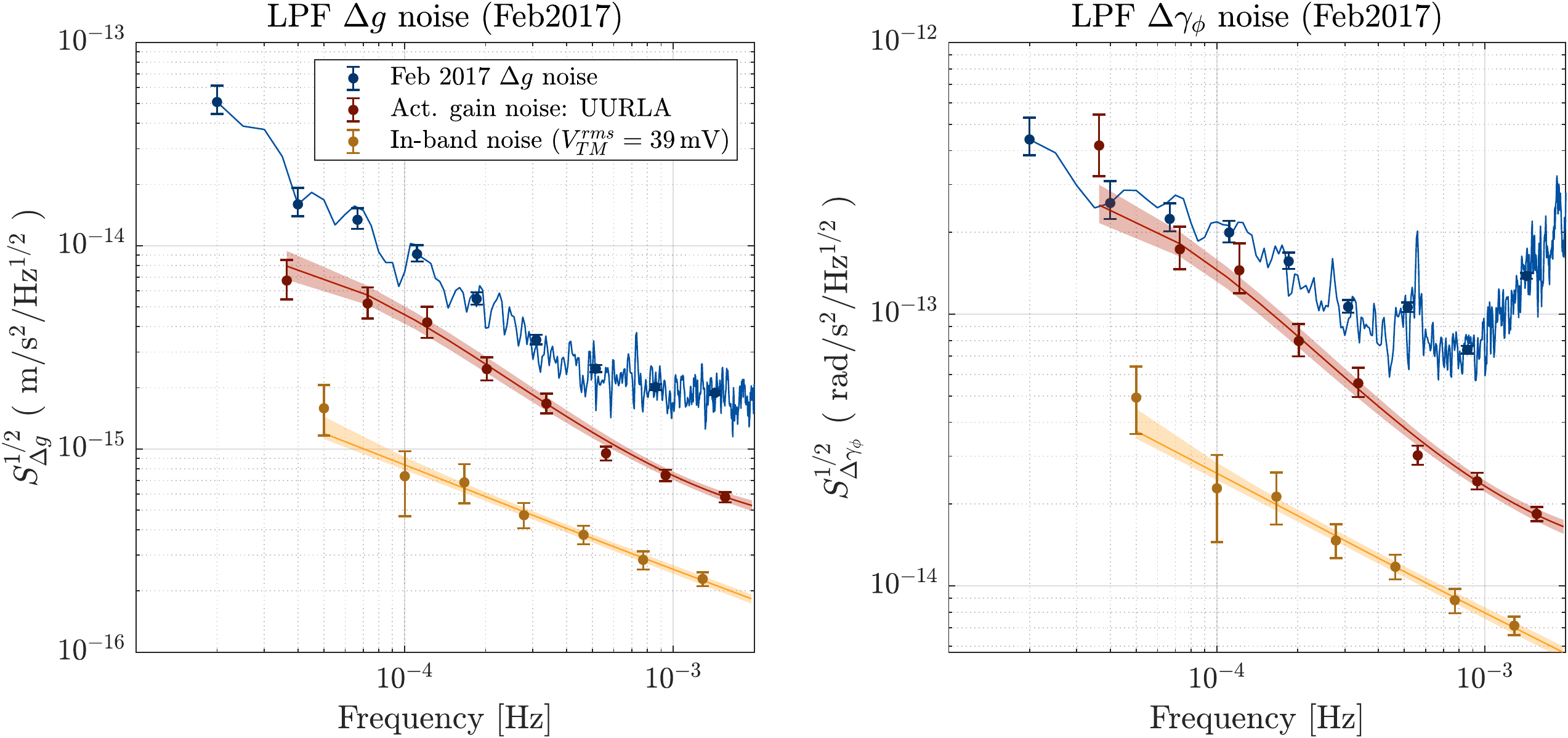} 
\caption{\label{fig_feb_2017}  
Differential acceleration noise measured 
by LPF across roughly 340~hours in February 2017~\cite{LPF_prl_2018}, shown with projections of the noise contributions
from actuation gain fluctuations and in-band additive voltage noise.  We note that the force /torque 
authorities $( \gamma_{\phi_{10}}, g_{20},\gamma_{\phi_{20}} ) $ were identical to those for 
UURLA in the actuation noise experiment (Table~\ref{tab_act_configs}) while the average applied 
forces were similar ( $\gamma_{\phi_{1\mathrm{c}}} = -970$~prad/s$^2$, $g_{2\mathrm{c}} = +5$~pm/s$^2$, and $\gamma_{\phi_{\mathrm{2c}}} = +150$~prad/s$^2$).
Time averaged RMS values for the TM potential were each found to be roughly 40~mV based on TM charge measurements performed before and after the long measurement.  }
\end{figure*}

All of these observations are consistent with coupling of charge into TM acceleration via  
uncorrelated actuation voltage noise, roughly with the same PSD in the 8 relevant $X$ electrodes for the two TM.  This is 
the simplest and perhaps most reasonable model -- considering the 8
nominally identical actuation circuits -- though not the only possible explanation of the measured data. 
In this model, the single electrode actuation voltage noise, 
%$S_v = \frac{1}{8} \left( S_{\DDr} + S_{\DDl} \right) 
%\approx  \frac{1}{4} S_{\Delta \left( \Delta_x \right) } 
%\approx  \frac{1}{4} S_{\Delta \left( \Delta_{\phi} \right) }$ 
$S_v \approx \frac{1}{4} S_{\DDr} \approx \frac{1}{4} S_{\DDl} 
\approx  \frac{1}{8} S_{\Delta \left( \Delta_x \right) } 
\approx  \frac{1}{8} S_{\Delta \left( \Delta_{\phi} \right) }$ 
and shown with the righthand axes in both plots of  Fig.~\ref{fig_Sv}, is roughly 
50~$\mu$V/Hz$^{1/2}$ at 0.1~mHz and 15~$\mu$V/Hz$^{1/2}$~at 1~mHz.  While we do not have detailed low frequency ground data with which to compare, the results at 1~mHz are roughly in line with the results of shorter pre-flight measurements.  

We note that the observed effective single electrode voltage noise will contribute to the noise in measurements of the TM charge, 
performed as in LPF\cite{LPF_prl_charge,lpf_charging} with a modulated voltage applied to the four~$X$ electrodes. This technique essentially detects the potential difference between the TM and the average DC potential of the modulating electrodes\cite{asr_dc_bias}.  However, the measured in-band actuation noise presented here is too small, by slightly more than an order of magnitude, to 
explain the measured TM charge fluctuations, equivalent to TM potential fluctuations of order 300~$\mu$V/Hz$^{1/2}$ at 0.1~mHz\cite{LPF_prl_charge}, which instead remain compatible with the ``shot noise'' of cosmic ray charging of the TM, with an effective Poissonian event rate of roughly 1200~/s.  This conclusion would remain valid even without the ``correction'' for the digitazion roundoff error mentioned above.

\section{Projections of acceleration noise from actuation in LPF and LISA}
\label{sec_projections}

A projection of the actuation gain and in-band voltage noise model and experimental parameters for the long (14 day) LPF benchmark differential acceleration noise measurement from February 2017 \cite{LPF_prl_2018} is shown, with  $\pm \sigma$ uncertainty bands, in Fig.~\ref{fig_feb_2017}.  Actuation gain noise is calculated using the models (Eqn.~\ref{eq:act_model}) and Markov chain parameter values of Sec.~\ref{sec_act_gain_noise}, considering the $A$ and $B$ coefficients (Eqn.~\ref{eqn_ABC_coeff}) as calculated from the commanded actuation force telemetry for the Feb.~2017 run.  The same is done for the in-band voltage noise mixing with the TM charge, using the model of Eqn.~\ref{eqn_S_DDx} with the noise parameters from the MCMC analysis of the charged TM experiment (Sec.~\ref{sec_inband_noise}) and the RMS TM charge values estimated from charge measurements before and after the Feb.~2017 runs.

%>> S_g_ACT = 20.5e-30; dS_g_ACT = 7e-30; % eyeball interp from Vittorio's data at 120 microHz and 70 microHz
%>> S_g_x = 65e-30; dS_g_x = 15e-30; 
%>> frac = S_g_ACT / S_g_x
%frac =
%    0.3154
%
%>> dfrac = sqrt( (dS_g_ACT/S_g_x)^2 + (dS_g_x*S_g_ACT/dS_g_x)^2)
%
%dfrac =
%
%    0.1077
%>> sqrt(27.5)
%    5.2440
%>> sqrt(13.5)
%    3.674

The actuation gain fluctuations were 
a sizable but not dominant contribution to the main science measurement of $\Delta g$; at 0.1~mHz this noise source 
is 3.5-5~fm/s$^2$/Hz$^{1/2}$ or roughly 20-40\% of the  total measured acceleration noise power. This contribution was 
limited by the exceptionally accurate 
gravitational (DC force) balancing along the LPF $x$ axis, which allowed lowering the actuation force authority to 
26~pm/s$^2$ for the mission science operations; had a larger DC force imbalance imposed the use of the 
``nominal'' authority (1140~pm/s$^2$), the acceleration noise due to actuation would have dominated the LPF noise floor and 
pushed it to the 60-70~fm/s$^2$ level at 0.1~mHz as in the ``nominal'' actuation test 
(Tab.~\ref{tab_act_configs} and light blue data in Fig.~\ref{fig_s_alpha_exper}). 
The in-band additive voltage noise is 
a small contributor, roughly 1~fm/s$^2$/Hz$^{1/2}$ at 0.1~mHz, at least with the relatively low, roughly 40~mV, RMS TM potential 
in this run.  

\begin{figure}[thb]
\includegraphics[width=3.3in]{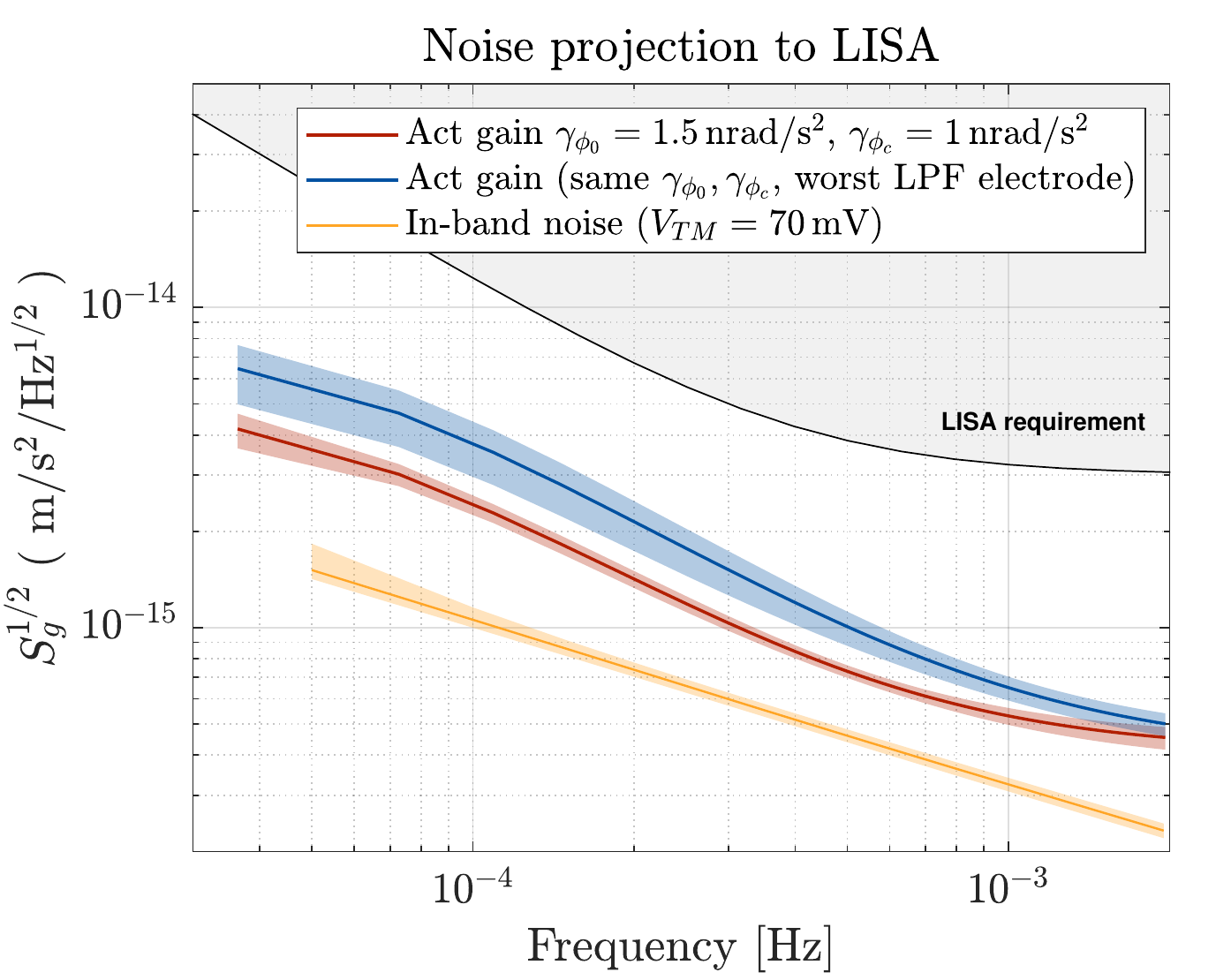} 
\caption{\label{fig_LISA}  
Projections of actuation noise contributions to LISA $x$ axis TM acceleration noise, from actuation 
gain noise and in-band voltage noise as estimated in this paper.  The LISA conditions assume 
$\phi$ actuation authority $\gamma_{\phi_0} = 1.5$~nrad/s$^2$ and commanded torque $\gamma_{\phi_c} = 1.0$~nrad/s$^2$, 
using the averaged gain noise measured here and, as a worst case, the values for the worst LPF electrode performance (TM1, electrode 2) assigned to all 4~$X$ electrode actuators.  For the coupling to in-band voltage noise, the TM potential is assumed to be 
at the discharge threshold of 70~mV. }
\end{figure} 

We can also project these results for actuation noise to the LISA mission, shown in Fig.~\ref{fig_LISA}. 
LISA requires no TM $x$ axis
actuation  forces, with the spacecraft ``drag-free'' controlled to follow the TM along 
this critical constellation interferometry axis.  With $g_{\mathrm{c}}=0$ in Eqn.~\ref{eqn_act_accel_coeff}, 
there is no coupling to fully correlated gain fluctuations of the 4~$X$ electrode channels ($A_i = 0$).  
However, the $\phi$ torque actuation with uncorrelated gain fluctuations in the 
4~$X$ actuation circuits will produce acceleration noise, with non-zero coefficients $A_{ij}$, 
\begin{eqnarray}
S_{g}^{ACT} 
& = &
R^{{\star}2}_{\phi}  \left( \gamma_{\phi_c}^2 +  \gamma_{\phi_0}^2 \right)
 S_{\alpha_{\text{UC}}} \left(  f \right) \nonumber \\
& \approx & 
\left[ 3.5 \: \mathrm{fm/s^2/Hz^{1/2}} \right]^2 
\times \left[ \frac{\gamma_{\phi_c}^2 + \gamma_{\phi_0}^2}
{\left( 1^2 + 1.5^2 \right) \: \left( \mathrm{nrad/s^2} \right)^2 }
\right]  \nonumber \\
& \times & \frac{S_{\alpha_{\text{UC}}}}{  \left(  50 \: \mathrm{ppm/Hz^{1/2}}  \right)^2 }   
\label{eqn_Sg_LISA_act_gain}
\end{eqnarray}
where 50~ppm/Hz$^{1/2}$ is a reference level at 0.1~mHz from the measurements presented in 
Sec.~\ref{sec_act_gain_noise} for the (noisier) $X+$
LPF actuators. The reference torque levels consider compensation of a $\phi$ DC angular acceleration 
of 1~nrad/s$^2$, with an actuator torque authority of 1.5~nrad/s$^2$, allowing some ``headroom'' for controller
dynamics.  This can be considered a realistic gravitational balance 
``requirement'' on board a LISA spacecraft, based on the LPF experience.  
Indeed the needed applied $\phi$ torques on the two LPF TM -- $\gamma_{\phi_{1c}}$ and $\gamma_{\phi_{2c}}$ 
in Tab.~\ref{tab_act_configs}, which were within 100~prad/s$^2$ of, respectively, 
-1000~prad/s$^2$ and +150~prad/s$^2$ over the course of the mission.  These are each 
compatible, magnitude and sign, with the preflight estimates from the gravitational model to within
200~prad/s$^2$\footnote[2]{Compare with values for angular accelerations $\alpha_{\phi}$ in Table A1 of 
Ref.~\cite{grav_valerio_cqg_2016}.} and further gravitational compensation could have reduced the residual 
values further, well within the 1000~prad/s$^2$ limit proposed for LISA.

For the coupling of the actuation ``in-band'' noise to TM charge, we would have 
\begin{eqnarray}
S_g^q 
& \approx & 
4 \left[\frac{1}{M} \pder{C_X}{x} \frac{q}{C_{T}} \right]^2 S_{v_n} \left( f \right)
\nonumber  \\
& \approx &  \left[ 1 \: \mathrm{fm/s^2/Hz^{1/2}} \right]^2  
\times \left( \frac{q}{1.5 \times 10^7 \: e} \right)^2 
\nonumber \\ 
& \times & \frac{S_{v_n}}{\left( 50 \:\mathrm{\mu V/Hz^{1/2}} \right)^2 }
\label{eqn_fluct_Delta_x}
\end{eqnarray} 
In Fig.~\ref{fig_LISA} we use a worst case TM charge of $\pm 15$~million elementary charges 
($\pm 70$~mV TM potential).  For the typical LPF charging conditions \cite{LPF_prl_charge} of +25~$e$/s, 
maintaining the TM charge in this interval would require discharging roughly every two weeks, with 
up to twice this rate depending on solar cycle variations \cite{imperial_charge_2005,catia_charge_2005}.  
Alternatively the TM charge can be held near zero with a continuous UV illumination of the TM, which was 
also demonstrated in a dedicated experiment by LPF \cite{continuous_tbp} and 
is under study for LISA \cite{sam_discharge}.  The in-band voltage fluctuations will introduce force noise 
of similar magnitude in the interaction with the average stray biases on each electrode, considered to be 
of order 20-50~mV RMS based on measurements in flight with LPF and on ground with various GRS prototypes \cite{LPF_prl_charge, 
charge_DC_bias_prl_2012}.

\section{Conclusions}

LISA Pathfinder allowed demonstration of the key electrostatic actuation system needed for LISA, 
demonstrating its compatibility with differential acceleration measurements at the 10~fm/s$^2$/Hz$^{1/2}$ 
level at 0.1~mHz, as needed for high precision measurement of tidal accelerations from super-massive black 
holes in the LISA observatory.  The conditions of LISA Pathfinder were actually slightly more challenging than those
envisioned for LISA from the standpoint of applied forces, requiring $x$ axis force actuation that will not be needed in LISA.

In the on-orbit measurement campaign presented in Sec.~\ref{sec_act_gain_noise}, 
we successfully detect acceleration noise from actuation gain fluctuations,
compatible with a simple parametric gain model with noise scaling with the applied forces and a clear detection of 
a dominant role, in LPF, of force noise from uncorrelated fluctuations in the different electrode actuation circuits rather 
than ``board-correlated'' actuator gain noise, as would instead result from a noisy DAC reference voltage.      
Measured levels of gain noise were in agreement with the limited ground testing measurements performed and used in LPF noise predictions \cite{LPF_noise_budget_cqg_2011}, even detecting the same chance variations in the noise levels for different groups
of the nominally identical eight relevant electrode actuator circuits.  

As shown in Fig.~\ref{fig_feb_2017}, actuation gain noise was an important but not dominant source of noise in LPF, 
responsible for roughly 20-40\% of the noise power for $\Delta g$ in the 0.1~mHz band. Our measurements clearly 
resolved the actuation gain fluctuations by increasing, compatibly with dynamic control constraints, the applied 
actuation forces well beyond their needed levels.    Similarly, in-band low frequency voltage
noise was indeed measurable, but only after increasing the TM charge by more than a factor 10
beyond the typically used values.  

The contribution of actuator gain noise in the LPF ``benchmark'' acceleration noise measurements was kept at a minimal
level because the gravitational balance was considerably better than the values that were ``budgeted'', or 650~pm/s$^2$ 
 \cite{LPF_noise_budget_cqg_2011} or 
even predicted as uncertainties in the preflight gravitational models \cite{grav_valerio_cqg_2016}, order $\pm 300$~pm/s$^2$.  
The residual $x$ axis acceleration noise contribution from the applied actuation forces was determined roughly
in equal measure from $\phi$ torques in addition to the TM2 applied $x$ forces.  

Additional LPF differential acceleration tests were also performed in an open-loop ``free-fall'' mode, without any TM2 $x$ force actuation except during brief control ``kicks'' that were excised from the data in post-processing.  Though originally designed to improve the LPF noise by removing a dominant actuation contribution \cite{LPF_noise_budget_cqg_2011}, the test results showed no measurable difference \cite{lpf_free_fall}: due to the better than expected gravitational balancing and consequent lower force levels, $x$-axis actuation noise was small enough that its removal produced essentially no resolvable improvement in the noise.   

%\color{red}
The remaining low frequency LPF noise is subject of ongoing research and will be addressed in a future 
publication \cite{lpf_noise_future}.   We cannot exclude that some of the remaining excess 
low frequency acceleration noise, in February 2017 and in the 
other low-force, low noise \cite{LPF_prl_2018, LPF_prl_2016} runs, comes from the actuation 
system by another mechanism \textit{not covered} by our gain noise model here.  To be quantitatively relevant, such an 
effect would have to scale ``more slowly'' with the applied forces than for gain noise, which scales as 
$\delta F \propto F$ and thus $S_g \propto g_c^2$.  An additive voltage noise with frequency dependence around the actuation 
carrier frequencies could give an effective low frequency amplitude-modulation noise, with a PSD that scales linearly with 
applied force authorities (such as in Eqn.~\ref{eqn_white_noise}).  Such an effect was detected with the actuation 
digitization accuracy issue mentioned earlier \cite{luigination_2020}, though the error is believed to have been successfully understood and corrected.  This deterministic correction is of order 10~$\mu$V RMS out of a total LSB of 153~$\mu$V; a possible residual inaccuracy, an effective DAC non-linearity, of 1-2~$\mu$V in the calculation of the true applied amplitudes, would be sufficient to account for the observed excess noise in the 0.1-1~mHz band. Such a level of error can not be ruled out by other in-flight measurements, such as the ``calibration tone'' experiment (Fig.~\ref{fig_caltone}) or the gain noise measurement campaign presented in Sec.~\ref{sec_act_gain_noise}.  We are currently investigating, with both analysis and measurement on available LPF-prototype FEE models, if such a residual error could be compatible  with the LPF actuation electronics.  
%\color{black}

Mixing of actuator in-band voltage noise with TM charge and other stray DC potentials is shown to have a small impact in 
LPF or, assuming appropriate charge management, in LISA, with the current measurements establishing a lower noise level with respect to earlier LPF results \cite{LPF_prl_charge} and confirming a scenario of stray field fluctuations dominated by uncorrelated additive actuation circuitry noise acting on individual electrodes.  However, even at the lower noise level measured here, stray electrostatics 
still drive the use of audio frequency carriers to apply DC or slowly varying electrostatic forces.  A ``DC drive'' electrostatic actuation 
for $\phi$ torques at the LISA-required 1.5~nrad/s$^2$ level would require applied voltages up to 0.7~V, 
an order of magnitude larger than the DC potential differences arising from the levels proposed for TM charge 
or stray DC biases.  With 50~$\mu$V/Hz$^{1/2}$ electrode voltage noise at 0.1~Hz, this would give 7~fm/s$^2$/Hz$^{1/2}$
and thus consume nearly half the entire LISA acceleration noise budget.   
  
The unique potential of the LPF $\Delta g$ dataset has been recognized beyond its value as a benchmark for LISA 
free-fall as a small force probe for, among other questions,  
setting upper limits on wavefunction collapse models \cite{collapse_andrea} and searching for ultralight dark matter \cite{luis_dark_lpf}.  
Our current article is the first to use the high precision LPF \textit{differential angular acceleration} measurement, 
$\Delta \gamma_{\phi}$, and the sensitivity reached merits a comment.  
While rotational acceleration noise is not intrinsically essential to the LISA low frequency gravitational 
wave sensitivity, the LPF $\Delta \gamma_{\phi}$, including an interferometric readout 
with sub-nrad/Hz$^{1/2}$ resolution\cite{prd_ifo}, proved invaluble in this investigation of 
actuator force noise, allowing us to disentangle effects that would have had degenerate signatures 
in the $\Delta g$ translational channel alone.  The differential 
torque resolution reached in the low force ``UURLA'' configuration (see the righthand plots in
Fig.~ \ref{fig_s_alpha_exper} or \ref{fig_feb_2017}) is below 0.1~fNm/Hz$^{1/2}$ 
across a decade of frequency (0.15-1.5~mHz), an improvement of nearly an order of magnitude 
beyond the best measurements on ground using a single, hollow LISA TM inside a prototype electrode housing with fused silica suspension 
torsion pendulums \cite{cqg_fused, utn_free_fall}.  
Actuation gain noise is a big portion for the LPF differential angular acceleration noise, 
responsibile for roughly half the noise power at 0.1~mHz and an even larger
fraction at lower frequencies.  The clean LPF system and environment on orbit thus also realized 
an improved ``fiberless'' torsion pendulum that proved essential for these tests of 
critical hardware for LISA, as well as for the overall understanding of low frequency force noise 
in the experimental challenge of free-falling reference test masses.

%In fact, this full differential acceleration noise 
%level serves as a true upper limit for the actuator force noise contribution across the entire LISA frequency band,
%down to the 20~$\mu$Hz frequencies that were not accessible in a statistically meaningful way in the dedicated actuation
%measurement campaigns, limited to roughly 2~day single measurements. 
% 
% sum ( 0.5 * dCx_dx_ACT * (VACT * sin omega t + n)^2 )
% sum ( dCx_dx_ACT * VACT * n(t) * sin omega t
% noise in n * sin omega t --> 2 S_n (f) * 
% 8 * dCx_dx_ACT^2 * VACT^2  * 2 Sn (f)   

% 800 +/- 200, −200 +/- 200
% feb 2017 applied ... +970, -150

%>> IZZ = 7e-4;
%>> S_Dg_phi = (3e-13)^2; % measured at 3 mHz
%>> f_phi = 3e-3; 
%>> S_Dphi = S_Dg_phi / (2*pi*f_phi)^4
%S_Dphi =
%      7.12915434353079e-19
%>> sqrt(S_Dphi)
%ans =
%      8.44343197019481e-10

% act stability projection for LISA, 50 ppm/sqrt(Hz) is really worst case ... 
% do we want to do different estimates of S_v single electrode?  
% correlazione / fit alla T? 
% fit DC luigination 

\vspace{1.5cm}

%Truly random unresolved thoughts ... 
%\begin{itemize}

%\item check and add comments from inband voltage noise measurements, measurement at 
%HEV (S2-HEV-RP-3042) 

%\item mention of T noise ... could this explain gain noise 

%\item do we need a figure with gain and/or stiffness measured over time (or with different
%applied forces)?
%\end{itemize}

%>> I = 7e-4; 
%>> gamma_phi_0 = 1.5e-9; 
%>> dCx_dphi = 1.2e-12 * 11/4; 
%>> dCx_dx = 3e-10; 
%>> M = 1.92;
%>> VDC = sqrt(I*gamma_phi_0/dCx_dphi); % for 2 electrodes applying a DC force
%>> S_V = (50e-6)^2;
%>> VDC
%VDC =
%         0.564076074817766
%>> S_g_ACT_DC = 2 * (VDC * dCx_dx / M )^2 * S_V ;
%>> sqrt(S_g_ACT_DC) =       6.23221902513645e-15

% V1 = V_{1x} \chi_{1x} + V_{1\phi} \chi_{1\phi} 
% V2 = -V_{1x} \chi_{1x} + V_{2\phi} \chi_{2 \phi} 
% V3 = V_{2x} \chi_{2x} - V_{1\phi} \chi_{1\phi} 
% V4 = -V_{2x} \chi_{2x} - V_{2\phi} \chi_{2\phi} 
% V_{TM} = 

\begin{acknowledgments}
This work has been made possible by the LISA Pathfinder mission, which is part of the space-science program of the European Space Agency.   

We acknowledge the work of the prime contractor for LPF and for the ``LISA Technology Package'', Airbus Defense and Space, for the industrial implementation of the electrostatic actuation suspension as part of the overall DFACS dynamic control under their responsibility.   

The Italian contribution has been supported by Istituto Nazionale di Fisica Nucleare (INFN) and Agenzia Spaziale Italiana (ASI), Project No. 2017-29-H.1-2020 “Attività per la fase A della missione LISA”. The UK groups wish to acknowledge support from the United Kingdom Space Agency (UKSA), the Scottish Universities Physics Alliance (SUPA), the University of Glasgow, the University of Birmingham, and Imperial College London. The Swiss contribution acknowledges the support of the Swiss Space Office via the PRODEX Programme of ESA, the support of the ETH Research Grant No. ETH-05 16-2 and the support of the Swiss National Science Foundation (Projects No. 162449 and No. 185051). The Albert Einstein Institute acknowledges the support of the German Space Agency, DLR. The work is supported by the Federal Ministry for Economic Affairs and Energy based on a resolution of the German Bundestag (No. FKZ 50OQ0501, No. FKZ 50OQ1601, and No. FKZ 50OQ1801). J. I. T. and J. S. acknowledge the support of the U.S. National Aeronautics and Space Administration (NASA). Spanish contribution has been supported by Contracts No. AYA2010-15709 (Ministerio de Ciencia e Innovación, MICINN), No. ESP2013-47637-P, No. ESP2015-67234-P, No. ESP2017-90084-P (Ministerio de Asuntos Económicos y Transformación Digital, MINECO), and No. PID2019–106515GB-I00 (MICINN). Support from AGAUR (Generalitat de Catalunya) Contract No. 2017-SGR-1469 is also acknowledged. M. N. acknowledges support from Fundacion General CSIC (Programa ComFuturo). F. R. acknowledges an FPI contract from MINECO. The French contribution has been supported by the CNES (Accord Specific de projet No. CNES 1316634/CNRS 103747), the CNRS, the Observatoire de Paris and the University Paris-Diderot. E. P. and H. I. would also like to acknowledge the financial support of the UnivEarthS Labex program at Sorbonne Paris Cité (No. ANR-10-LABX-0023 and No. ANR-11-IDEX-0005-02). N. K. would like to thank for the support from the CNES Fellowship

\end{acknowledgments}
\subsection{Appendix A: Model of electrostatic forces and force noise}
\label{app_estat_model}
This section summarizes the electrostatic model used, which is then applied to 
describe both the nominal actuation design and the two main acceleration noise sources 
addressed by this paper.  

We use the electrostatic force model developed in Refs.~\cite{charge_DC_bias_prl_2012}, 
\cite{nico_lisa_symp}, and \cite{cap_herrera}, considering
the test mass (TM) and surrounding sensor (S) surfaces as a patchwork 
of equipotential domains.  The resulting instantaneous force on the TM along the sensor $x$ axis is given by  
\begin{equation}
 F_x = \frac{1}{2} 
\sum_{m,n<m}{\pder{C_{nm}}{x} \left( V_m - V_{n} \right)^2 } \, ,
\label{estat_force}
\end{equation}
where $C_{nm}$ is the capacitive element between surface domains $m$ and $n$.  

For actuation / sensing electrode $j$, we can consider the potential on domain $m(j)$
as a sum of the actuation voltage $V_j$ applied homogeneously to the entire electrode 
and a local ``patch potential'' $\delta V_m$, 
\begin{equation}
V_{m(j)} = V_j + \delta V_{m}
\label{estat_Vmj}
\end{equation}

The sensor surfaces also include grounded electrode housing ``guard rings'' (H) with possible
stray surface potentials, $V_{m(H)}=\delta V_m$.  

For a TM domain $n(\mathrm{TM})$, we have $V_{n(\mathrm{TM})} = V_{\mathrm{TM}} + \delta V_m$, with the average
TM potential given by   
\begin{eqnarray}
V_{\mathrm{TM}} & = & \frac{q}{C_T} + \frac{\sum\limits_{n(S)}{ C_{n(S)} \delta V_n }}{C_T} 
+  \frac{\sum\limits_{j}{ C_j V_j }}{C_T}
\nonumber \\
& = & V_{\mathrm{TM}0} + V_{\mathrm{TM}}^{ACT} 
\label{estat_vm}
\end{eqnarray} 
In the second line we divide into an ``intrinsic'' TM potential $V_{\mathrm{TM}0}$ from EH stray surface 
potentials and an induced electrostatic potential from the actuation voltages $V_{\mathrm{TM}}^{ACT}$.  
Here $C_{n(S)}$ is the capacitance of sensor domain $n$ to all TM domains, $C_j$ is the total
capacitance of electrode $j$ to the TM, and $C_T \equiv \sum\limits_{n(S)} C_{n(S)}$ is the 
total TM capacitance to all electrode housing surfaces.  

To isolate the electrostatic force contribution from applied actuation voltages, we can 
consider the terms in Eqn.~\ref{estat_force} that involve domains on the sensing / actuation
electrodes.  We neglect here the interaction between domains on the same electrode -- for which 
the applied actuation voltage cancels in the potential differences in Eqn.~\ref{estat_force} --
and between domains on different electrodes, as the capacitance between these is quite small
and shielded by the ground ring surfaces.  
Thus considering interactions between domains on actuation electrodes and those on the TM and 
on the grounded guard ring surfaces, we find,  
summing over the electrostatic domains $m(j)$ on each actuation electrode $j$, 
\begin{eqnarray}
F_x^{ACT} & = &  \sum_{j} \left\{ \frac{1}{2}  {\pder{C_j^{\star}}{x}   V_j^2  } \right.
\nonumber \\
& & + V_j \left[ - \pder{C_j}{x} \left( V_{\mathrm{TM}0} + V_{\mathrm{TM}}^{ACT} \right) \right.
\nonumber \\ 
& & \left. \left. + \sum_{m_j, n \left(TM,H \right)}{ \pder{C_{m_j n}}{x} \left( \delta V_{m_j} - \delta V_n \right)} \right] \right\}
%F_x & = &  
%\frac{1}{2} \sum_{j} {\pder{C_j^{\star}}{x}   V_j^2  } - V_{\mathrm{TM}0} \sum_{j} {\pder{C_j}{x}   V_j  }  
%- V_{\mathrm{TM}}^{ACT} \sum_{j} {\pder{C_j}{x}   V_j  }
%\nonumber \\
%& & + \sum_j { V_j  \left[ 
%\sum_{m_j, n \left(TM,H \right)}{ \pder{C_{m_j n}}{x} \left( \delta V_{m_j} - \delta V_n \right)} \right] }
\label{eqn_estat_f_act}
\end{eqnarray}
This equation for the instantaneous electrostatic force considers the capacitive 
derivatives evaluated for the true TM position.  We have introduced the ``total capacitance'' of 
an electrode to the TM and to the adjacent grounded housing, $C_j^{\star} \equiv \sum{C_{m(j),n(TM,H)}}$. 

Analogous expressions govern the force and torques on the other axes.  We note that, while only the 4 electrodes 
on the $X$ faces of the EH have a significant derivative, \pder{C_j}{x}, the single TM average potential is relevant 
to the force on any degree of freedom, such that the third term involving $V_{\mathrm{TM}}^{ACT}$ will couple 
the actuation forces and torques on different axes.  

\subsubsection{Actuation design, nominal forces and force gradients}
The guiding principles to the actuation concept proposed early in LPF development \cite{grs_SPIE_2002}
follow from Eqn.~\ref{eqn_estat_f_act}: 
\begin{itemize}
\item exploit the quadratic dependence, $F \propto V^2$, to produce DC or slowly varying control forces with zero-mean audio-frequency voltages.  The control force is the low-frequency component $\langle V_j^2 \rangle$ in the 1$^{st}$ term in Eqn.~\ref{eqn_estat_f_act} -- $\langle \, \rangle$ denoting a time average over the actuation carrier period -- and thus independent of TM potential from charge $q$ and of stray surface potentials, which enter in the 2$\mathrm{^{nd}}$ ($V_{\mathrm{TM}0}$) and 4$\mathrm{^{th}}$ ($\delta V_m$) terms in Eqn.~\ref{eqn_estat_f_act}.  Forces from these uncontrolled and noisy quasi-DC potentials are spectrally shifted out of band, around the AC carrier frequencies (60-270~Hz in LPF), as is any ``self-mixing'' between the actuation carrier and low frequency ``in-band'' actuation noise.   

\item reduce the coupling between degrees of freedom (DOF) by applying actuation voltages with symmetry such that the induced actuation potential $V_{\mathrm{TM}}^{ACT}$ -- the 3$\mathrm{^{rd}}$ term in Eqn.~\ref{eqn_estat_f_act} for $x$ but also relevant to the actuation force on all DOF -- is zero, at least for a centered TM.  This is done by requiring that the sum of the applied actuation voltages on the 
4 $X$ (or $Y$ or $Z$) electrodes is zero at all times. 

\item further decouple the DOF with actuation waveforms chosen to be ``orthogonal'' between different degrees of freedom.  In LPF this was implemented with sinusoids at 6 different actuation frequencies.  

\item make TM control dynamics more simple and predictable, by maintaining the force gradients introduced by the actuation 
electrostatic fields -- through the position dependence of the capacitive derivatives (1$^{\mathrm{st}}$ term) 
and the induced $V_{\mathrm{TM}}^{ACT}$ (3$^{\mathrm{rd}}$ term) -- independent of the commanded force and torque.   
How to do this will be shown shortly.  
\end{itemize}

As an example of the actuation algorithm and nominal forces, torques and their gradients, we consider the $x$ and $\phi$ degrees of freedom, actuated using the 4~$X$-face electrodes.  We consider the following actuation scheme, generalized from Eqn.~\ref{eqn_applied_V} 
\begin{eqnarray}
V_{1c} \left( t \right) & = & 
V_{1x} \chi_{1x}  \left( t \right) + V_{1\phi} \chi_{1 \phi} \left( t \right) 
\nonumber \\
V_{2c} \left( t \right) & = &  
-V_{1x} \chi_{1x}  \left( t \right) + V_{2\phi} \chi_{2 \phi} \left( t \right) 
\nonumber \\
V_{3c} \left( t \right) & = & 
+V_{2x} \chi_{2x}  \left( t \right) - V_{1\phi} \chi_{1 \phi} \left( t \right) 
\nonumber \\
V_{4c} \left( t \right) & = & 
- V_{2x} \chi_{2x}  \left( t \right) - V_{2\phi} \chi_{2 \phi} \left( t \right) 
\label{eqn_applied_V_gen}
\end{eqnarray}
Each waveform $\chi$ is zero mean -- for instance $\langle \chi_{1x} \left( t \right) \rangle = 0$, with ``sine-equivalent'' amplitudes such that   $\langle \chi_{1x}^2 \rangle = 0.5$.  Additionally, the waveforms are orthogonal between different degrees of freedom ($\langle \chi_{1x} \chi_{1\phi}\rangle = 0 $ for instance), which was achieved in LPF with 6 different actuation frequencies for the 6 DOF.  

In evaluating the capacitances relevant to the electrostatic forces, we consider an expansion to first order to the relevant displacements in the $xy$ plane, 
\begin{eqnarray}
C_j & = & C_{X0} \: \pm \pder{C_X}{x} x  \: \pm \pder{C_X}{\phi} \phi 
\nonumber \\
\pder{C_j}{x} & = & \pm \left| \pder{C_X}{x} \right| \: + \left|  \pdder{C_X}{x} \right| x 
\: \pm \left| \frac{\partial ^2 C_X}{\partial x \, \partial \phi} \right|  \phi  
\nonumber \\
\pder{C_j}{\phi} & = & \pm \left| \pder{C_X}{\phi} \right|  \: + \left|  \pdder{C_X}{\phi} \right|  \phi \: 
\pm \left|  \frac{\partial ^2 C_X}{\partial x \, \partial \phi} \right|  x  
\end{eqnarray}
with the signs ($\pm$) changing with the positions of the different electrodes.  

Neglecting any stray DC fields ($\delta V_m$ and $q$) and considering only the DC force component, 
considering higher order harmonics as out of band, we obtain a force
\begin{eqnarray}
F_x & = & \frac{1}{2}  \left| \pder{C_X^{\star}}{x} \right| \left( V_{1x}^2 - V_{2x}^2 \right) 
\nonumber \\ 
& & +  \frac{1}{2}  \left| \pdder{C_X^{\star}}{x} \right| \: x  \: \left[ V_{1x}^2 + V_{2x}^2 + V_{1 \phi}^2 + V_{2 \phi}^2 \right]
\nonumber \\ 
& & -  2 \frac{ \left| \pder{C_X}{x} \right|^2 }{C_T}  \: x  \: 
\left[  V_{1 \phi}^2 + V_{2 \phi}^2  + V_{1\phi} V_{2\phi} \langle \chi_{1 \phi} \chi_{2 \phi}  \rangle \right]
\label{eqn_F_act}
\end{eqnarray}

Additionally requiring that the two $\phi$ 
waveforms be ``orthogonal'', with $\langle \chi_{1 \phi} \chi_{2 \phi}  \rangle =0$ -- obtained using 
sine and cosine waveforms in LPF, see Eqn.~\ref{eqn_applied_V} -- eliminates 
the crossterm dependent on the product $V_{1 \phi} \times V_{2 \phi}$.  We thus obtain 
\begin{equation}
F_x = M g_c - M\omega_{xx}^2 x 
\label{eqn_F_act_simp}
\end{equation}
with a  nominal -- TM centered -- applied force per unit mass $g_c$ 
and a force gradient described by  ``stiffness'' ~$\omega_{xx}^2$
\begin{eqnarray}
g_c & = & \frac{1}{2 M}  \left| \pder{C_X^{\star}}{x} \right| \left( V_{1x}^2 - V_{2x}^2 \right)
\nonumber \\
\omega_{xx}^2  & = & - \frac{1}{2 M}  \left| \pdder{C_X^{\star}}{x} \right| \times 
\nonumber \\ 
& & 
\left[ \left( V_{1x}^2 + V_{2x}^2 \right) + \left( V_{1\phi}^2 + V_{2\phi}^2 \right) 
\left( 1 -\frac{ 4 \left| \pder{C_X}{x} \right|^2 }{C_T \pdder{C_X^{{\star}}}{x}} \right) \right]
\end{eqnarray}

Under the orthogonality condition $\langle \chi_{1x} \chi_{2x}  \rangle =0$, the torque and its 
gradient are given analogously
\begin{equation}
N_{\phi} = I \gamma_{\phi_c} - I\omega_{\phi \phi}^2 \phi 
\label{eqn_N_act_simp}
\end{equation}
with 
\begin{eqnarray}
\gamma_{\phi_c} & = & \frac{1}{2 M}  \left| \pder{C_X^{\star}}{\phi} \right| \left( V_{1\phi}^2 - V_{2\phi}^2 \right)
\nonumber \\
\omega_{\phi \phi}^2  & = & - \frac{1}{2 I}  \left| \pdder{C_X^{\star}}{\phi} \right| \times 
\nonumber \\ 
& & 
\left[ \left( V_{1\phi}^2 + V_{2\phi}^2 \right) + \left( V_{1x}^2 + V_{2x}^2 \right) 
\left( 1 - \frac{ 4 \left| \pder{C_X}{\phi} \right|^2 }{C_T \pdder{C_X^{{\star}}}{\phi}} \right) \right]
\end{eqnarray}

From these equations, we foresee an actuation scheme where the stiffness is held constant, both in $x$ and $\phi$ 
by holding constant both $V_{1x}^2 + V_{2x}^2 = V_{\mathrm{MAX}_x}^2$ and  $V_{1 \phi}^2 + V_{2 \phi}^2 = V_{\mathrm{MAX}_{\phi}}^2$. 
This simplifies the control dynamics, with a predictable elastic coupling that is independent of the commanded forces 
and torques, $g_c$ and $\gamma_{\phi_c}$.  This also avoids conversion of a varying torque command into force noise for a translated TM -- the force just depends on the TM translation offset, not on the torque command. 
 
 The  commanded forces per unit mass, sensitive to the difference $\left(  V_{1x}^2 - V_{2x}^2 \right) $ and programmed by the force to voltage conversion in Eqn.~\ref{eqn_F_to_V}, are bounded by the range $\pm g_0$, with force authority $g_0$ given by 
\begin{equation}
g_0  =  \frac{1}{2m} \left| \pder{C_X^{\star}}{x} \right| V_{\mathrm{MAX}_x}^2
\end{equation} 
The maximum positive (or negative) force is obtained for $V_{1x} = V_{\mathrm{MAX}_x}$ (or $V_{2x} = V_{\mathrm{MAX}_x}$) and 
the other voltage $V_{2x}$ or $V_{1x}$ set to zero, essentially pulling only on the positive (or negative) $X$ face 
of the TM. A null force is commanded by setting $V_{1x} = V_{2x} = \frac{V_{\mathrm{MAX}_x}}{\sqrt{2}}$.  
The angular accelerations are similarly bounded, with authority 
\begin{equation}
\gamma_{\phi_0}  =  \frac{1}{2I} \left| \pder{C_X^{\star}}{\phi} \right| V_{\mathrm{MAX}_{\phi}}^2
\end{equation} 
The stiffnesses, translational and rotational, 
can be expressed in terms of these force and torque authorities,  
\begin{eqnarray}
\omega_{xx}^2  = - \frac{ \left| \pdder{C_X^{\star}}{x} \right|}{ \left| \pder{C_X^{\star}}{x} \right|} 
\left[ g_0  + R^{\star}_{\phi} \gamma_{\phi_0} 
\left( 1 - \frac{ 4 \left| \pder{C_X}{x} \right|^2 }{C_T \pdder{C_X^{{\star}}}{x}} \right) \right]
\nonumber 
\\ 
\omega_{\phi \phi}^2  = - \frac{ \left| \pdder{C_X^{\star}}{\phi} \right|}{ \left| \pder{C_X^{\star}}{\phi} \right|} \times 
\left[ \gamma_{\phi_0} + \frac{g_0}{R^{\star}_{\phi}}   
\left( 1 - 4 \frac{ \left| \pder{C_X}{\phi} \right|^2 }{C_T \pdder{C_X^{{\star}}}{\phi}} \right) \right]
\label{eqn_stiffness}
\end{eqnarray}
The prefactor setting the relationship between the $x$ axis authority $g_0$ and the stiffness $\omega_{xx}^2$ 
can be approximated geometrically, $\frac{ \left| \pdder{C_X^{\star}}{x} \right|}{ \left| \pder{C_X^{\star}}{x} \right|} 
\approx \frac{2}{d_x}$, 
roughly $6 \times 10^{-7}$/s$^2$ per nm/s$^2$ of force authority.  The contributions of both force and torque authority to the translational stiffness were measured in flight by LPF \cite{cal_paper} and confirmed finite element analyses \cite{nico_lisa_symp} at the level of several percent.  

To limit both the force gradients and the force noise arising from actuation, the strategy in both LPF and LISA is that of reducing the authorities $g_0$ and $\gamma_{\phi_0}$ to the minimum levels allowing compensation of the intrinsic DC forces -- in both cases mainly the residual error from spacecraft gravitational balancing \cite{grav_valerio_cqg_2016} -- with some margin for any controller dynamics.

% cross talk ... cite result from LPF + PETER result
% 
% cross talk ... voltage imbalance (phi) geometric impurities 
% scheme avoids that, with a rotated TM, that a change in commanded toruqe does not give a change in force for rotated TM

\subsubsection{Remaining actuation noise and imperfections}

Deviations of the applied actuation voltages from the desired values create force noise in LISA Pathfinder
and LISA.  We model the true applied voltage $V_j \left( t \right)$ in terms of the commanded waveform 
$V_{jc} \left( t \right)$, 
\begin{equation}
V_{j} = V_{jc} \left( 1 + \alpha_j \left( t \right) \right) + v_j \left( t \right) 
\label{eqn_app_Vapp}
\end{equation}
where $\alpha_j$ represents a gain fluctuation in the electrode $j$ actuation circuit and noise $v_j \left( t \right)$
is the actuator additive voltage noise, independent of the applied amplitude.   

From Eqns.~\ref{eqn_estat_f_act} and \ref{eqn_app_Vapp}, we can identify the force noise terms which are the 
subject of this article: 
\begin{itemize}
\item a non-zero average (DC) value of $\alpha$ will result in miscalibration of the actuation force.  
Referring to Eqn.~\ref{eqn_Dg} and considering a \textit{common mode} miscalibration $\alpha^{DC}$ 
of all 4 $X$ electrodes, we will have 
\begin{equation}
\lambda = 1 + 2 \alpha^{DC}
\nonumber
\end{equation}   
This static gain deviation was observed in LPF, with a measured difference of nearly 4\% in the nominal voltage commands ($\lambda \approx 1.08$ \cite{cal_paper}), though this difference 
is understood and expected, given the actuation circuit detailed design.  

\item a \textit{differential} gain offset between different electrodes can create actuation crosstalk.  For instance a difference $\Delta \alpha$ in the gains between electrodes 1/4 relative to electrodes 2/3 would result in a spurious acceleration of roughly  $ R_{\phi}^{\star} \gamma_{\phi_c} \times \Delta \alpha$, cross-coupling applied $\phi$ torque into an $x$ force. In LPF such a crosstalk was observed, differentially between the two TM, at a level implying $\Delta \alpha \approx 0.5\%$\cite{LPF_prl_2018}, though this apparent actuation crosstalk was degenerate with other interferometric readout geometric crosscouplings between the apparent differential acceleration $\delta g$ and the SC rotational jitter.  Additional actuation crosstalk terms can arise in geometric imperfections and TM rotations.   

\item fluctuations in the various $\alpha$ will create a low frequency force error, per electrode, of 
\begin{equation}
\delta F \approx \alpha_j \pder{C_j^{\star}}{x} \langle V_{jc}^2  \rangle
\nonumber
\end{equation}
This is the actuator gain noise described in Sec.~\ref{sec_act_gain_noise}A and subject of the measurement campaign reported in Sec.~\ref{sec_act_gain_noise}C.  

\item additive noise $v_j$ at the actuation carrier frequency will mix with the carrier waveform $V_{jc} \left( t \right)$ to ``down-convert'' into force in the LISA / LPF measurement band, part of the first term in Eqn.~\ref{eqn_estat_f_act} and described in Sec.~\ref{sec_act_gain_noise}A

\item additive noise $v_j$ in the LISA / LPF band will mix with DC voltages -- $V_{\mathrm{TM}0}$, the 2$^{\mathrm{nd}}$ term in Eqn.~\ref{eqn_estat_f_act} and various stray surface potentials $\delta V_m$, 4$^{\mathrm{th}}$~term in Eqn.~\ref{eqn_estat_f_act} -- to give in-band force noise.  The first contribution, coupling to TM potential and thus charge, is addressed in the measurement campaigns presented in Sec.~\ref{sec_inband_noise}. 
\end{itemize}

The role of applied actuation voltages, and their noise, in their contribution to the coupling to TM charge merits a comment here.  Following the notation of Refs.~\cite{charge_DC_bias_prl_2012} and \cite{LPF_prl_charge}, we define
\begin{equation}
\pder{F_x}{q} \equiv  - \frac{1}{C_T} \left| \pder{C_X}{x} \right| \Delta_x 
\label{eqn_Delta_x}
\end{equation} 
$\Delta_x$ essentially represents the average electrostatic field acting on the TM, normalized to an electrostatic potential applied to a single electrode housing $X$ electrode.   Considering Eqn.~\ref{estat_force} and both contributions from applied electrode voltages (covered by Eqn.~\ref{eqn_estat_f_act}), and from stray DC surface potentials (not covered by \ref{eqn_estat_f_act}), we find
\begin{eqnarray}
\Delta_x & \approx & \left[ \frac{1}{\left| \pder{C_X}{x} \right|}  
\sum_{m \left( S \right) , n \left( TM \right) }
{ \pder{C_{mn}}{x} \left( \delta V_m - \delta V_n \right)  }  \right]
\nonumber \\
  & &  +  \left[ V_1^{DC} + V_2^{DC} - V_3^{DC} - V_4^{DC} \right]
\label{eqn_Delta_x_calc}
\end{eqnarray} 
for a centered TM, where $V_j^{DC}$ is the DC actuation voltages on $X$ electrode $j$.  

We can thus see that the $X$ electrode applied DC voltages can be used to balance any ``intrinsic'' bias $\Delta_x$, in the top line of Eqn.~\ref{eqn_Delta_x_calc}, to null the effective coupling to the TM charge noise, as used in LPF\cite{LPF_prl_charge}. Likewise, any ``in-band'' noise in the same electrodes will create noise in the relevant average electrostatic field, 
\begin{equation}
\delta \Delta_x^{ACT} = v_1 + v_2 - v_3 - v_4
\label{eqn_dDelta_x_calc}
\end{equation}
which is object of the LPF charged TM experiment presented in Sec.~\ref{sec_inband_noise}.

\subsection{Appendix B}
	\subsubsection{Model of force noise from actuation gain fluctuations and data analysis technique}
\label{app_gain_noise}

This Appendix contains a detailed description of the actuation noise model and the MCMC fitting procedure. 

As anticipated in Eqns.~\ref{eqn_act_accel_coeff} and \ref{eqn_act_ang_accel_coeff}, we model the actuation contribution to the acceleration fluctuations in $\Delta g$ and $\Delta \gamma_{\phi}$ in terms of correlated/uncorrelated gain fluctuations $\alpha,\,\alpha_i,\,\alpha_{ij}$ through coefficients $a,\,a_i,\,a_{ij}$ for $\Delta g$ and with $b,\, b_i,\,b_{ij}$ for rotational acceleration $\Delta \gamma_{\phi}$.  The values of these coefficients are:
\begin{align}
	&a  =  2  \Delta g \nonumber \\
	&a_1  =  2 g_{\mathrm{c1}} \nonumber \\
	&a_2  =  -2 g_{\mathrm{c2}} \nonumber \\
	&a_{i1}  =\frac{(-1)^{i+1}}{2}
	\left( \bar{g}_{i\mathrm{c}} + g_{i0} + R^* \bar{\gamma}_{\phi_ic} 
	+ R^* \gamma_{\phi_{i0}}  \right) \nonumber \\
	&a_{i2}  = \frac{(-1)^{i+1}}{2}
	\left( \bar{g}_{i\mathrm{c}} + g_{i0} + R^* \bar{\gamma}_{\phi_ic} 
	- R^* \gamma_{\phi_{i0}}  \right) \nonumber \\
	&a_{i3}  = \frac{(-1)^{i+1}}{2}
	\left(- \bar{g}_{i\mathrm{c}} - g_{i0} + R^* \bar{\gamma}_{\phi_ic} 
	- R^* \gamma_{\phi_{i0}}  \right) \nonumber \\
	&a_{i4}  = \frac{(-1)^{i+1}}{2}
	\left(- \bar{g}_{i\mathrm{c}} - g_{i0} + R^* \bar{\gamma}_{\phi_ic} 
	+ R^* \gamma_{\phi_{i0}}  \right) \nonumber \\
	&	b  =  2  \Delta \gamma_{\phi} \nonumber \\
	&b_1  =  2 \gamma_{\phi_{\mathrm{c1}}} \nonumber \\
	&b_2  =  -2 \gamma_{\phi_{\mathrm{c2}}} \nonumber \\
	&b_{i1}  =  \frac{a_{i1}}{R^{*}} \qquad \qquad \,\,\,\,  b_{i4}  =  \frac{a_{i4}}{R^{*}} \nonumber \\
	&b_{i2}  =  -\frac{a_{i2}}{R^{*}} \qquad\qquad b_{i3}  =  -\frac{a_{i3}}{R^{*}} \label{eqn_ABC_coeff}
\end{align}
In our chosen actuation gain noise parametrization, discussed in Sec.~\ref{subsec_act_gain_model}, 
 $S_{\alpha}$, $S_{\alpha_i}$, $S_{\alpha_{ij}}$ are considered as mutually uncorrelated.  
 The PSD of $\Delta g$ and $\Delta \gamma_{\phi}$ at frequency $f$ can thus be expressed 

\begin{align}
	S_{\Delta g}^{ACT}(f)&=  A S_{\alpha}(f) + \displaystyle \sum_{i=1,2} A_i S_{\alpha_i}(f) + \displaystyle\sum_{ij} A_{ij} S_{\alpha_{ij}}(f)\notag \\
	S_{\Delta \gamma_{\phi}}^{ACT}(f)&=  B S_{\alpha}(f) + \displaystyle \sum_{i=1,2} B_i S_{\alpha_i}(f) + \displaystyle\sum_{ij} B_{ij} S_{\alpha_{ij}}(f)\notag \\
	S_{\Delta g,\Delta \gamma_{\phi}}^{ACT}(f)&=  C S_{\alpha}(f) + \displaystyle \sum_{i=1,2} C_i S_{\alpha_i}(f) + \displaystyle\sum_{ij} C_{ij} S_{\alpha_{ij}}(f)
	\label{eq:act_model}
\end{align} 

with
\begin{align}
	&A=a^2\,\quad A_i=a_i^2\,\quad A_{ij}=a_{ij}^2\notag \\ 
	&B=b^2\,\quad B_i=b_i^2\,\quad B_{ij}=b_{ij}^2 \notag\\
	&C=ab,\quad C_{i}=a_ib_i,\quad C_{ij}= a_{ij} b_{ij}\,.
\end{align} 
Equation~\ref{eq:act_model} holds in any configuration of forces and authorities for LPF, including the four experiments with enhanced actuation: each experiment has \textit{its own $A,\,B,\,C$ coefficients}, calculated from the averaged 
force/torque commands and authorities.  We introduce the index $q$ to label the four different experiments (UURLA, nominal, big, big+offset). Furthermore,
it is useful to re-write all the previous quantities in matrix formulation, for which we introduce
\begin{equation}
	\textbf{S}^{ACT}(f,q) = \begin{bmatrix*}[l]
		%	\textbf{M}(f,q)=\begin{bmatrix}
			S_{\Delta g}^{ACT}(f,q) & \quad S_{\Delta g,\Delta \gamma_{\phi}}^{ACT}(f,q) \\
			&\\
			S_{\Delta g,\Delta \gamma_{\phi}}^{ACT}(f,q) & \quad S_{\Delta \gamma_{\phi}}^{ACT}(f,q) 
		\end{bmatrix*} \,,
\end{equation}
% a non-zero imaginary part would be associated with the presence of common signals which appear delayed in one of the two time series. 
% Since it is highly reasonable to assume that a torque is the result of the application of a force with a certain arm, 
% there is no reason to assume that a force signal could leak in the angular measurement with a delay. 
Note that the off-diagonal terms (cross-correlations) are assumed to be real: in the hypothesis that net torques and forces arise from the same forces acting locally, then any force-torque correlation would be free of delays and thus result in a real value of the cross-spectrum.

We then also introduce the coefficients matrix
		\begin{equation}
	\textbf{a}(q) = \begin{bmatrix*}[l]
		%	\textbf{M}(f,q)=\begin{bmatrix}
			a & b \\
			a_1&b_1\\
			a_2&b_2\\
			a_{11}&b_{11}\\
			...&...\\
		    a_{24}&b_{24}
		\end{bmatrix*} \,,
	\end{equation}
and the noise generators matrix
		\begin{equation}
	\textbf{S}_{\alpha}(f) = \begin{bmatrix*}[l]
			%	\textbf{M}(f,q)=\begin{bmatrix}
				S_{\alpha} &  & && &\\
				&	S_{\alpha_1} &  & &0&\\
				& &	S_{\alpha_2} &  && \\
				& & & 	S_{\alpha_{11}} & &\\
				& 0& & &... &\\
				& & & & &S_{\alpha_{24}}\\
			\end{bmatrix*} \,,
		\end{equation}
actuation noise model (\ref{eq:act_model}) is then rewritten as
		\begin{equation}
			\textbf{S}^{ACT}(f,q) = \textbf{a}^T(q) \textbf{S}_{\alpha}(f)\textbf{a}(q)\,.
			\label{eq:mat_act_model}
		\end{equation}
$\textbf{a}(q)$ is a 11$\times$2 matrix, containing all the known coefficients in (\ref{eqn_ABC_coeff}), that depend only on the actuation forces (index q), while  $\textbf{S}_{\alpha}(f)$ is a diagonal 11$\times$11 matrix which contains the actuator gain noise PSDs at frequency $f$, which we will then want to extract by fitting the experimental data. 
		 
		We fit our experiments to the actuation noise model in Eqn.~\ref{eq:mat_act_model}, adding a possible background for force and torque noise, that is independent of the changes in actuation forces.  This background actually absorbs the first actuation term -- with coefficients $A$, $B$, and $C$ -- because the coupling to the relevant ``fully correlated'' gain noise $S_{\alpha}$ is essentially unchanged across these four tests.   The background acceleration noise term is described
		\begin{align}
			\textbf{S}^{bg}(f) &= \begin{bmatrix*}[c]
				S_{\Delta g}^{bg}(f)& 	\xi(f) \sqrt{	S_{\Delta g}^{bg}(f) S_{\Delta\gamma_{\phi}}^{bg}(f)}\\	\xi(f) \sqrt{	S_{\Delta g}^{bg}(f) S_{\Delta\gamma_{\phi}}^{bg}(f)}&
				S_{\Delta\gamma_{\phi}}^{bg}(f) 
			\end{bmatrix*} \notag 
		\end{align}
		This allows for backgrounds in $\Delta g$ and $\Delta \gamma$ that can have non-zero cross-coherence, which is modeled by parameter $\xi\in[-1, 1]$. The full modeled noise is summarized 
		\begin{equation}
			 \textbf{M}(f,q)=	\textbf{S}^{ACT}(f,q) + \textbf{S}^{bg}(f)\,.\label{eq:mod_bg}
		\end{equation}
%		\textcolor{blue}{
		
		The experimental data are $\sim$2 day long noise time series, sampled every $T=0.1\,$s, and indicated with $\Delta g[n,q]=\Delta g(t=nT,q)$ and $\Delta \gamma_{\phi}[n,q]=\Delta \gamma_{\phi}(t=nT,q)$, one couple for each of the four experiments. We divide these in $N_s(q)$, 50\% overlapping stretches (110$\times10^3\,$s, labeled with $s$) of length $N$ and multiply by a Blackman-Harris window $w[n]$. Then, we calculate the \textit{modified periodograms} as in Ref.~\cite{LorenzPhD}, for example for $\Delta g$ we have
		\begin{align}
		  X_{\Delta g}[k,q,s]&=\notag \sqrt{\frac{T}{N}} \sum_{n=0}^{N-1} \Delta g[n,q,s] w[n] e^{-i2\pi k\, n/N  }\\ &\equiv X_{\Delta g}(f=k/NT,q,s)	\,.
		\end{align}
		  The two periodograms are then grouped in the following complex column vector
		\begin{equation}
			\textbf{X}_s=\sqrt{2}\begin{bmatrix*}[c]
				X_{\Delta g}(f,q,s)\\ X_{\Delta \gamma_{\phi}}(f,q,s)
			\end{bmatrix*}\label{eq:periodograms}\,,
		\end{equation}
		 and the experimental estimate $\hat{S}$ of the \textit{single-sided cross spectral density} (CPSD) matrix  can be expressed as 
			\begin{align}
				\hat{\textbf{S}}(f,q)&=\begin{bmatrix*}[l]
					\hat{S}_{\Delta g}(f,q) &\quad \hat{S}_{\Delta g,\Delta \gamma_{\phi}}(f,q) \\
					&\\
					\hat{S}^*_{\Delta g,\Delta \gamma_{\phi}}(f,q) &\quad \hat{S}_{\Delta \gamma_{\phi}}(f,q) 
				\end{bmatrix*} =\langle \textbf{X}_s \, \textbf{X}_s^{\dagger}\rangle
				\label{eq:exp_data}
%				&=\bigg\langle\begin{bmatrix*}[l]
%					\Delta \tilde{g}(f,q) \\
%				    \Delta \tilde{\gamma}_{\phi}(f,q)
%				\end{bmatrix*}\begin{bmatrix*}[l]
%				\Delta \tilde{g}^*(f,q) &
%						\Delta \tilde{\gamma}^*_{\phi}(f,q)
%				\end{bmatrix*}\bigg\rangle_s \notag
				%&=\langle Z \, Z^{\dagger}\rangle_s \notag
		\end{align}
		where $\langle\rangle$ indicates an average over all periodograms.
		The way the stretches are chosen and weighted at each frequency allows to obtain different estimates of the spectra. In this study, we do it in two ways:
		\begin{itemize}
			\item ``Standard Welch method'': as in the standard Welch method \cite{welch}, the spectrum is calculated at each frequency with the same number of stretches $N_s=N_s(q)$, 3 for UURLA, 2 in other tests. The frequencies that can be considered uncorrelated using a Blackman-Harris window are $nf_{min}$, with $f_{min}=4/\Delta T$ and $n$ an integer \cite{fit_noise_prd}.  Data from the standard Welch technique are shown as continuous lines in various spectral plots, and we use these data for the smooth frequency-dependent fits described below (around Eqn.~\ref{eqn_smooth_functions}).   
			% \textbf{We anticipate here that PSDs calculated in this way are used for the MCMC fit to smooth functions: the considered frequencies are $f_n= nf_{min}$ to ensure that there is no correlation between PSDs at different frequencies (introduced by BH window)}.
			\item ``Minimally correlated frequencies method'': this technique is used to calculate $\hat{\textbf{S}}$ with a number of data stretches that increases with frequency, having therefore $N_s=N_s(f,q)$.  This allows us to have data points with roughly equal spacing in $\log \left( f \right)$, maintaining  minimal correlation.   The frequencies are given by (see Supp.~Material, Ref.~\cite{LPF_prl_2018}) 
				\begin{equation}
					f_1=\frac{4}{\Delta T},\, f_2=2f_1,\, f_3=\left(\frac{5}{3}\right)f_2,\, f_4=\left(\frac{5}{3}\right)f_3, \, ...
				\end{equation}  	
			All of our single-frequency fits, described below, use CPSD data calculated with this method.
		\end{itemize}  
		%$S_{\Delta g}^{exp,q}(f)$, $S_{\Delta \gamma_{\phi}}^{exp,q}(f)$ and $S_{\Delta g,\Delta \gamma_{\phi}}^{exp,q}(f)$ quantities do not contain just actuation noise, but also other sources (contributing significantly only in UURLA run), which are kept into account in our model adding some background terms:
		%\begin{align}
		%	S_{\Delta g}^{mod,q}(f) = S_{\Delta g}^{ACT,q}(f) + S_{\Delta g}^{bg}(f) \notag\\
		%	S_{\Delta \gamma_{\phi}}^{mod,q}(f) = S_{\Delta \gamma_{\phi}}^{ACT,q}(f) + S_{\Delta \gamma_{\phi}}^{bg}(f) \notag \\
		%	S_{\Delta g, \Delta \gamma_{\phi}}^{mod,q}(f) = S_{\Delta g,\Delta \gamma_{\phi}}^{ACT,q}(f)\,.
		%	\label{eqn_model_fit} 
		%\end{align}
		%where the superscript $q$ was added also to $S^{ACT}$ terms in order to stress that they depend on the experiment.
		%It can be noticed that the background terms $S_{\Delta g}^{bg}$, $S_{\Delta \gamma_{\phi}}^{bg}$ are assumed to be uncorrelated, therefore they do not appear in the cross correlation, which is consequently determined by actuation only in this model. 

Our experimental data (Eqn.~\ref{eq:exp_data}) are described by the model in Eqn.~\ref{eq:mod_bg}, at a given frequency $f$ by 13 parameters: actuation noises PSD $S_{\alpha_i}$ and $S_{\alpha_{ij}}$, plus acceleration noise backgrounds $S_{\Delta \gamma_{\phi}}^{bg}$ and $S_{\Delta g}^{bg}$ with the background correlation coefficient and $\xi$. As mentioned above fully correlated noise $S_{\alpha}$ is considered as a contribution to the unchanging background acceleration noise.\\
%\textcolor{red}{ The quantities in Eqn.~\ref{eq:exp_data} are distributed as $\chi^2$ and are strictly positive, therefore we can not use a simple gaussian fitting procedure to estimate the parameters: we use a Bayesian approach instead.}
		
We estimate these parameters with a Bayesian approach. We indicate with $\hat{\textbf{D}}$ the collection of all data on which we want to perform the parameter estimation with, for example, all the CPSD estimates at a single frequency $\hat{\textbf{D}}(f)=\{\hat{\textbf{S}}(f,q)\}_{q}$.  
The posterior distribution of our model conditioned by the observed data can then be derived using Bayes theorem
	\begin{equation}
			p(\textbf{M}|\hat{\textbf{D}})= \frac{L(\hat{\textbf{D}}|\textbf{M})P(\textbf{M})}{P(\hat{\textbf{D}})}\,.\label{eq:Bayes_theorem}
		\end{equation}
The likelihood $L(\hat{\textbf{D}}|\textbf{M})$ is obtained following Refs.~\cite{LorenzPhD} and \cite{buonuomo}. Considering a single realization of $\textbf{X}_s$ as defined in (\ref{eq:periodograms}), this follows a  a 2-variate complex Gaussian probability distribution 
			\begin{equation}
				p_{f,q}(\textbf{X}_s|\textbf{M}) = \frac{1}{\pi^2|\textbf{M}|}\exp\left\{-\textbf{X}_s^{\dagger} \textbf{M}^{-1}\textbf{X}_s \right\}\,. \label{eq:2variate}
			\end{equation}
		Then, in Ref.~\cite{buonuomo} it is shown that the CPSD matrix defined in (\ref{eq:exp_data}) is distributed according to a \textit{Complex Wishart distribution}
			\begin{equation}
			p_{f,q}(\hat{\textbf{S}}|\textbf{M} ) = \frac{N_s^{N_s}|\hat{\textbf{S}}|^{N_s-2}}{\widetilde{\Gamma}_2(N_s)|\textbf{M}|^{N_s}} \text{etr}\left[-N_s\textbf{M}^{-1}\hat{\textbf{S}}\right] 
			\label{eqn_60}
		\end{equation}	
		where $\text{etr}[\cdot]=\text{exp}(\text{tr}[\cdot])$ and $\widetilde{\Gamma}$ is the multivariate complex Gamma function.
			Summarizing, Eqn.~\ref{eqn_60} gives the probability of observing a certain value for matrix $\hat{\textbf{S}}$ given a theoretical CPSD $\textbf{M}$, at a single frequency, using $N_s$ periodograms of $\Delta g$ and $\Delta \gamma_{\phi}$. 
		Finally, the total probability of observing $\hat{\textbf{D}}(f)=\{\hat{\textbf{S}}(f,q)\}_{q}$ at a single frequency is
		\begin{equation}
			L_f(\hat{\textbf{D}}|\textbf{M})=  \prod_q	\label{eqn_likeli} p_{f,q}(\hat{\textbf{S}}|\textbf{M})
		\end{equation}
		Starting from this likelihood, we perform two types of Markov Chain Monte Carlo sampling (MCMC) to estimate the posterior distributions of our model parameters using Eqn.~\ref{eq:Bayes_theorem}: first at discrete frequencies -- without assuming any specific model of frequency dependence of noise -- and second with a phenomenological analytic model of frequency dependence.
		In the first fit we consider data at a single frequency $f$ of the ``minimally correlated frequencies'' introduced above. The ten elements in $\textbf{S}_{\alpha}(f)$, plus background parameters $S_{\Delta \gamma_{\phi}}^{bg}(f)$, $S_{\Delta g}^{bg}(f)$ and $\xi(f)$  are taken as the parameters in model Eqn.(\ref{eq:mod_bg}) and samples from their joint posterior distribution are drawn using a standard Metropolis-Hastings MCMC algorithm with adaptive covariance. Priors used for these parameters are discussed in the following.
	To be fully agnostic in the \textit{a priori} assumptions used in the fit, an ideal non-informative prior \cite{jeffrey} for each of the twelve noise PSD could consist in a uniform (flat) prior on their logarithm -- eg $\log{S_{\alpha_{11}}}$ -- essentially allowing \textit{any} order of magnitude of the noise spectra. However, various parameters in our fit are essentially unresolved in our data.  This is due to a combination of a relatively large ratio of parameters (total: 13) to the number of experimental data (12, for 3 channels between $\Delta g$, $\Delta \gamma_{\phi}$ and their cross-correlation, in 4 different measurements), to some near degeneracies in the coefficients \textbf{A}(q) and to the small values of some of the spectra.  As such, the MCMC fit admits solutions in which some parameters are compatible with zero, or with logarithmic values that can diverge to $-\infty$ -- when the parameter becomes \textit{small}, the likelihood in Eqn.~\ref{eqn_likeli} becomes insensitive to just \textit{how small} and the Markov chain convergence becomes an issue.  
		
		Different approaches exist for solving this problem, including reducing the number of model parameters and inserting lower limits (cutoffs) in the \textit{a priori} assumptions on the parameters.  We have chosen a physically-motivated approach that reparametrizes the actuation noise in terms of just two \textit{average} noise PSD, for board correlated gain noise ($S_{\alpha_\mathrm{C}}$) and uncorrelated single electrode gain noise ($S_{\alpha_{\mathrm{UC}}}$), and a series of parameters $\mu_{m}$ that describe the \textit{difference} in the noise between different noise generators: 
		\begin{align}
			&S_{\alpha_1}= S_{\alpha_{\mathrm{C}}} (1+\mu_c)\notag\\ 
			&S_{\alpha_2}= S_{\alpha_{\mathrm{C}}} (1-\mu_c)\notag\\
			&S_{\alpha_{11}}= S_{\alpha_{\mathrm{UC}}} (1+\mu_{\pm})(1+\mu_{tb+})(1+\mu_{1})\notag\\ 
			&	S_{\alpha_{12}}= S_{\alpha_{\mathrm{UC}}} (1+\mu_{\pm})(1-\mu_{tb+})(1+\mu_{2})\notag\\
			&		S_{\alpha_{13}}= S_{\alpha_{\mathrm{UC}}} (1-\mu_{\pm})(1-\mu_{tb-})(1+\mu_{3})\notag\\
			&					S_{\alpha_{14}}= S_{\alpha_{\mathrm{UC}}} (1-\mu_{\pm})(1+\mu_{tb-})(1+\mu_{4})\notag\\
			&					S_{\alpha_{21}}= S_{\alpha_{\mathrm{UC}}} (1+\mu_{\pm})(1+\mu_{tb+})(1-\mu_{1})\notag\\ 
			&				S_{\alpha_{22}}= S_{\alpha_{\mathrm{UC}}} (1+\mu_{\pm})(1-\mu_{tb+})(1-\mu_{2})\notag\\
			&				S_{\alpha_{23}}= S_{\alpha_{\mathrm{UC}}} (1-\mu_{\pm})(1-\mu_{tb-})(1-\mu_{3})\notag\\
			&				S_{\alpha_{24}}= S_{\alpha_{\mathrm{UC}}} (1-\mu_{\pm})(1+\mu_{tb-})(1-\mu_{4})\,.\label{eq:reparametrization}
		\end{align}
		where all parameters $\mu_m$ are constrained to the interval $\mu_m \in \left[-1,1 \right]$. For instance, $\mu_{\pm} = 0$ would imply that the $+ X$ and $-X$ actuators are equally noisy, while $\mu_{\pm}=\pm 1$ implies that the total sum of the uncorrelated gain noise comes from the $+X$ (or -$X$) actuators, with the other group completely noiseless.  We keep the uninformative log-flat prior in the PSDs $S_{\alpha_{\mathrm{C}}}$ and $S_{\alpha_{\mathrm{UC}}}$, with however a uniform (flat) prior for the $\mu_m$ across the interval $\left[ -1 , 1 \right]$.  This allows exploring orders of magnitude in the average noise, but makes it improbable that, for instance, a single electrode actuator might be many orders of magnitude noisier than another electrode (or that the board correlated gain noise for TM1 be many orders of magnitude quieter than that for TM2).  As the electrode actuation circuits are nominally identical, we consider this a physically reasonable hypothesis.
		%, in addition to a simplification that also renders the MCMC fit convergence quicker and more reliable.  
		
		In this parametrization we write the combined actuator quantities that are presented in Figs.~\ref{fig_S_alpha} and \ref{fig_cross}, considering their definitions in Eqns.~\ref{eqn_S_UC_ave} and \ref{eqn_S_UC_14_23}:
		\begin{small}
			\begin{align}
				& S_{\alpha_{\mathrm{UC}+}} = S_{\alpha_{\mathrm{UC}}} \left( 1+\mu_{\pm} \right) \\
				& S_{\alpha_{\mathrm{UC}-}} = S_{\alpha_{\mathrm{UC}}} \left( 1-\mu_{\pm} \right) \notag \\
				& S_{\alpha_{\mathrm{UC}14}} = S_{\alpha_{\mathrm{UC}}} 
				\left[ \left( 1 + \mu_{\pm} \right) \left( 1 + \mu_{tb+} \right) + \left( 1 - \mu_{\pm} \right) \left( 1 + \mu_{tb-} \right)  \right] \notag \\
				& S_{\alpha_{\mathrm{UC}23}} = S_{\alpha_{\mathrm{UC}}} 
				\left[ \left( 1 + \mu_{\pm} \right) \left( 1 - \mu_{tb+} \right) + \left( 1 - \mu_{\pm} \right) \left( 1 - \mu_{tb-} \right)  \right] \notag 
			\end{align}
		\end{small}

\begin{figure}[t]
			\includegraphics[width=3.2in]{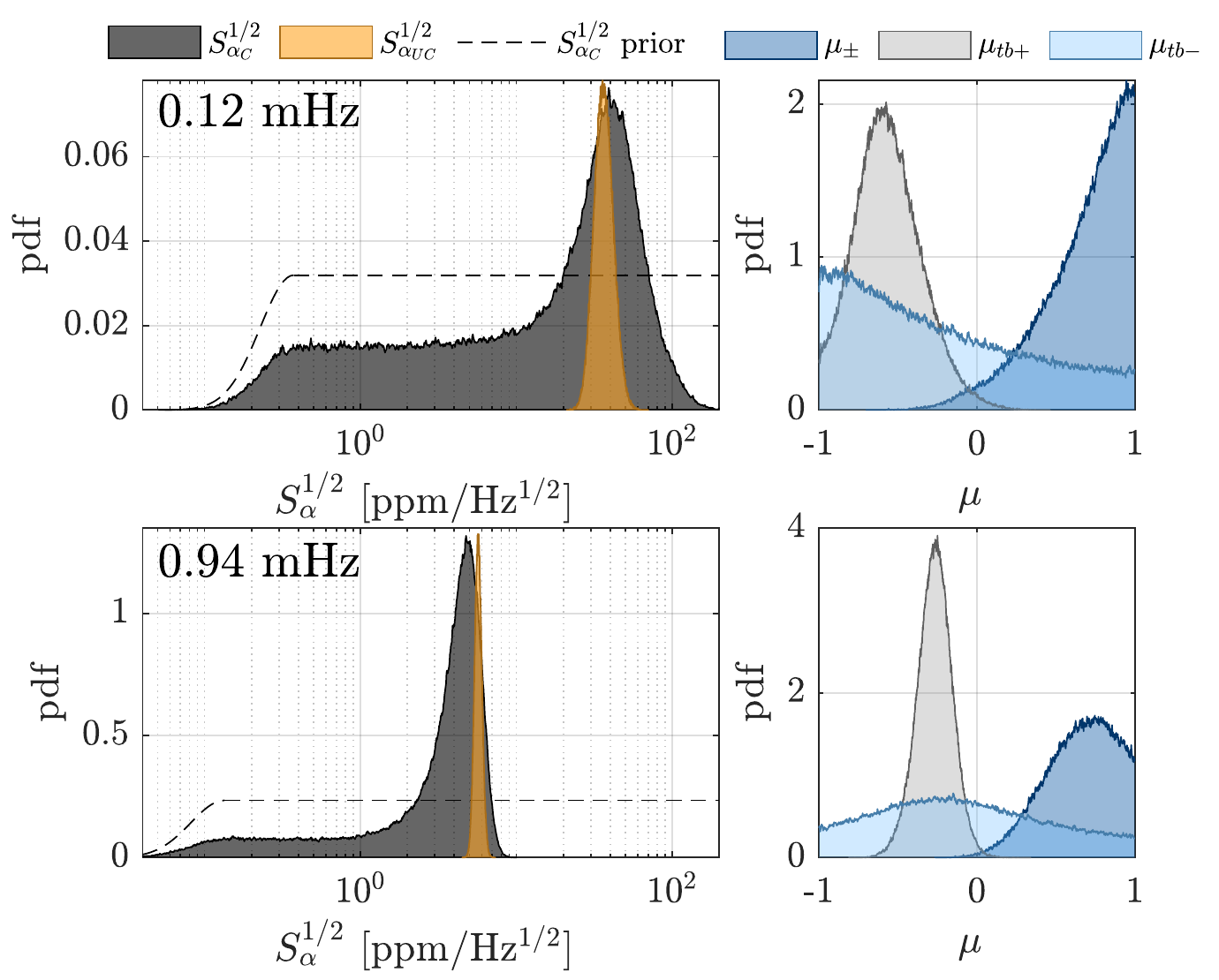}
			\caption{Examples of distributions of $S_{\alpha_{\mathrm{UC}}}$ (orange) and $S_{\alpha_{\mathrm{C}}}$ (black) obtained from the MCMC fit.  At both frequencies the uncorrelated electrode gain noise $S_{\alpha_{\mathrm{UC}}}$ is robustly resolved from the experimental 
data and largely independent to any prior assumptions on the distribution, while the lower tail of the posterior for  ``board correlated'' gain noise $S_{\alpha_{\mathrm{C}}}$ is only limited, in the lower values, by the prior. To allow better visualization, the former is normalized to 1, the latter to 4. The dashed line instead gives an idea of the $S_{\alpha_{C}}$ cutoff: its values have no particular meaning in this plot and should only help visualizing its position and shape.  At right are distributions, at the same two 
frequencies, for the partitioning of the levels of uncorrelated gain noise between different groups of electrodes. }
			\label{fig_distr_Salpha} 
		\end{figure}

		Figure \ref{fig_distr_Salpha} shows an example of distributions of $S_{\alpha_{\mathrm{C}}}$ and $S_{\alpha_{\mathrm{UC}}}$, raw parameters in the parametrization of Eqn.~\ref{eq:reparametrization} obtained from the MCMC at two relevant frequencies.
		
		Even though the group of experiments is sensitive to $S_{\alpha_{\mathrm{C}}}$, the data are compatible with the hypothesis that $S_{\alpha_{\mathrm{C}}}$ is zero (and that the overall acceleration noise is explained by uncorrelated fluctuations). Also $S_{\Delta\gamma_{\phi}}^{bg}$ seems to be compatible with zero at the two lowest frequencies. As a result, for these parameters a lower cutoff was still needed to ensure convergence; this was introduced using a smooth, improper prior of the form
		\begin{equation}
			P \left( S_{\alpha_{\mathrm{C}}} \right) \propto
			\begin{cases}
				\exp\left[-\frac{(\log S_{\alpha_{\mathrm{C}}} - \log \bar{S}_{co})^2}{2\sigma_{co}^2}\right]\quad &\text{if}\quad S<\bar{S}_{co}\\
				1\quad &\text{if}\quad S>\bar{S}_{co}
			\end{cases}
		\end{equation}
		where we chose $\sigma_{co}\approx 1$ and $\bar{S}_{co}$ as 1/100 of the noise (power) declared on the data-sheet for the components associated with $S_{\alpha_{\mathrm{C}}}$ \cite{lt1021}. A cutoff of the same form was introduced for $S_{\Delta\gamma_{\phi}}^{bg}$, with the corresponding $\bar{S}_{\Delta\gamma_{\phi}}^{bg}$ taken as 1/100 of the observed residual gas Brownian noise level, which is very well understood and constitutes a solid lower limit \cite{PRL2009}. We chose to take 1/100 of these physically motivated lower limits to be as agnostic as possible in the parameter estimation. Figure \ref{fig_distr_Salpha} again shows that the the lower tail of the distribution associated with these parameters is completely determined by the prior (therefore not constrained by observations), as the likelihood (Eqn.~\ref{eqn_likeli}) becomes insensitive to $S_{\alpha_{\mathrm{C}}}$ when this is too small.  While lower bounds are not resolved, upper bounds are instead very well constrained, as demonstrated by the complete independence of the distribution peak position (and width) on the cutoff choice. 		
	%	\textcolor{blue}{We take as an upper limit on $S_{\alpha_C}$ the value where the distribution is reduced of a factor $1/e$ with respect to the peak value, which is equally robust.} 
		Figure \ref{fig_distr_Salpha} also shows how $S_{\alpha_{\mathrm{UC}}}$ is instead very well resolved (left panel, in orange), and some examples of relevant $\mu$ parameters (right panel). $\mu_{\pm}$ expresses the difference between $+X$ and $-X$ actuators, while $\mu_{tb+}$ between $+X$ actuators which contribute positively/negatively to cross-correlation: they are both well constrained. $\mu_{tb-}$ instead (analogous of $\mu_{tb+}$ for $-X$), is almost flat, in first place due to the fact that $-X$ electrodes happened to be less noisy than $+X$; then, also because we lack a second "big+offset" experiment enhancing only negative electrodes. One last mention to the $\mu_1\,,\mu_3\,,\mu_3\,,\mu_4$ parameters, which represent the difference between two electrodes in the same position of different TMs: these are basically not resolved, mainly due to the degeneracy of their $\textbf{A}(q)$ coefficients.
		In this paper, estimates of the parameters are often reported by points with error bars (the central point is the 50 percentile while bars indicate the equally-tailed 68\% confidence interval), that can be seen for example in Figs.~\ref{fig_s_alpha_exper} and \ref{fig_S_alpha}. While we consider our priors to be both conservative and physically motivated, we give a visual indication of the data points for model parameters in the cases where the lower limit of the posterior distribution is strongly dependent on the prior assumptions and not strongly constrained by the experimental data. 
We can define an effective lower limit $S_{ll} \left( \bar{S}_{co}\right) $ as the PSD value at the $-1 \sigma$~level
(15.9\% percentile) of the distribution.   In presence of important tails like $S_{\alpha_{\mathrm{C}}}$ in Figure \ref{fig_distr_Salpha}, if the prior cutoff ($\bar{S}_{co}$) is lowered of a factor 10, the lower limit $S_{ll}\left( \bar{S}_{co} \right)$ behaves almost exactly as if the posterior distribution were uniform (in log), decreasing by $\sim 86$\% in linear scale. We adopted, as a rule of thumb, the criteria for marking with a dash-dot line, to indicate sensitivity to the lower-cutoff prior, all the parameters estimates for which
		\begin{equation}
			\frac{ S_{ll}(\bar{S}_{co}/10)}{ S_{ll}(\bar{S}_{co})}< \frac12\,.
		\end{equation}
		For these parameters, the distribution upper limit is instead much less sensitive to the prior cutoff choice, varying typically by order several percent and never more than 10\% for a factor 10 increase in the upper-cutoff.
		
	% (Chiavegato's 3dB per decade rule)	
	
		The regularity of the results obtained analysing each frequency independently suggest that the whole behaviour can be safely described also using smooth analytical functions of the frequency $f$, specifically
		\begin{align}
			S_{\alpha_{i}}(f) &= \frac{\epsilon_i}{f^2+f_{\mathrm{cut}}^2} +\frac{\rho_i}{f}\notag \\
			S_{\alpha_{ij}}(f) &= \frac{\epsilon_{ij}}{f^2+f_{\mathrm{cut}}^2} +\frac{\rho_{ij}}{f}\notag \\
			S_{\Delta g}^{bg}(f) &=  \frac{\epsilon_{\Delta g}^{bg}}{f^2}+ \rho_{\Delta g}^{bg}\notag \\
			S_{\Delta \gamma_{\phi}}^{bg}(f) &=  \frac{\epsilon_{\Delta \gamma_{\phi}}^{bg}}{f}+ \rho_{\Delta \gamma_{\phi}}^{bg} f^2\, .
			\label{eqn_smooth_functions}
		\end{align}
		The analytic form was chosen empirically based on the measured PSD. After re-parametrizing $\epsilon,\,\rho$ in Eqn.~\ref{eqn_smooth_functions} similarly to Eqn.~\ref{eq:reparametrization}, the MCMC runs on a total of 26 parameters. While the previous fit was performed at a single frequency, here all bins are fit at once, namely $\hat{\textbf{D}}=\{\hat{\textbf{S}}(f,q)\}_{f,q}$ at the frequencies, roughly 50, set by the standard Welch periodogram length. The likelihood function is exactly the same as Eqn.~\ref{eqn_likeli}, but multiplied over all frequencies, since they are all independent:
\begin{equation}
	L(\hat{\textbf{D}}|\textbf{M})=  \prod_{f,q}	\label{eqn_likeli_smooth} p_{f,q}(\hat{\textbf{S}}|\textbf{M})
\end{equation}
Also in this case, logarithmic uniform priors were put on $\epsilon, \rho$ parameters related to average uncorrelated noise, correlated noise and background, while uniform priors were used for $\mu$'s expressing difference between different noise generators, and $\xi$ (cross-background parameter). A lower cutoff was needed just for the two parameters associated with correlated noise,  and was fixed to 1/100 of data-sheet values. \\

To test  the goodness of our fits, we employed a posterior predictive check \cite{gelman2020bayesian}. Assuming the distribution of the model parameters ($\textbf{M}$) obtained from the MCMC, we derive the distributions for $\hat{\textbf{D}}$ and then evaluate the compatibility 
with the the experimental data. To do this, for each value of $\textbf{M}$ accepted in the Markov Chain we sample values of $\textbf{X}_s$ -- the 
Fourier components in $\Delta g$ and $\Delta \gamma_{\phi}$ extracted randomly for a finite length experiment -- from Eqn.~\ref{eq:2variate} by diagonalizing $M^{-1}$ and expressing the elements of $\textbf{X}_s$ in terms of independent variables, which can be easily sampled \footnote[4]{This 2-variate complex Gaussian distribution can be equivalently sampled by extracting real and imaginary parts of $\textbf{X}_s$ from a 4-variate (real) Gaussian distribution \cite{buonuomo}.}, then $\hat{\textbf{S}}$ is calculated through Eqn.~\ref{eq:exp_data}. Once we have distributions for $\hat{\textbf{S}}$, we calculate the $\pm\sigma$ intervals for $\hat{S}_{\Delta g}$, $\hat{S}_{\Delta \gamma_{\phi}}$, $\hat{S}_{\Delta g,\Delta \gamma_{\phi}}$ at all frequencies, for each experiment. If the fit is indeed "good", then we should find that $\approx 68$\% of the experimental data fall within the posterior predicted intervals. Considering the fit performed at single frequencies, we find that $\approx 85$\% of experimental data fall in the predicted intervals; this may indicate that we are slightly over-fitting by using 13 parameters at each frequency, with 12 experimental data (four separate experiments with three data, with PSD estimates for $\Delta g$, $\Delta \gamma_{\phi}$, and their cross-spectrum). With regard to the fit to smooth functions instead, we find 68\% of experimental data falling in the posterior prediction: this result is consistent even when not considering the whole data-set, for example when estimating the goodness of fit in a single experiment, or only considering noise in the translational acceleration $\Delta g$, indicating that the model is able to properly reproduce the observations.

\subsubsection{Model of force noise from in-band fluctuations and data analysis technique}
The analysis of "in-band" voltage fluctuations is carried out analogously to the previous subsection. We report here a quick summary of the models used. We have only two experiments, with $V_{TM}\approx 0\,$V and then with $V_{TM}\approx 1\,$V, which we label again with index $q$. In general our model for acceleration/torque PSDs and cross spectra can be written as in Eqn.~\ref{eqn_S_DDx}
\begin{align}
	S_{\Delta g} &= S_{\Delta g}^{bg} +\left(\frac{\frac{\partial C_X}{\partial x}V_{TM}}{M}\right)^2 S_{\Delta\left(\Delta_x\right) } \notag\\
	S_{\Delta \gamma_{\phi}} &= S_{\Delta \gamma_{\phi}}^{bg} +\left(\frac{\frac{\partial C_X}{\partial\phi}V_{TM}}{I}\right)^2 S_{\Delta\left(\Delta_{\phi}\right) } 	\label{eqn:inband_PSD}\\
	S_{\Delta g,\Delta \gamma_{\phi}} &= S_{\Delta g,\Delta \gamma_{\phi}}^{bg} +\frac{\frac{\partial C_X}{\partial x}\frac{\partial C_X}{\partial\phi}V_{TM}^2}{IM} S_{\Delta\left(\Delta_x\right),\Delta\left(\Delta_{\phi}\right) } \notag
\end{align}
As described in the text, an estimate of $S_{\Delta(\Delta_x)}$ can be extracted by simply fitting the first line in Eqn.~\ref{eqn:inband_PSD} to the observed $S_{\Delta g}$ using $S_{\Delta(\Delta_x)}$ and $S^{bg}_{\Delta g}$ as parameters, and using the one dimensional version of likelihood (\ref{eqn_likeli}). Same holds for $S_{\Delta(\Delta_{\phi})}$ and the second line of Eqn.~\ref{eqn:inband_PSD}. This was done at single frequencies and also fitting smooth functions: in the first case, given the very small amount of periodograms available (therefore wide PSD distributions), a lower cutoff was needed on  $S_{\Delta(\Delta_x)}$ and  $S_{\Delta(\Delta_{\phi})}$, fixed to the \textit{very} conservative 1/100 thermal limit of actuation circuits.
  In the hypothesis of stray electrostatics noise dominated by uncorrelated
fluctuations in the electrode actuation
voltages, we can summarize the PSD model in Eqn.~\ref{eqn:inband_PSD}, similarly to Eqn.~\ref{eq:mod_bg}, as
\begin{equation}
	\textbf{M}(f,q)= \textbf{v}^T(q)	\textbf{S}_{\Delta(\Delta)}(f)\textbf{v}(q) + \textbf{S}^{bg}(f)\,
\end{equation}
where $\textbf{S}^{bg}(f)$ is the same as that in Eqn.~\ref{eq:mod_bg}, while
\begin{small}
	\begin{align}
		&\textbf{v}(q)=V_{TM}(q)\begin{bmatrix*}[c]
			\frac{\partial C_X}{\partial x}/M & \frac{\partial C_X}{\partial\phi}/I \\
		-	\frac{\partial C_X}{\partial x}/M& \frac{\partial C_X}{\partial\phi}/I & 
		\end{bmatrix*}\notag\\
		&\textbf{S}_{\Delta(\Delta)}(f) = \begin{bmatrix*}
			S_{\Delta(\Delta_{14})} (f)& 0 \\
			0&S_{\Delta(\Delta_{23})} (f)
		\end{bmatrix*}
		\label{eqn:stary_spectra}
	\end{align}
\end{small}
Finally, consistently with the approach for gain noise, we again use the hypothesis of ``statistically identical'' electrodes, writing
\begin{align}
	S_{\Delta(\Delta_{14})} = S_{\Delta} (1+\mu_{b}) \notag\\
	S_{\Delta(\Delta_{23})} = S_{\Delta} (1-\mu_{b})
	\label{eqn:mu_band}
\end{align}
At each minimally correlated frequency, we use the same MCMC described in the previous section with likelihood (Eqn.~\ref{eqn_likeli}) to estimate the distributions of the parameters $S_{\Delta}$, $S^{bg}_{\Delta g}$, $S^{bg}_{\Delta \gamma_{\phi}}$ with a prior uniform in the logarithm, while $\mu_{b}$ and $\xi$ (expressing the correlation between backgrounds as in Eqn.~\ref{eq:mod_bg}) with a uniform prior distribution across the interval $[-1, 1]$. 

We also perform a fit describing the relevant quantities as smooth functions of frequency. While backgrounds are written as in (\ref{eqn_smooth_functions}), we write the other quantities as
\begin{align}
	S_{\Delta(\Delta_{14})} = \frac{\epsilon_{14}}{f^2} +\frac{\rho_{14}}{f} +\eta_{14}\notag\\
	S_{\Delta(\Delta_{23})} = \frac{\epsilon_{23}}{f^2} +\frac{\rho_{23}}{f} +\eta_{23}
\end{align}
and MC estimate $\epsilon,\,\rho,\,\eta$ after a a re-parametrization similar to Eqn.~\ref{eqn:mu_band}. The results indicate that only the $1/f$ component is significantly different from zero.

\bibliography{act_paper_bib.bib}
\end{document}